\documentclass[graybox,natbib,norunningheads]{svmult}

\usepackage{mathptmx}       
\usepackage{helvet}         
\usepackage{courier}        
\usepackage{type1cm}        
\usepackage{makeidx}         
\usepackage{graphicx}        
\usepackage{multicol}        
\usepackage[bottom]{footmisc}

\makeindex             

\def\ion#1#2{{\rm#1#2}}
\def\Msun{\dim{M}_{\odot}}
\def\Ht{{\ion{H}{}}}
\def\HI{{\ion{H}{I}}}
\def\HII{{\ion{H}{II}}}
\def\HeI{{\ion{He}{I}}}
\def\HeII{{\ion{He}{II}}}
\def\HeIII{{\ion{He}{III}}}
\def\H2{{{\rm H}_2}}
\def\CVI{\ion{C}{VI}}
\def\CO{\ion{CO}{}}

\def\Zsun{Z_\odot}

\def\ga{\mathrel{\hbox{\rlap{\hbox{\lower4pt\hbox{$\sim$}}}\hbox{$>$}}}}
\def\la{\mathrel{\hbox{\rlap{\hbox{\lower4pt\hbox{$\sim$}}}\hbox{$<$}}}}

\def\dim#1{\mbox{\,#1}}

\begin{document}

\title*{Modeling Physical Processes at Galactic Scales and Above}
\author{Nickolay Y.\ Gnedin}
\institute{Nickolay Y.\ Gnedin \at Particle Astrophysics Center, 
Fermi National Accelerator Laboratory, Batavia, IL 60510, USA, \email{gnedin@fnal.gov};
Kavli Institute for Cosmological Physics and Department of Astronomy \& Astrophysics, The University of Chicago, Chicago, IL 60637 USA}
%
%
\maketitle

\tableofcontents

\section{In Lieu of Introduction}

What should these lectures be? The subject assigned to us is so broad that many books can be written about it. So, in planning these lectures I had several options. 

One would be to focus on a narrow subset of topics and to cover them in great detail. Such a subset necessarily would be highly personal and useful to a few readers at best. Another option would be to give a very shallow overview of the whole field, but then it won't be very much different from a highly compressed version of a university course (which anyone can take if they wish so).

So, I decided to be selfish and to prepare these lectures as if I was teaching my own graduate student. Given my research interests, I selected what the student would need to know to be able to discuss science with me and to work on joint research projects. So, the story presented below is both personal and incomplete, but it does cover several subjects that are poorly represented in the existing textbooks (if at all).

Some of topics I focus on below are closely connected, others are disjoint, some are just side detours on specific technical questions. There is an overlapping theme, however. Our goal is to follow the cosmic gas from large scales, low densities, (relatively) simple physics to progressively smaller scales, higher densities, closer relation to galaxies, and more complex and uncertain physics. So, we (you - the reader, and me - the author) are going to follow a ``yellow brick road'' from the gas well beyond any galaxy confines to the actual sites of star formation and stellar feedback. On the way we will stop at some places for a tour and run without looking back through some others. So, the road will be uneven, but I hope that some readers find it useful.

\newcounter{btcount}
\def\bt#1{
\stepcounter{btcount}
\begin{svgraybox}
{\bf Brain teaser \#\thebtcount}: #1
\end{svgraybox}
}

\abstract*{The main emphasis of this chapter is on modeling the IGM (in particular, Lyman-$\alpha$ forest. For that purpose we will review some of the basic physics of low density photo-ionized gas and its observational diagnostics. We will explorer how thermal pressure makes gas distribution deviate from the dark matter, how the gas temperature evolves in time under the effect of cosmic ionizing background, how observed spectra emerge from the interplay of density, temperature, and velocity. We will briefly review most important observations of the IGM and will end up with a mystery.}

\section{Physics of the IGM}
\label{igm}

Most of the volume of the universe is occupied by gas outside galaxies, the so-called intergalactic medium (IGM). It may seem this gas is located far from galaxies, and should not be relevant to formation of galaxies and stars. Wrong! - IGM is the gas that eventually gets accreted by galaxies and turns into stars. After all, before the first galaxy formed, the whole universe was just IGM.

Hence, as we follow the ``yellow brick road'' to our goal of modeling star formation in galaxies, we pass through the IGM land first...

\subsection{Linear Hydrodynamics in the Expanding Universe}

Linear dynamics of the non-relativistic cold dark matter is almost trivial, density fluctuation $\delta_X(t,k)$ with a spatial wavenumber $k$ satisfies a simple ordinary differential equation (ODE),
\begin{equation}
  \frac{d^2}{dt^2}\delta_X(t,k) + 2H\frac{d}{dt}\delta_X(t,k) = 4\pi G\bar\rho\delta_{\rm tot}(t,k),
  \label{eq:linx}
\end{equation}
where $a(t)$ is the cosmic scale factor, $H(t)\equiv \dot{a}/a$ is the Hubble parameter and $\bar\rho$ is the mean density of the universe. If the universe only contained cold dark matter, then $\delta_{\rm tot}=\delta_X$. A second order  ODE has two solutions, one of them is always growing with time,
\begin{equation}
  \delta_X(t,k) = D_+(t) \delta_0(k),
  \label{eq:linxsol}
\end{equation}
where $D_+$ is called "the linear growing mode".

In reality, the universe contains gas, which is also subject to pressure forces. Hence, in the linear regime the evolution of the dark matter and gas fluctuations $(\delta_X,\delta_B)$ is described by a system of two coupled equations,
\begin{eqnarray}
  \frac{d^2\delta_X}{dt^2} + 2H\frac{d\delta_X}{dt} & = & 4\pi G\bar\rho\left(f_X\delta_X+f_B\delta_B\right),\\ 
  \frac{d^2\delta_B}{dt^2} + 2H\frac{d\delta_B}{dt} & = & 4\pi G\bar\rho\left(f_X\delta_X+f_B\delta_B\right) - \frac{c_S^2}{a^2}k^2\delta_B,
  \label{eq:linxb}
\end{eqnarray}
where $f_X\approx 0.84$ and $f_B\approx 0.16$ are the mass fractions of dark matter and baryons respectively, and $c_S$ is the speed of sound in the gas.

This system of equations is coupled, but if high precision is not required, one can assume $f_B\ll f_X$ and ignore the baryonic contribution in the gravity terms in both equations. In that case the solution for the dark matter fluctuation is still given by equation (\ref{eq:linxsol}), while the equation for the baryonic fluctuation reduces to
\begin{equation}
  \frac{d^2\delta_B}{dt^2} + 2H\frac{d\delta_B}{dt} = 4\pi G\bar\rho\delta_X - \frac{c_S^2}{a^2}k^2\delta_B.
  \label{eq:linb}
\end{equation}
Notice the difference between this equation and an equation for baryonic fluctuations in a static reference frame ($a=\mbox{const}$, no expansion of the universe) in the absence of dark matter:
\[
  \frac{d^2\delta_B}{dt^2} = 4\pi G\bar\rho\delta_B - \frac{c_S^2}{a^2}k^2\delta_B.
\]
We know that in the latter case the characteristic scale over which baryonic fluctuations are suppressed by the pressure force is the Jeans scale,
\[
  k_J \equiv \frac{a}{c_S}\sqrt{4\pi G\bar\rho}.
\]

Equation (\ref{eq:linb}) cannot be solved analytically in a general case, but the important physics we are after is how baryonic fluctuations deviate from the dark matter ones. Hence, a quantity of interest is the ratio of two fluctuations, which can be expanded in the Taylor series of powers of $k^2$,
\begin{equation}
  \frac{\delta_B(t,k)}{\delta_X(t,k)} = r - \frac{k^2}{k_F^2} + O(k^4),
  \label{eq:kfdef}
\end{equation}
where $r=\mbox{const}$ and we will call $k_F(t)$ a \emph{filtering scale}. Because dark matter is expected to be more clustered that baryons (it is not a subject of the pressure force in the linear regime), we can expect that, in a general case $k_F>k_J$ (in the presence of dark matter baryonic fluctuations are less suppressed than in a purely baryonic case).

In the following we will only consider the case of $r=1$ (baryons trace the dark matter on large scales), since this is an excellent approximation for $z<10$. However, at higher redshifts this is not the case any more \citep{nb07}, as the different evolution of baryons and dark matter during the recombination epoch is not completely forgotten at these high redshifts (for example, $r\ll1$ at $z>1000$).

Substituting equation (\ref{eq:kfdef}) into (\ref{eq:linb}), it is possible to obtain an expression for $k_F$ in a closed form \citep{ng:gh98},
\[
  \frac{1}{k_F^2(t)} = \frac{1}{D_+(t)} \int_0^t dt^\prime a^2(t^\prime)
  \frac{\ddot{D}_+(t^\prime)+2H(t^\prime)\dot{D}_+(t^\prime)}{k_J^2(t^\prime)}
  \int_{t^\prime}^t \frac{dt^{\prime\prime}}{a^2(t^{\prime\prime})}.
\]
While this expression is long and ugly, for reasonable thermal histories of the universe a good rule of thumb at $z\sim2-4$ is $k_F\approx 2\times k_J$ (the filtering scale is about half the Jeans scale).

\begin{figure}[t]
\sidecaption[t]
\includegraphics[width=0.64\hsize]{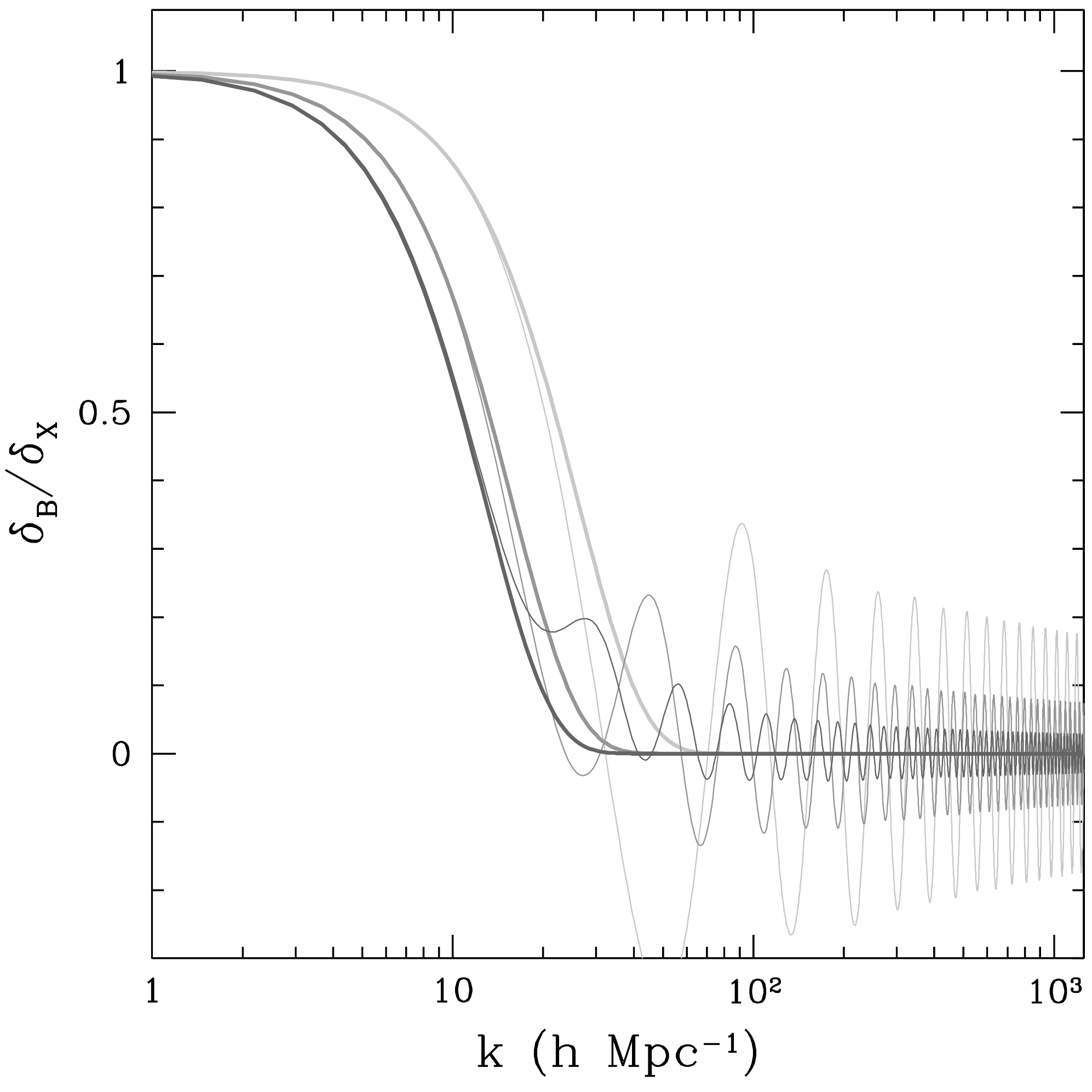}%
\caption{Solutions to equations (\protect\ref{eq:linxb}) for a representative thermal history of the universe at $z=4$ (light gray), $z=1.5$ (medium gray), and $z=0$ (dark gray); thin lines show the exact solutions, thick lines give the approximation $\delta_B/\delta_X=\exp(-k^2/k_F^2)$ (adopted from \protect\citet{ng:gbbd03}).\label{fig:kj}}
\end{figure}

Figure \ref{fig:kj} gives an example of scale-dependence of $\delta_B(t,k)/\delta_X(t,k)$ for a representative thermal history of the universe at several redshifts \citep[see][for details]{ng:gbbd03}. Fluctuations on small scales, where the pressure force dominates, are simple sound waves, and the transition to the baryons-trace-the-dark-matter regime is well described by the filtering scale.

\bt{Pressure generates sound waves, and sounds waves in the ideal gas do not dissipate. Why, then, are fluctuations "suppressed" by the pressure force?}

\subsection{Lyman-$\alpha$ Forest}

A well known empirical fact is that the IGM is highly ionized at low and intermediate redshifts, $z<6$ (we will come back to that fact). To keep the cosmic gas ionized, the universe must be filled with ionizing radiation, the so-called ``Cosmic Ionizing Background'' (CIB). 

Since most of the IGM is hydrogen, let us consider hydrogen first. The ionization balance equation for hydrogen in the expanding universe is simple,
\[
  \dot{n}_\HI = -3Hn_\HI -n_\HI\Gamma + R(T)n_en_\HII,
\]
where $n_\HI$, $n_\HII$, and $n_e$ are number densities of neutral hydrogen, ionized hydrogen, and free electrons respectively, $\Gamma$ is the \emph{photoionization rate} and $R(T)$ is a (temperature-dependent) recombination coefficient. 

Often it is more convenient to consider not the actual number density of neutral or ionized hydrogen, but the \emph{neutral fraction} $x \equiv n_\HI/n_\Ht$, because then the Hubble expansion term cancels out,
\begin{equation}
  \dot{x} = -x\Gamma + R(T)n_e(1-x).
  \label{eq:ionbal}
\end{equation}
In the ionization equilibrium $\dot{x}=0$, hence
\[
  x_{\rm eq} = \frac{R(T)}{\Gamma}n_e(1-x_{\rm eq}),
\]
and since the IGM is highly ionized ($x\ll1$),
\[
  x_{\rm eq} = \frac{R(T)}{\Gamma}(\bar{n}_\Ht+2\bar{n}_{\rm He})(1+\delta),
\]
where we assumed that Helium is fully ionized, $\bar{n}_e=\bar{n}_\Ht+2\bar{n}_{\rm He}$ (denser gas is more neutral).

Let us now consider a light source somewhere in the universe (a quasar, a galaxy, a gamma-ray burst, etc); the light source is at redshift $z_e$ in our reference frame. Let us also imagine that a photon with wavelength $\lambda_e$ is emitted by the source. As it propagates through the universe, the photon is going to be redshifted. At a redshift $z_a<z_e$ (from our reference frame) the photon has a wavelength
\[
  \lambda_e\frac{1+z_a}{1+z_e}.
\]
Hence, for any $1216\mbox{\AA}(1+z_e) < \lambda_e < 1216\mbox{\AA}$ there is such $z_a$ that
\[
  \lambda_e\frac{1+z_a}{1+z_e} = 1216\mbox{\AA}.
\]
When a photon with wavelength of $1216\mbox{\AA}$ (= Lyman-$\alpha$) hits a neutral hydrogen atom, it can get absorbed and excite the atom to $n=2$ level.

\begin{figure}[t]
\includegraphics[width=1.0\hsize]{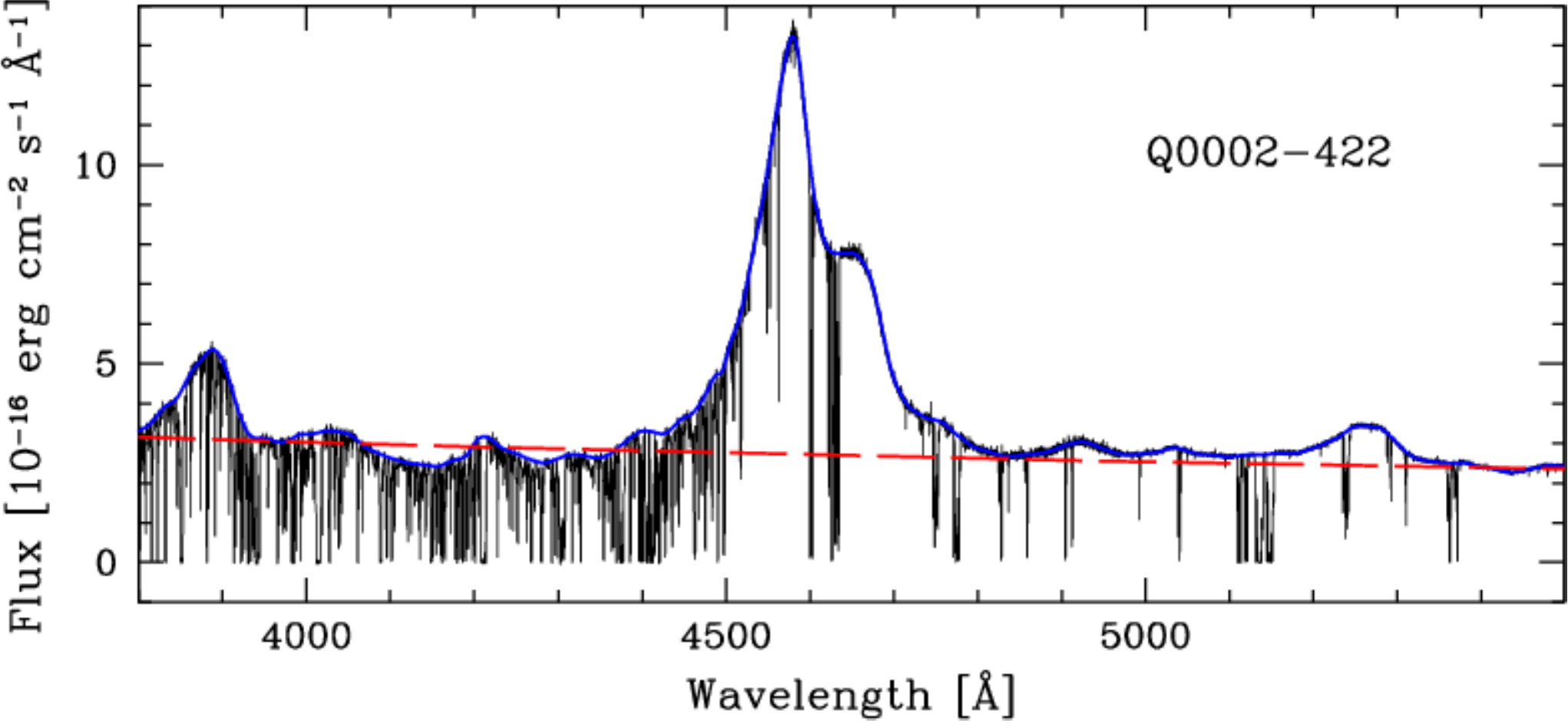}%
\caption{Typical $z\sim3$ quasar spectrum together with the power law continuum fit (dashed red line) and the local continuum fit (blue line; adopted from \protect\citet{dww08}).\label{fig:lya}}
\end{figure}

Indeed, this is exactly what happens in the real universe. Figure \ref{fig:lya} shows a spectrum of a typical $z\sim3$ quasar. The broad emission line in the middle is the Lyman-$\alpha$ of the quasar itself, and blue envelope for the observed spectrum is the continuum - i.e.\ the light that the quasar itself emitted. Black absorption lines come from the gas between us and the quasar, and the numerous forest of them at shorter wavelength is the hydrogen Lyman-$\alpha$ absorption from the neutral gas in the IGM, the so-called \emph{Lyman-$\alpha$ Forest}.

\begin{figure}[t]
\includegraphics[width=1\hsize]{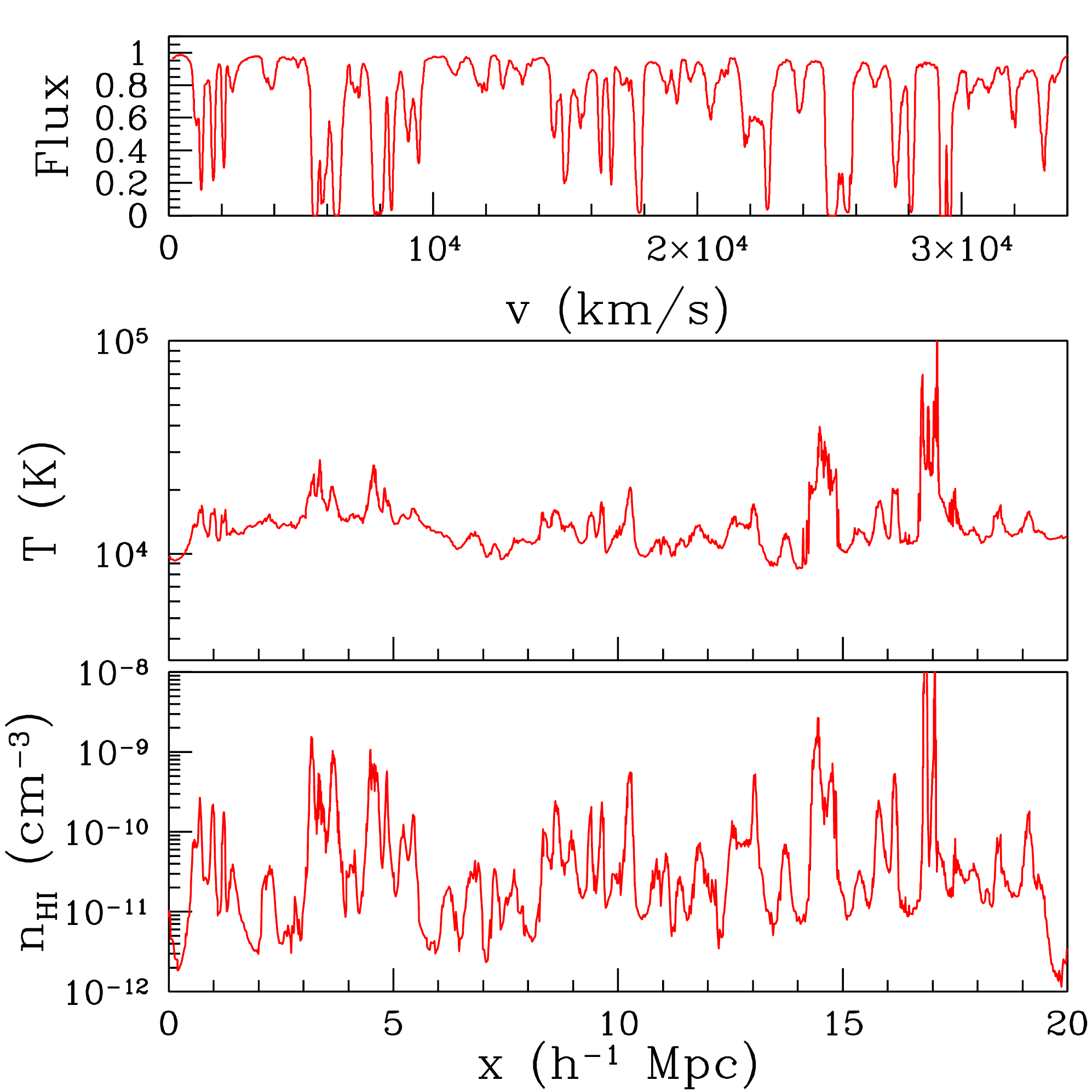}%
\caption{Runs of neutral hydrogen density (bottom) and gas temperature (middle) along one line of sight in a numerical simulation of Lyman-$\alpha$ forest at $z\sim3$. The resultant absorption spectrum is shown in the top panel.\label{fig:flux}}
\end{figure}

Figure \ref{fig:flux} illustrates how fluctuations in the neutral hydrogen density and in the gas temperature combine to produce the Lyman-$\alpha$ forest of absorption features in the spectrum. In order to understand how one goes from the lower two panels to the top one in that figure, we need to refresh the basics of resonant line absorption in the expanding universe.

\bt{Hydrogen atoms do not sit forever in $n=2$ state, they decay back into $n=1$ and a Lyman-$\alpha$ photon is re-created back. How can there be any Lyman-$\alpha$ absorption?}

\subsubsection{Introduction To Resonant Line Absorption}

The cross-section for an atom at rest to absorb a photon in the frequency range from $\nu$ to $\nu+\Delta\nu$ to the energy level with the energy $h\nu_0$ is
\[
  \sigma(\nu) = \frac{\pi e^2}{m_ec\nu_0}f\phi(\nu) \equiv \sigma_0\phi(\nu),
\]
where $f$ is the oscillator strength for the transition and 
\[
  \phi(\nu) = \frac{1}{\pi}\frac{w\nu_0}{(\nu-\nu_0)^2+w^2} \approx \nu_0\delta(\nu-\nu_0),
\]
where $w$ is the \emph{natural line width} in frequency units. For hydrogen Lyman-$\alpha$ the combination of fundamental constants
\[
  \sigma_0 = \frac{\pi e^2}{m_ec\nu_0}f = 4.5\times10^{-18}\dim{cm}^2.
\]

Atoms, though, are social creatures and rarely live alone. For a cloud of gas of density $n$, size $L$, and temperature $T$ we need to integrate over all atoms to compute the optical depth of the transition at any frequency $\nu$,
\[
  \tau(\nu) = nL\int\sigma_0\phi(\nu^\prime) \frac{1}{\sqrt{\pi}b}e^{\displaystyle-\frac{(u_\nu-u^\prime)^2}{b^2}}du^\prime,
\]
where $\nu^\prime=\nu_0(1+u^\prime/c)$ and $u_\nu$ is defined via the expression $\nu=\nu_0(1+u_\nu/c)$. The quantity \[
 b = \left(2\frac{k_BT}{m_H}\right)^{1/2}
\]
is called the \emph{Doppler parameter} and the product $nL$ is the \emph{column density}.

In an expanding universe it is not enough just to multiply by the cloud size $L$, since different locations along the line-of-sight are redshifted relative to the observer and project to different locations in the velocity (or frequency) space. Hence, we must integrate along the line-of-sight,
\begin{equation}
  \tau(\lambda) = \sigma_0\int n(x) \frac{c}{\sqrt{\pi}b_x} e^{\displaystyle-\frac{(u_\lambda-u_x)^2}{b_x^2}}\frac{dx}{1+z_x},
  \label{eq:tau}
\end{equation}
where we switch to from the frequency to the wavelength (as almost all observers tend to live in the wavelength space), and we integrate over the comoving distance $x$ (as almost all theorists tend to live in the comoving space); both $b_x$ and $u_x$ are, in general, functions of position, since the temperature and velocity vary in space. The wavelength is related to the velocity along the line-of-sight through the usual Doppler effect,
\[
  \lambda = \lambda_0\left(1+z_x+\frac{u_\lambda}{c}\right)
\]
and $z_x$ is the redshift of location $x$ along the line-of-sight.

The spectrum shown on the top panel of figure \ref{fig:flux} is just $\exp(-\tau(\lambda))$ with $\tau(\lambda)$ computed with equation (\ref{eq:tau}) from the two bottom panels of the same figure.

Now we are ready to figure out why the IGM must be highly ionized at $z<6$. From figure \ref{fig:lya} we notice that the forest absorbs about 50\% of the quasar flux, so the average  optical depth is $\tau\sim0.5-1$. Considering one absorption system stretching for about $\Delta x \sim 100\dim{kpc}$ and having temperature of $10^4\dim{K}$ (or $b\sim10\dim{km/s}$), an crude estimate for $\tau$ is
\begin{eqnarray}
  \tau & \sim & \sigma_0 \frac{c}{\sqrt{\pi}b} x_\HI n_\Ht a\Delta x  \nonumber\\
       & = & 4.5\times10^{-18}\dim{cm}^{2} \frac{3\times10^5\dim{km/s}}{\sqrt{\pi}10\dim{km/s}} x_\HI 1.3\times10^{-5}\dim{cm}^{-3} (4a)^{-3} 0.75\times10^{23}\dim{cm} (4a)  \nonumber\\
       & = & 7\times10^4\frac{x_\HI}{(4a)^2} \nonumber
\end{eqnarray}
at the cosmic scale factor $a$. To get $\tau \sim 1$ the neutral fraction $x_\HI$ must be $x_\HI \sim 10^{-5}$.

\subsubsection{Temperature}

The final component in modeling the IGM is to know what the temperature of the gas is.

Since the IGM is highly ionized, a process of photo-heating (heating by ionizing radiation) is important. When a high energy photon hits a hydrogen atom, $13.6\dim{eV}$ of its energy goes into ionizing the atom, the rest goes into the energy of the ejected electron. If the electron is not super-energetic (less that $\sim40\dim{eV}$), it thermalizes and adds its energy to the thermal energy of the gas. For more energetic electrons the situation may be more complex, as it can ionize another atom by colliding with it (a so-called \emph{secondary ionization}). That, in turn, produces an energetic electron which may ionize another atom etc. Usually, these secondary ionizations are only important if the gas is substantially (more than a few percent) neutral; for the low redshift IGM with $x_\HI\sim10^{-5}$ secondary ionizations are completely unimportant.

If all the excess energy of an ionizing photon goes into heat, the rate of internal energy increase in the gas due to photo-heating is
\[
  \left.\frac{3}{2}\frac{d}{dt}\left(nk_BT\right)\right|_{\rm PH} = cn_\HI\int_{E_0}^\infty (E-E_0)\sigma_\HI(E) n_E dE,
\]
where $E_0=13.6\dim{eV}$ is the hydrogen ionization threshold, $\sigma_\HI(E)$ is the hydrogen ionization cross-section, and $n_E$ is the radiation spectrum measured in photons per unit volume per unit energy.

The photoionization rate of hydrogen is 
\[
  \Gamma = c\int_{E_0}^\infty \sigma_\HI(E) n_E dE,
\]
hence
\[
  \left.\frac{3}{2}\frac{d}{dt}\left(nk_BT\right)\right|_{\rm PH} = n_\HI\Gamma\langle\Delta E\rangle,
\]
where $\langle\Delta E\rangle$ is the average excess energy (over $13.6\dim{eV}$) of an ionizing photon,
\begin{equation}
  \langle\Delta E\rangle \equiv \frac{\displaystyle\int_{E_0}^\infty (E-E_0)\sigma_\HI(E) n_E dE}{\displaystyle\int_{E_0}^\infty \sigma_\HI(E) n_E dE}.
\label{eq:deltae}
\end{equation}

Let us ignore helium for a moment: $n_e=(1-x)n_\Ht$, $n=n_\Ht+n_e=(2-x)n_\Ht$. Then the thermal balance equation together with the ionization balance equation become
\begin{eqnarray}
  \frac{3}{2}\frac{d}{dt}\left((2-x)k_BT\right) = x\Gamma\langle\Delta E\rangle,\label{eq:t}\\
  \frac{d}{dt}x = -x\Gamma + R(T)n_\Ht(1-x)^2.\label{eq:x}
\end{eqnarray}

Let us start with cold neutral IGM ($x=1$, $T=0$, like at very high redshift, before cosmic reionization), and assume that the ionizing radiation pops out of nowhere instantaneously at a cosmic time $t_R$ (a favorite approximation of your CMB friends),
\[
  \Gamma \propto \theta(t-t_R)
\]
($\theta(x)$ is a Heaviside function). In the ionization equilibrium
\[
  x_{\rm eq} = \frac{Rn_\Ht}{\Gamma}(1-x_{\rm eq})^2.
\]
Hence, until the ionization equilibrium is established (i.e. while $x\gg x_{\rm eq}$) $x\Gamma \gg Rn_\Ht(1-x)^2$. In that limit equation (\ref{eq:x}) becomes simply
\[
  \frac{d}{dt}x = -x\Gamma
\]
and its solution for $t>t_R$ is
\[
  x(t) = e^{-\Gamma (t-t_R)}.
\]
That solution is valid until $x$ becomes small enough ($\sim Rn_\Ht/\Gamma$) for the ionization equilibrium to get established.

Equation (\ref{eq:t}) can also be solved easily in the same limit,
\[
  (2-x)k_BT = \frac{2}{3} \langle\Delta E\rangle \left(1-e^{-\Gamma (t-t_R)}\right),
\]
and in the limit of small $x$ the gas temperature becomes constant (i.e.\ gas becomes \emph{isothermal}),
\begin{equation}
  T_\infty = \frac{\langle\Delta E\rangle}{3k_B}
  \label{eq:trei}
\end{equation}
and independent of density or the photoionization rate. This is an important lesson: {\bf if a region of space is ionized rapidly, its temperature does not depend on the strength of the radiation field}. I.e., you cannot heat up the IGM by cranking up the ionizing source, only by making the source spectrum \emph{harder}.

It is also instructive to plug some numbers into equation (\ref{eq:trei}). For example, for a power-law energy spectrum for ionizing photons, $n_E \propto E^{-\alpha}$, and using the fact that just beyond the ionization edge $\sigma_\HI(E)\propto E^{-3}$, we find
\[
  \langle\Delta E\rangle = \frac{\displaystyle\int_{E_0}^\infty (E-E_0)\sigma_\HI(E) n_E dE}{\displaystyle\int_{E_0}^\infty \sigma_\HI(E) n_E dE} = \frac{E_0}{1+\alpha}
\]
and
\[
  T_\infty = \frac{52{,}000\dim{K}}{1+\alpha} = \left[
  \begin{array}{rl}
  26{,}000\dim{K} & (\alpha=1) \\
   5{,}000\dim{K} & (\alpha=9) \\
   \end{array}
   \right.
\]
In other words, the temperature of the photo-ionized gas is about $10{,}000\dim{K}$, give-or-take a factor of 2.

Let us now consider what happens next. A region of space was ionized to $x=x_{\rm eq}$ at $t=t_R$ ($a=a_R$), and the temperature of the gas is at this moment constant at $T=T_\infty$. Another important effect is plain adiabatic cooling due to the expansion of the universe, so that the full equation that governs the temperature evolution after ionization equilibrium is established is
\begin{equation}
  \frac{dT}{dt} = T \frac{2\dot{n}_\Ht}{3n_\Ht} + T_\infty x_{\rm eq}\Gamma =
T \frac{2\dot{n}_\Ht}{3n_\Ht} + T_\infty Rn_\Ht.
  \label{eq:tlate}
\end{equation}
The recombination coefficient can be well approximated as a power-law function of gas temperature, $R(T)\approx 4.3\times10^{-13} T_4^{-0.7} \dim{cm}^3/\dim{s}$ ($T_4 \equiv T/10^4\dim{K}$). It is easy to solve equation (\ref{eq:tlate}) for the temperature $T_0$ at the cosmic mean density, $\bar{n}_\Ht \propto a^{-3}$,
\begin{equation}
  T_0(a) = T_\infty \left(\frac{a_R}{a}\right)^2\left[1+1.34R(T_\infty)\bar{n}_{\Ht,R}t_R\left(\left(\frac{a}{a_R}\right)^{1.9}-1\right)\right]^{1/1.7}.
  \label{eq:t0evol}
\end{equation}
At late times ($a\gg a_R$) the asymptotic behavior of the temperature at the mean density is
\[
  T_0(a) \propto \left(\frac{a_R}{a}\right)^{1.5/1.7}.
\]
It is less rapid than pure adiabatic expansion $T\propto a^{-2}$ because photo-heating off the residual neutral hydrogen fraction remains non-negligible at all times.

It turns out that for densities other than the mean a power-law ansatz provides a decent approximation for moderate overdensities, $\delta \la 10$,
\begin{equation}
  T(\rho) \approx T_0(1+\delta)^{\gamma-1},
  \label{eq:tdrel}
\end{equation}
where both $T_0$ (as is given above) and $\gamma$ are functions of time \citep{ng:hg97}. Expression for $\gamma(a)$ is rather ugly, but its main important properties are that $\gamma=1$ right after instantaneous reionization and $\gamma\rightarrow 1.62$ at late times (notice, it is $1.62$ and \emph{not} $2/3$).

\begin{figure}[b]
\includegraphics[width=0.5\hsize]{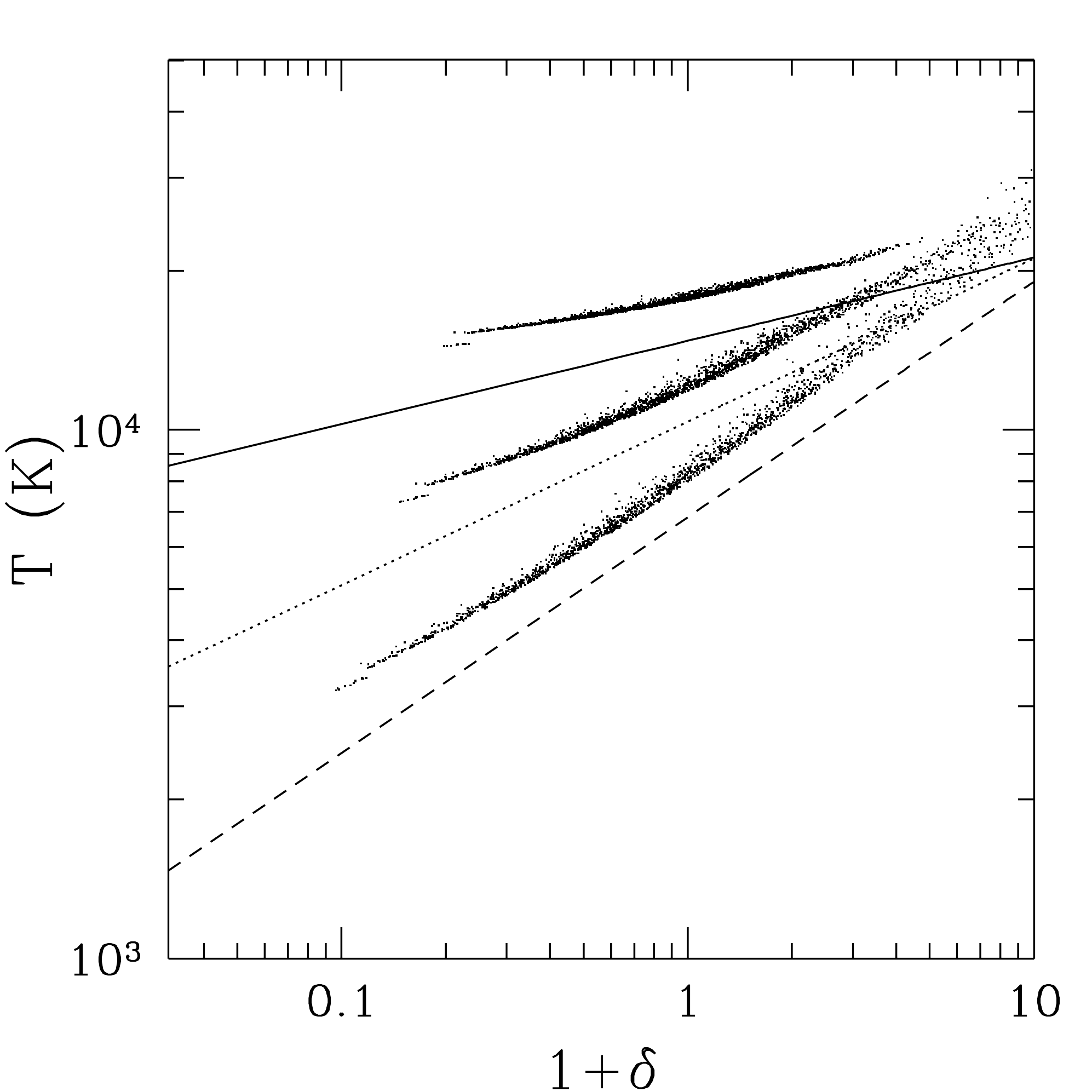}%
\includegraphics[width=0.5\hsize]{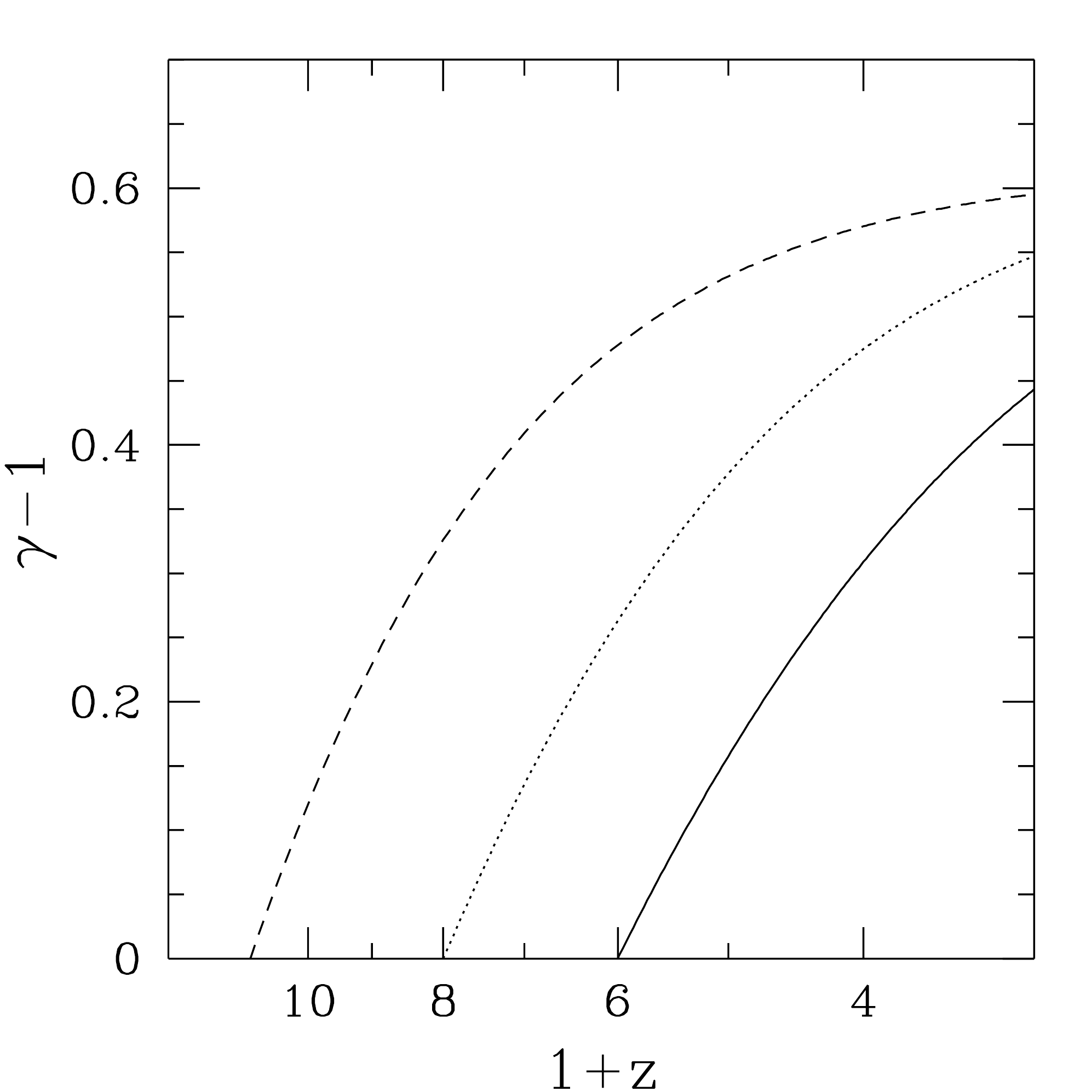}%
\caption{(Left) Temperature-density relation for a sudden reionization model at $z=6$: dots show the  the results of a full calculation at $z=4,3,2$ (from top down) while lines are approximation (\ref{eq:tdrel}) with $T_0$ given by equation (\ref{eq:t0evol}); the approximate solution slightly underestimates the temperature because it ignores the heat input from helium ionizations. (Right) Time evolution of $\gamma$ for $z_R=10,7,5$ respectively.\label{fig:td}}
\end{figure}

Figure \ref{fig:td} compares the temperature-density relation in the IGM from a real calculation \citep[by following heating and cooling of individual fluid elements,][]{ng:hg97} with the approximate solution above. One effect that we ignored is photo-heating of helium - heat input from the ionizations of the residual neutral helium will heat the gas a bit more than is given by equation (\ref{eq:t0evol}), but, overall, our analytical calculation does rather well.

``Hey'', a meticulous reader will exclaim, ``what about radiative cooling?'' After all, gas does cool by emitting radiation. A story of gas cooling, with all its gory details, awaits us in the future, but here let us estimate how important radiative cooling actually is in the Lyman-$\alpha$ forest.

In a highly ionized gas the dominant radiative cooling mechanism is recombination cooling,
\[
  \left.\frac{dU}{dt}\right|_{\rm RC} = -\frac{3}{4}k_BT R(T) n_en_\HII.
\]
If we compare this term to photoionization heating in ionization equilibrium,
\[
  \left.\frac{dU}{dt}\right|_{\rm PH} = n_\HI \Gamma \langle\Delta E\rangle = \langle\Delta E\rangle R(T) n_en_\Ht,
\]
we see that the radiative cooling is lower by a factor of
\[
  \frac{3}{4}\frac{k_BT}{\langle\Delta E\rangle} = \frac{T}{4 T_\infty}.
\]
Hence, radiative cooling makes at most a 25\% correction, and well after reionization ($T < T_\infty$) the correction is even smaller.

Is this the complete story? Alas, no, the reality is always more complicated than we are ready to accept and you need to be aware of several caveats when using the temperature-density relation.
\begin{itemize}
\item The temperature-density relation is an \emph{approximation}, with 5-10\% scatter at low densities and progressively larger scatter as one moves up the density axis, because it misses a major hydrodynamic effects - shocks. Gas motions in the IGM will cause shock waves that will lead to additional gas heating.
\item There may exist other heating and cooling mechanisms. For example, \citet{pps12} argued that heating of the Lyman-$\alpha$ forest by ultra-high energy gamma rays from a population of blazars is important at very low densities. The jury is still out on whether such an effect is important or not, but we should always be aware that we do not known everything.
\item Our analysis assumed that the gas is optically thin to ionizing radiation. While this is the case for hydrogen at $z\la 6$, helium is believed to be reionized the second time (from $\HeII$ to $\HeIII$) at $z\sim3$, in which case gas is \emph{not} optically thin to helium ionizing radiation at $z \ga 3$. Non-trivial opacity to ionizing radiation normally leads to increasing photo-heating rates in the gas.
\end{itemize}

\bt{The temperature-density relation is sometimes called an ``equation of state'' (occasionally even without quotes). Do not fall into that trap - it relates the gas temperature and density, but it is \emph{not an equation of state}. Can you explain why?}

\subsection{Modeling the IGM}

The most straightforward model of Lyman-$\alpha$ forest is a hydrodynamic simulation with ionization balance. In the 1990-ties several approximate methods have been used, such as a log-normal approximation, Zel'dovich approximation, a pure N-body simulation, Hydro-Particle-Mesh (HPM) approximation. None of these methods is competitive any more and their use can be hardly justified.

The assumption of the ionization equilibrium is very good in the IGM, but it does break down in a few special cases (quasar proximity zones, helium reionization, etc). Hence, the most accurate simulation of the IGM includes (a model of) Cosmic Ionizing Background (CIB), radiative transfer, non-equilibrium ionization, separate fields for each of ionizing species ($\HI$, $\HII$, $\HeI$, $\HeII$, ...). Such a simulation, however, is usually an overkill, except when it is used for a special purpose like modeling non-equilibrium effects in quasar proximity zones.

A standard approach is to include CIB and ionization equilibrium and follow radiative heating (and, optionally, cooling) ``on-the-fly'' (a-la equation \ref{eq:tlate}). For simulations of the Lyman-$\alpha$ forest alone the temperature-density relation may be assumed, but the computational savings in that case will be modest and it is rarely worth it.

\begin{figure}[b]
\includegraphics[width=0.5\hsize]{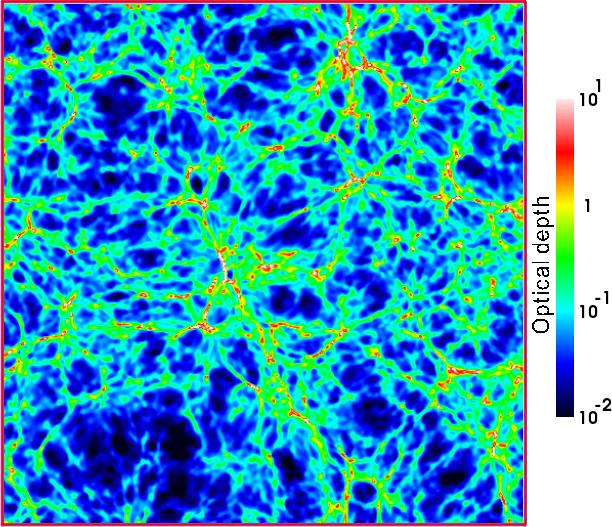}%
\includegraphics[width=0.5\hsize]{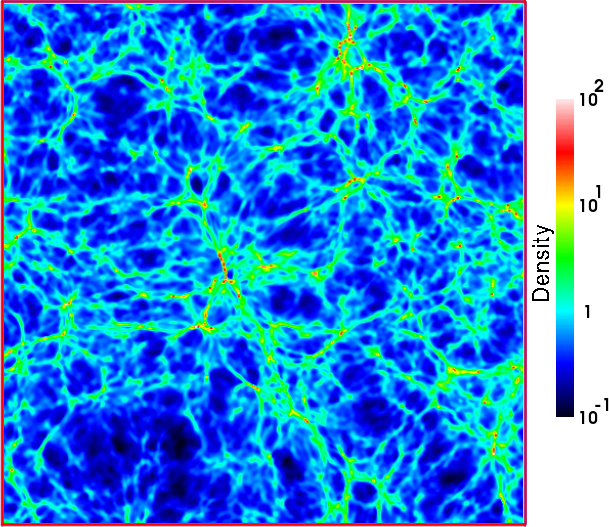}%
\caption{Slices of optical depth (left) and gas density (in units of the cosmic mean, right) in a simulation of Lyman-$\alpha$ forest at $z\sim3$. The box size is $20h^{-1}$ comoving Mpc.\label{fig:slice}}
\end{figure}

Example of a numerical simulation of the forest is shown in figure \ref{fig:slice}. The right panel shows the gas density, and looks like a usual image of large-scale structure. The left panel shows the Lyman-$\alpha$ optical depth that would be observed in the corresponding position along the absorption spectrum towards a high redshift quasar. The main thing to take from that figure is that the actual absorption lines we see clearly in the spectra (those with $\tau\ga 0.5$) come from filaments: weaker ones tend to cluster around stronger ones, although a few of the weakest ones do occur in the voids. The higher optical depth systems, those that lead to saturated lines with $\tau\ga 2$ tend to occur at the intersections of filaments, nearer to galaxies.

\subsubsection{Density - Column Density Correlation}

What is clear from figure \ref{fig:slice} is that the gas density and the optical depth of the corresponding absorption feature are well correlated. Crudely, the relation is 
\[
  \tau \sim (1+\delta)^{1.5},
\]
although the slope and normalization of this correlation are redshift dependent.

This correlation is so good, especially on large scales, that it is often used to match directly the gas density into the opacity along the line of sight - such an \emph{ansatz} is called \emph{Fluctuating Gunn-Peterson Approximation}, or FGPA. FGPA is useful for modeling the forest on large scales, but one has always keep in mind that the absorption spectrum is in the velocity space, while the density is sampled in real, physical space, hence FGPA breaks down on sufficiently small scales (roughly less than $1\dim{Mpc}$).

\subsection{What Observations Tell Us}

For a long time since the Lyman-$\alpha$ forest was discovered in the 60-ties, it was treated by observers as just another absorption spectrum - as a collection of individual absorption lines, each having a fixed column density $N$ and the Doppler parameter $b$, as if the absorption was coming from discrete clouds. Now we know that this is not a good description - the density, temperature, and neutral fraction fields are continuous, and it is impossible to decompose the realistic spectrum into a set of $(N,b)$ pairs uniquely.

The modern view of the forest is that $\tau(\lambda)$ is a continuous field and should be treated as such. However, there is one application where the $(N,b)$ decomposition is still useful - measuring the temperature-density relation. If we think about a segment of the spectrum that has an "absorption line", the width of the feature is determined by the temperature of the gas plus any velocity gradient across the region that may exist. In some cases that velocity gradient will be very small, so the narrowest features at each column density should be those that are broadened by temperature alone. Hence, looking at the distribution of fitted $b$ parameters at given column density, one can measure $T(N)$ in the forest and, by virtue of the strong correlation between $\rho$ and $\tau$ (and, hence, $N$), translate that measurements into the measurement of $T-\rho$ correlation.

\begin{figure}
\sidecaption[t]
\includegraphics[width=0.64\hsize]{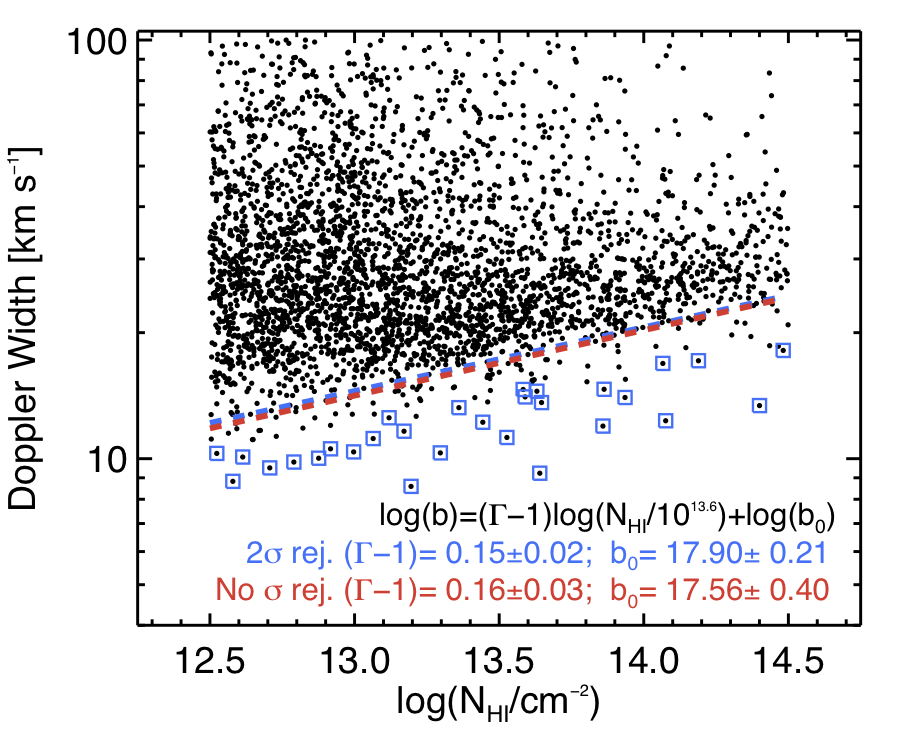}%
\caption{Example of the $(N,b)$ distribution for a quasar spectrum at $z=2.4$. The measurement points labeled by blue squares are contamination from heavy elements. A relatively sharp edge of the distribution of Doppler parameters at a given $N$ is apparent in the figure \protect\citep[adopted from][]{igm:rsp12}.\label{fig:bndist}}
\end{figure}

Figure \ref{fig:bndist} shows an example of such distribution from a single high resolution quasar spectrum \citep{igm:rsp12}. The cutoff in the distribution of Doppler parameters for a given $N$ is clearly visible, and the value of the cutoff is well fit by the power-law in $N$, demonstrating the fact that the power-law temperature-density relation is indeed a good approximation.

\begin{figure}[b]
\sidecaption[t]
\includegraphics[width=0.64\hsize]{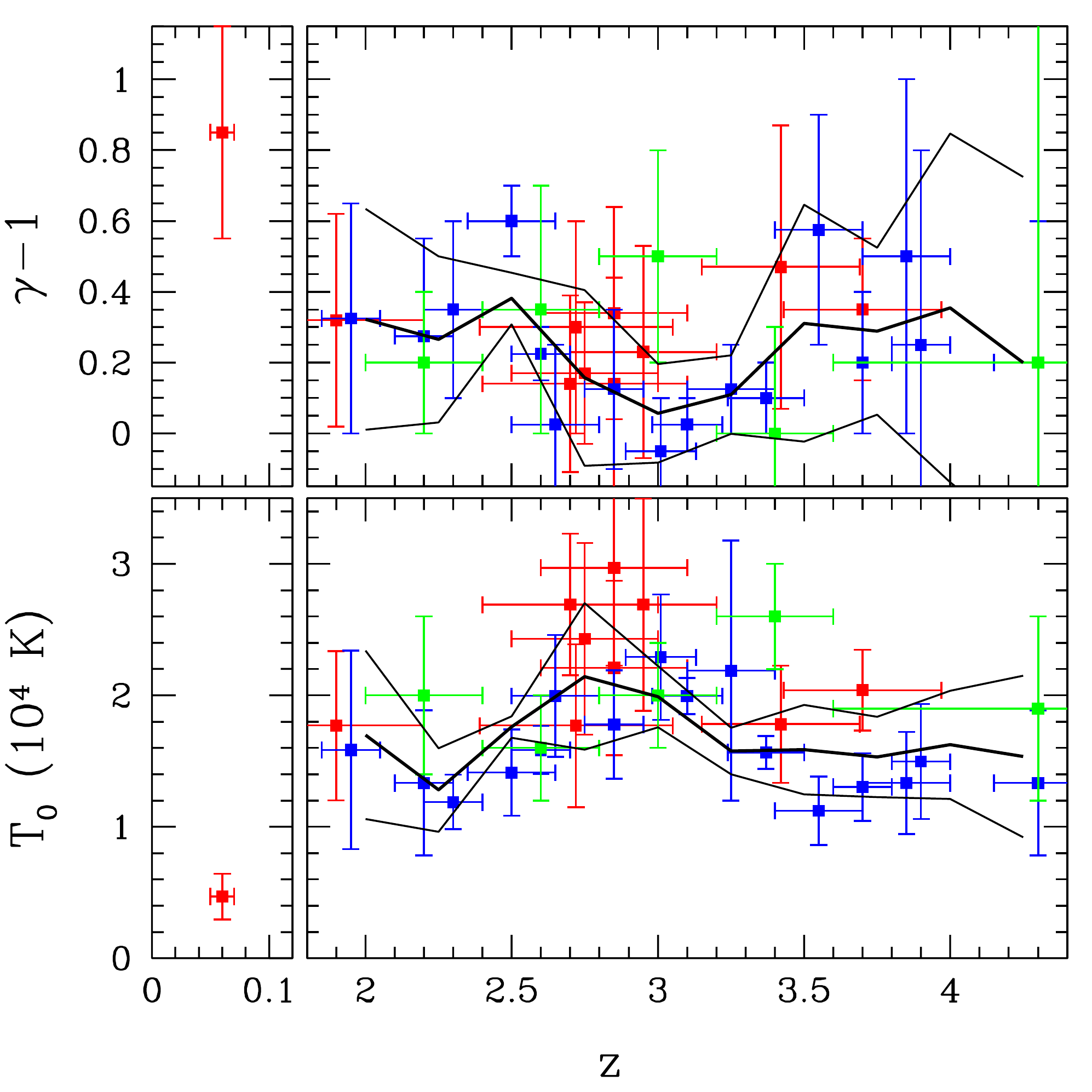}%
\caption{Evolution of the temperature-density relation $T\approx T_0(1+\delta)^{\gamma-1}$. Red, blue, and green points show individual measurements from \protect\citet{ng:rgs00,igm:stre00,igm:lfd10} respectively; thick and thin black lines shows the average values and $1\sigma$ dispersion for all measurements. Possible increase in temperature and dip in $\gamma$ at $z\sim3$ is attributed to $\HeII\rightarrow\HeIII$ reionization.\label{fig:gamt0}}
\end{figure}

A compilation of the majority of existing measurements is shown in figure \ref{fig:gamt0} \citep{ng:rgs00,igm:stre00,igm:mmrs01,igm:lfd10,igm:rsp12}. The data seem to indicate (albeit rather vaguely) an increase in the temperature and  a decrease in $\gamma$ at $z\approx3$ - a behavior reminiscent of cosmic reionization (equation \ref{eq:trei}). Indeed, this may correspond to the second reionization of helium ($\HeII$ goint into $\HeIII$), thought to occur at redshifts around 3.

\begin{figure}
\includegraphics[width=1\hsize]{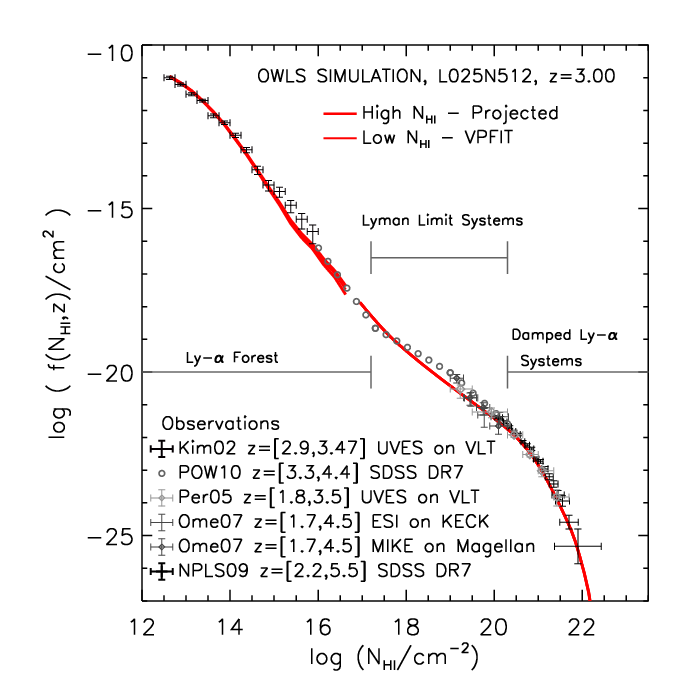}%
\caption{Distribution of column densities of Lyman-$\alpha$ absorbing systems (adopted from \protect\citet{igm:ats11}). \label{fig:altay}}
\end{figure}

An even simpler quantity is the \emph{column density distribution} - a distribution of all $N$ values irrespectively of what their $b$ values are. \citet{igm:ats11} show how modern cosmological simulations can match the observed distribution over 10 orders of magnitude in column density (see Figure \ref{fig:altay}).

The column density distribution is a useful observational measurement for other types of hydrogen absorbing systems, such as \emph{Lyman-limit system} ($10^{17}\mbox{cm}^{-2}<N_\HI<10^{20}\mbox{cm}^{-2}$) and \emph{Damped Lyman-$\alpha$ systems} ($10^{20}\mbox{cm}^{-2}<N_\HI$), but has not been particularly constraining for the forest.

\bt{The photo-ionization cross-section for neutral hydrogen at the ionization edge ($13.6\dim{eV}$) is $\sigma_{\rm ion}=6.3\times10^{-18}\mbox{cm}^2$. Hence, a column density of $N_\HI = 1.7\times10^{17}\mbox{cm}^{-2}$ has an optical depth of $\tau_{\rm ion} = \sigma_{\rm ion} N_\HI = 1$. Never-the-less, Lyman-$\alpha$ absorbers remain ionized almost all the way to Damped Lyman-$\alpha$ systems, $N_\HI \ga 10^{19}\mbox{cm}^{-2}$ ($\tau_{\rm ion} \sim 500$). Can you explain why?}

\subsubsection{Lyman-$\alpha$ Power Spectrum}

Perhaps the most important use of Lyman-$\alpha$ forest in cosmology is in measuring the evolution of the matter power spectrum. Observations of the forest cover a wide redshift range, from $z\ga 2$ to $z\la5$; since the observed optical depth is well correlated with the gas density, which, in turn, traces the matter density on large scales (above the filtering scale), the observed spectra of the forest contain hidden information about the clustering of matter and its evolution over the redshift range $2\la z\la 5$. 

\begin{figure}
\includegraphics[width=1\hsize]{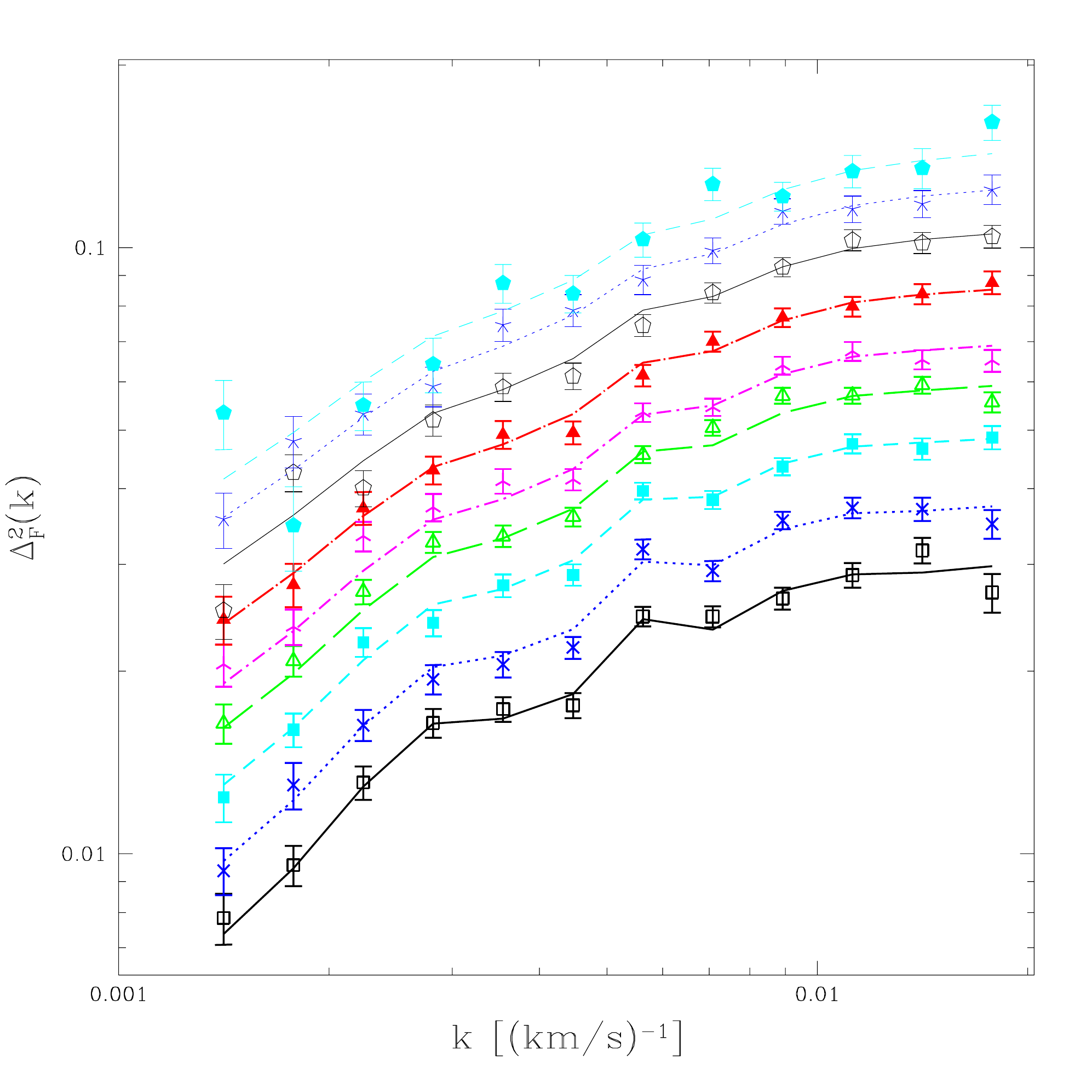}%
\caption{Matter power spectra measured from SDSS Lyman-$\alpha$ measurements at a range of redshifts from $z=2.2$ (bottom) to $z=4.2$ (top) (adopted from \protect\citet{igm:msbs06}). \label{fig:lyaps}}
\end{figure}

Measuring the matter power spectrum is exactly the application for which the Fluctuating Gunn-Peterson Approximation (FGPA) is most suitable. In the theory of large scale structure formation there is a theorem that states that if a locally non-linear field is a function of matter density only ($f=f(\rho)$), then on sufficiently large scales the field $f$ is \emph{linearly biased} with respect to the density field, i.e.\ for sufficiently small $k$
\[
  P_f(k) = b_f^2 P(k),
\]
where the bias factor $b_f$ is independent of $k$. Hence, one can measure the matter power spectrum $P(k)$ in a few simple steps:
\begin{enumerate}
\item measure the 1D power spectrum of the transmitted Lyman-$\alpha$ flux, $P_{1D}(k)$, directly from the observed spectra;
\item convert from a 1D to a 3D flux power spectrum,
\[
  P_F(k) = -\frac{2\pi}{k}\frac{dP_{1D}}{dk};
\]
\item determine the flux bias factor, $b_F$, from numerical simulations,
\item and, finally, compute the matter power spectrum
\begin{equation}
  P(k) = \frac{P_F(k)}{b_F^2}.
  \label{eq:pofklya}
\end{equation}
\end{enumerate}
Such a program was first completed by \citet{igm:cwk98} and later repeated many times with better data. For example, the largest set of observed Lyman-$\alpha$ spectra was obtained as part of the Sloan Digital Sky Survey (SDSS), and is shown in figure \ref{fig:lyaps}. A little bit of nuisance is that the flux power spectrum is measured in the velocity space, so the units of $k$ in equation (\ref{eq:pofklya}) are $(\mbox{km/s})^{-1}$. That makes it hard to compare with other measurements of matter power spectrum without knowing cosmological parameters. But a good piece of news is that the power spectrum grows (or the plotted quantity, $\Delta(k)^2 = k^3P(k)/2\pi^2$, decreases) with redshift with the rate prescribed by the standard cosmology, so you have not studied your Introduction To Cosmology in vain... 

On a more serious note, this measurement provides extremely powerful constraints on the matter power spectrum at the smallest scales - in fact, the forest probes the smallest scales currently accessible to any observational measurement. Many important cosmological and physical studies use these measurements, from determining cosmological parameters to constraining neutrino masses (but that is a field I am not going to review in these lectures).

\subsubsection{Where the Forest Ends}

The Lyman-$\alpha$ forest is a small-scale scale counterpart of the large-scale structure - but how small is "small"? In other words, what are the smallest spatial scales on which there is structure in the IGM?

This question is not moot - indeed, the filtering scale tells us where the baryonic fluctuations lag behind the dark matter, but it only applies to linear evolution. The forest is nonlinear, and nonlinear evolution may drive new fluctuations on a variety of scales. 

\begin{figure}[b]
\sidecaption[t]
\includegraphics[width=0.64\hsize]{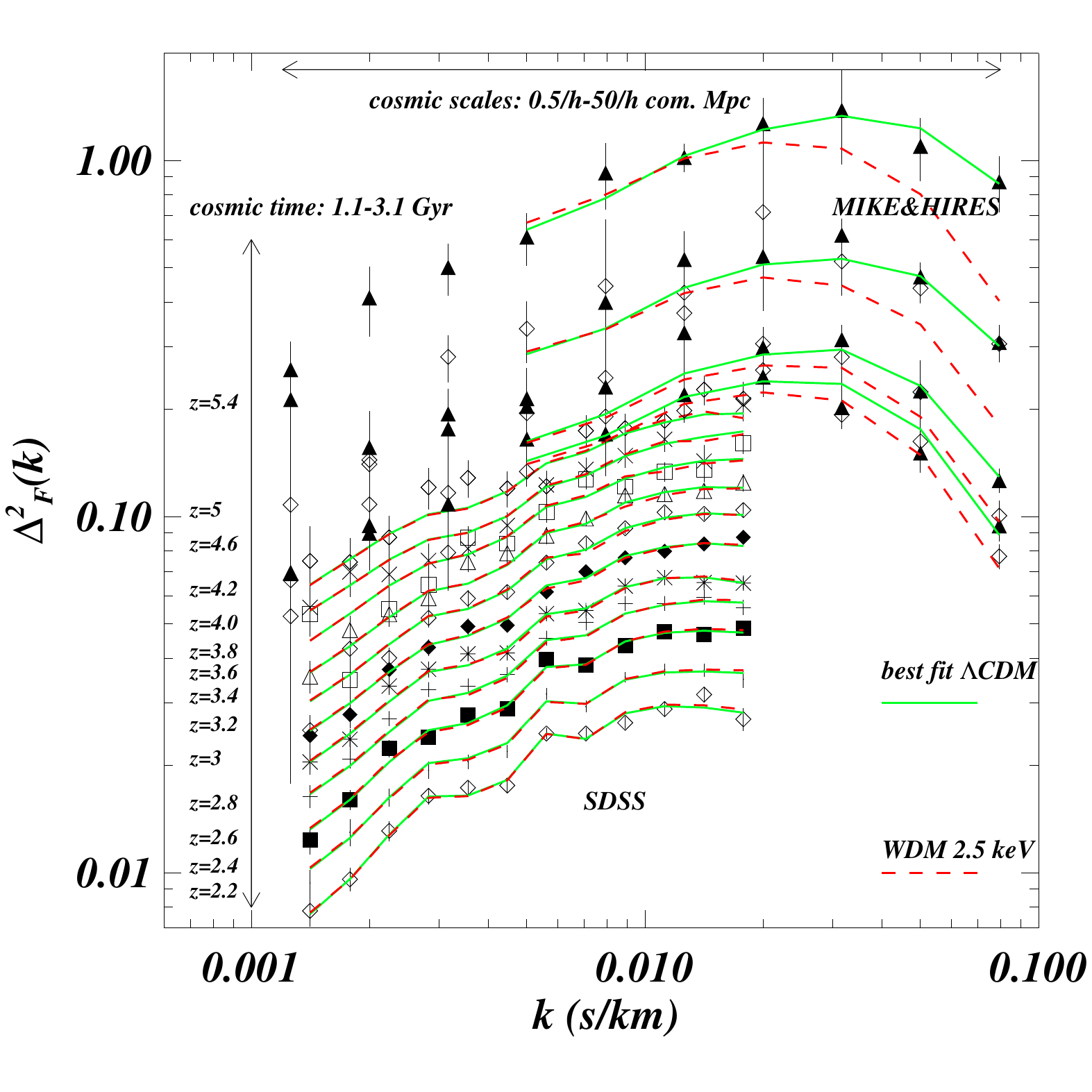}%
\caption{Matter power spectra measured from SDSS Lyman-$\alpha$ measurements (as in figure \protect\ref{fig:lya}) combined with data from high resolution spectra of several quasars (adopted from \protect\citet{vbb13}). \label{fig:lyaps2}}
\end{figure}

One way to measure structure in any distribution is the, familiar to us already, power spectrum. Using high resolution spectra from 8m-class telescopes one can extend the SDSS measurement to much smaller scales, as is illustrated in figure \ref{fig:lyaps2}. The decrease in the clustering amplitude is clearly visible at $k>0.03\dim{s/km}$, but is it really the end of the forest? The answer is "unfortunately, no" - unfortunately, because the roll over in the flux power spectra has nothing to do with the actual matter clustering - it is merely an artifact of the thermal broadening of the spectra (the exponential factor in equation \ref{eq:tau}). Alternatively, one can think of it as the break up of the linear biasing approximation (equation \ref{eq:pofklya}).

\begin{figure}[t]
\sidecaption[t]
\includegraphics[width=0.64\hsize]{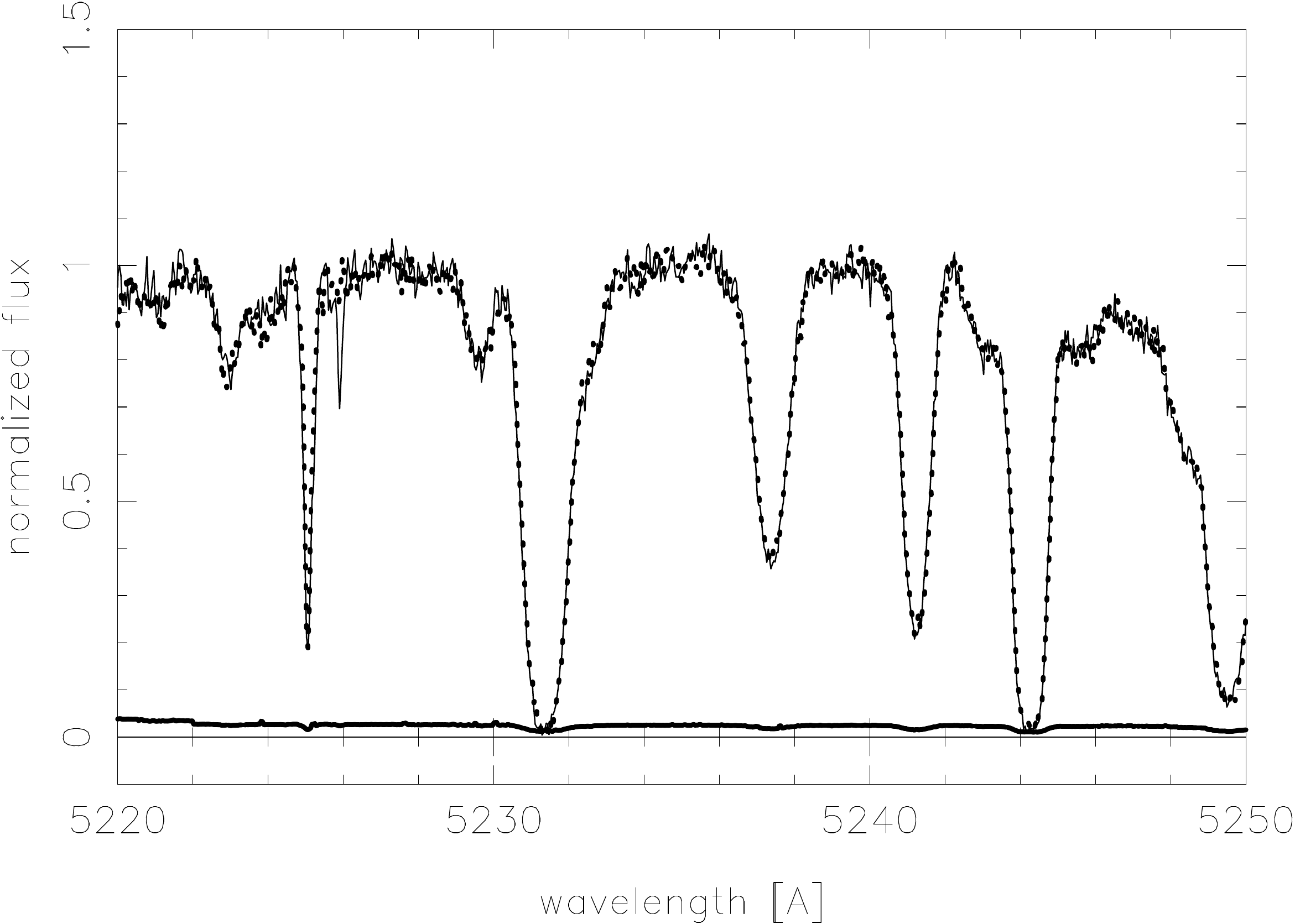}%
\caption{Lyman-$\alpha$ spectra along lines of sight to the gravitationally lenses quasar Q1422+231 (images A and C). One spectrum is shown with the solid line, another one with the dotted line beaded with dots. The two spectra are identical to within the observational errors (adopted from \protect\citet{rsb01}). \label{fig:lyass}}
\end{figure}

So, how would one approach the question of studying the smallest scale structure in the forest? One option is offered by spectra of double or gravitationally lensed quasars - if the two quasar images are not too far on the sky, their sightlines probe small spatial scales. Unfortunately, this approach has not been particularly popular among observers - in the only study I am aware of \citet{rsb01} demonstrated that, in fact, there is not that much structure in the forest on scales below a kpc. For example, figure \ref{fig:lyass} shows Lyman-$\alpha$ spectra along two lines of sight to two images of a gravitationally lensed quasar separated by about 0.5 comoving kpc at $z\sim3$.

Using this measurement, \citet{rsb01} placed a strict constraint on the density variation in the forest on small scales,
\[
  \sqrt{\langle(\Delta\ln\rho)^2\rangle} < 3\times10^{-2}\mbox{ for }\langle\Delta x\rangle = 0.6\dim{kpc},
\]
or, alternatively,
\[
  \sqrt{\left\langle\left(\frac{\Delta\ln\rho}{\Delta x}\right)^2\right\rangle} < 0.05\,\mbox{kpc}^{-1}.
\]
A scientifically interesting question is whether the IGM is turbulent on small scales. The \citet{rsb01} constraint implies that either the forest is \emph{not} turbulent on these small scales, or that any turbulence that is present is highly sub-sonic (i.e. incompressible). The latter option is possible but is not too likely - density fluctuations in the sub-sonic turbulence scale as Mach number squared, with the flow in the forest becoming transonic at scales $100-200\dim{kpc}$. If we take a Kolmogorov-like scaling law,
\[
  \sqrt{\langle(\Delta\ln\rho)^2\rangle} \approx 1 \left(\frac{\Delta x}{200\dim{kpc}}\right)^{1/3},
\]
then on scale of $0.6\dim{kpc}$ we find the rms density fluctuation of $\sqrt{\langle(\Delta\ln\rho)^2\rangle}\approx0.15$, 5 times higher than the actual observed upper limit. Of course this is not a formal derivation, and factors of several may be lurking here and there, but the estimate serves to demonstrate that the forest is remarkably quiet on scales below a kpc. 

\bt{It is well known in classical hydrodynamics that any flow with Reynolds number in excess of about 1000 becomes turbulent. The viscosity in the IGM is very small, and Reynolds number in the forest is of the order of $10^6$. Hence, the naive expectation is that the IGM must be very turbulent on small scales, but the \citet{rsb01} observations suggest it is not. Can you think of an explanation?}

\abstract*{We continue our journey (together with cosmic gas) through the dense filaments of large scale structure to galactic halos. We will explore various ways the gas takes to accrete onto halos, and how hot gas in the halo uses radiative cooling to continue its journey to the galactic disk, becoming the ISM of galaxies. On our way we meet wonderful "cool streams", explore various ways to account for cooling and heating in the gas, and will end up with another mystery story of high velocity clouds.}

\section{From IGM to CGM}
\label{cgm}

Circumgalactic medium, or CGM, is often understood as the gas within the galactic dark matter halo. I am taking a broader view here, since some of the structures in the universe, like filaments, fall in the border zone between the IGM and CGM, they are not always considered to be part of the Lyman-$\alpha$ forest, but they also are not related to galaxies. They do produce absorption lines in the quasar spectra, but they also stream gas into galactic halos.

\subsection{Large Scale Structure}

Probably everyone has seen a picture of the large-scale structure of the universe by now (if you have not, check out excellent visualizations of the Millennium simulation at {\tt www.mpa-garching.mpg.de/galform/virgo/millennium}). Since the large-scale structure forms as a result of gravitational clustering from the linear Gaussian fluctuations, it is fully characterized by the linear matter power spectrum. Hence, various scales that we see in the pictures are all related to features in the power spectrum. For example, typical size of voids corresponds to about $1/2$ of the scale at which the power spectrum peaks (which is about $100h^{-1}\dim{Mpc}$ in comoving units). Hence, in comoving reference frame void sizes do not change - they are as large at $z=10$ as they are at $z=0$ (although, these largest voids are, of course, not nearly as empty at $z=10$ as they are at $z=0$). Filaments that surround voids are highly non-linear structures and their width is controlled by the the nonlinear scale at each epoch, i.e. the scale at which the amplitude of linear fluctuations reaches unity. Finally, material that makes the largest objects at any time (clusters of galaxies today, galaxies at $z\gg2$) is assembled from regions roughly the nonlinear scale in size, so masses of these objects are about $4\times\mbox{(mean density)}\times\mbox{(nonlinear scale)}^3$.

\begin{figure}[t]
\includegraphics[width=1\hsize]{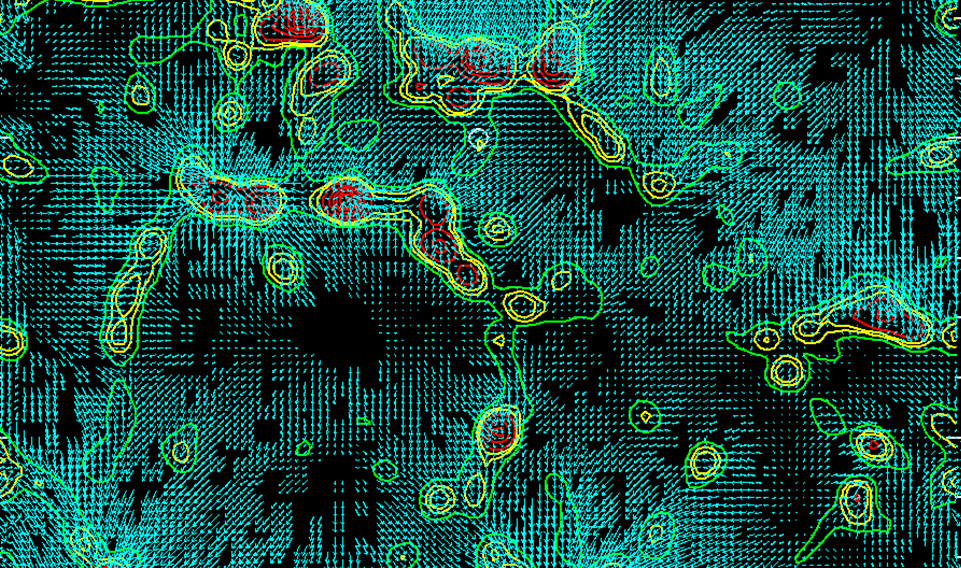}%
\caption{Large-scale flows (cyan arrows) on top of the density contours (green, yellow, and red). Flows of matter onto (almost all of the) filaments are clearly visible in this visualization of a numerical simulation (adopted from \protect\citet{khk03}). \label{fig:lss}}
\end{figure}

Since our main interest is how gas flows from low to high density regions, the actual motion of matter is of particular importance to us. With time voids become deeper as matter (both dark and gaseous) flows from them onto filaments, and then along the filaments into the galaxies. This pattern of flows is illustrated in figure \ref{fig:lss} from a numerical simulation of a local region around the Local Group by \citet{khk03}. 

As gas flows into a filament from opposite directions, it gets shocked, and the gas temperature is expected to rise above that maintained by photo-heating and adiabatic expansion/contraction - a complication that eventually destroys nice and tight density-temperature relation that exists in the lower density IGM.

\begin{figure}[b]
\sidecaption[t]
\includegraphics[width=0.64\hsize]{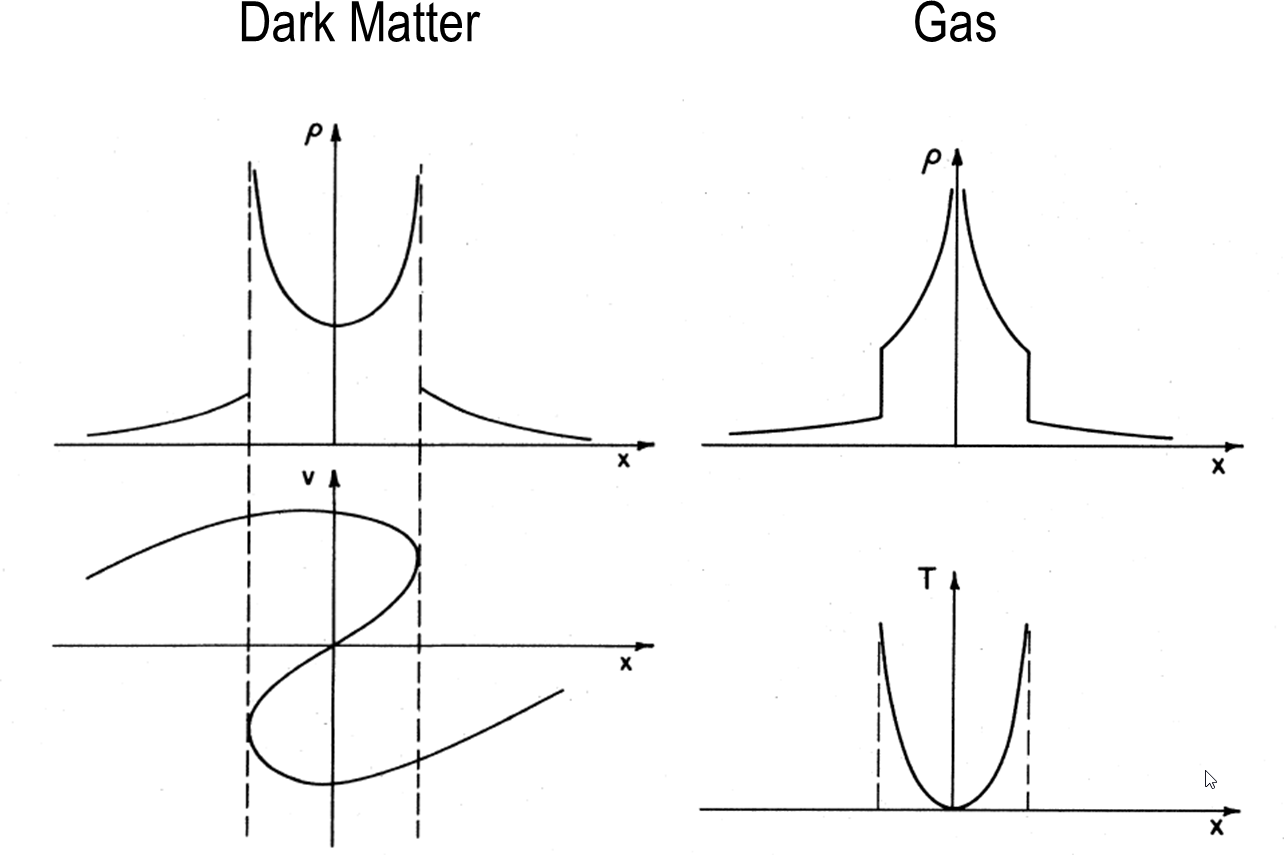}%
\caption{Profiles of one-dimensional collapse of dark matter and gas onto a 2D pancake (adopted from \protect\citet{sz89}). \label{fig:sz89}}
\end{figure}

The actual structure of the filaments received surprisingly little attention in the literature. In a classical review \citet{sz89} showed the profiles of one-dimensional collapse onto a 2D pancake (figure \ref{fig:sz89}). Collapse onto a 1D filament is qualitatively similar, because the physics is the same - gas gets piled up at the center, where the entropy is the lowest, while the dark matter from each side flows through the upcoming stream, creating density caustics on the outside. What happens next is determined by whether the filaments are self-gravitating - but since they have widths comparable to the nonlinear scale, we know that they, on average, \emph{are} self-gravitating. In a self-gravitating filament the dark matter will stop streaming, turn around, and fall on itself once again and again, increasing the number of intersecting streams as time goes on.
 
\begin{figure}[t]
\includegraphics[width=0.5\hsize]{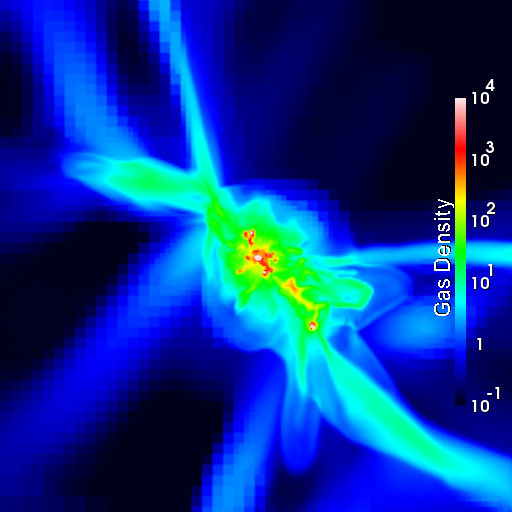}%
\includegraphics[width=0.5\hsize]{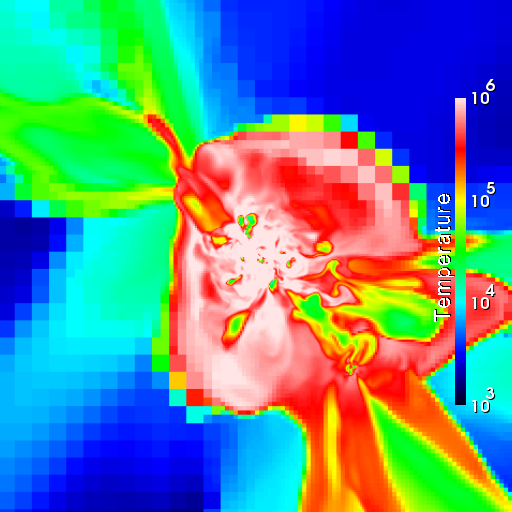}%
\caption{Thin slices through cosmological simulation that show the gas density (left) and the gas temperature (right) around a typical galaxy at $z\approx2$. \label{fig:galz2}}
\end{figure}

In order to illustrate the large- (and not-so-large-) structure further, I will use a cosmological simulation from \citet{ng:gk10}. This simulation is not very big, and focuses on the environs of a single, Milky-Way like galaxy, but it will suffer for our purpose. Figure \ref{fig:galz2} shows the gas density and the gas temperature around the main galaxy at $z=2$. 

There are a few features to note. First of all, the gas filaments do appear to be denser and cooler in the middle, similarly to the 1D collapse. Second, in the temperature plot we see really hot (million degrees) gas.  Most of that hot gas is concentrated around the galaxy, in the dark matter halo and beyond, but some of it extends way into the filaments - those are the temperature spikes that we see in figure \ref{fig:sz89}.

\subsection{How Gas Gets Onto Galaxies}

Everyone knows that dense enough regions of the large-scale structure will collapse and \emph{virialize} (i.e. reach, or, at least, approach, the virial equilibrium). The simplest model of such collapse is a \emph{top-hat},
\[
  \rho(\vec{x}) = 
	\left\{
    \begin{array}{ll}
	  \bar\rho(1+\delta_i), & r<r_i\\
	  \bar\rho, & r>r_i \\
	\end{array}
	\right.
\]
where $r_i$ and $\delta_i$ are the initial radius and amplitude of the perturbation.
The overdense perturbation collapses, and the evolution of the radius of the perturbation can be solved analytically in the matter-dominated regime ($a\propto t^{2/3}$), albeit parametrically with a parametric variable $\theta$:
\begin{eqnarray}
  r & = & \frac{GM}{\delta_i\dot{r}^2_i}(1-\cos\theta),\nonumber\\
  t & = & \frac{GM}{\delta_i^{3/2}\dot{r}^3_i}(\theta-\sin\theta).\nonumber
\end{eqnarray}
The moment of collapse is defined as $r=0$, which occurs at the time when $\theta=2\pi$. A remarkable property of the top-hat solution is that at the moment of collapse $t_f$ the linear density fluctuation
\[
  \delta_L(t) = \frac{D_+(t)}{D_+(t_i)}\delta_i
\]
is just a number, independent of the initial overdensity, size, or the mass of the overdense region,
\[
  \delta_L(t_f) = \frac{3}{5}\left(\frac{9\pi^2}{4}\right)^{1/3} = 1.69.
\]

A perturbation cannot collapse to a point - that would be even less likely than making a pencil stand on a sharp end. A standard assumption is that the collapsing perturbation virializes - i.e.\ reaches the virial equilibrium - at around the time $t_f$. In that case the average overdensity $\delta_v$ of the final virialized object is $1+\delta_v = 18\pi^2 \approx 178 \approx 180 \approx 200$.

The virial radius of the dark matter halo in figure \ref{fig:galz2} is roughly the green roundish region in the density panel (overdensity $\ga$ 100), while the million-degree gas extends well beyond it. The virial radius serves as a good approximation of a boundary beyond which any, even imaginable, resemblance of spherical symmetry totally vanishes! As gas falls into potential wells of dark matter halos, it gets shocked and heated to around the virial temperature (also deviations can easily be a factor of 2-3 in each direction). Shocks never stand still (in the reference frame of the gas behind them), so the accretion shock propagates outward. For typical cosmological objects, be it star-forming galaxies at $z\sim2$ or galaxy clusters at $z=0$ (or anything in between), it is not uncommon to find the accretion shock extending to 3 virial radii. Since it goes so much beyond the quasi-spherical region, it is highly asymmetric and non-spherical, with some of its protrusions reaching well into voids, up to $\sim10$ virial radii, while along filaments the accretion shock may not even exist (or do not reach to even a modest fraction of the virial radius).

\subsection{Cool Streams}

A story of the "cold streams" is a real-life story of an elephant-in-the-room. In a gesture of non-conformity, I am going to call them "cool streams", because in the ISM-speak (which we are going to use for the most of this course) the term "cold" refers to truly cold gas, below $100\dim{K}$. Strictly speaking, they should be called "warm streams", since $10^4\dim{K}$ gas is "warm" in the ISM-speak, but that would confuse too many people...

Every practicing simulator knew about cool streams, but no one paid any attention to them until in 2005 in an influential paper \citet{igm:kkw05} showed that at intermediate redshifts - the epoch where galaxies make most of their stars - cool streams deliver significant, or even dominant, fraction of gas onto the galactic disks, where stars actually form. Hence, from the point of view of a galaxy as a gas consumer, cool streams are the primary consumption channel.

\begin{figure}[t]
\includegraphics[width=1\hsize]{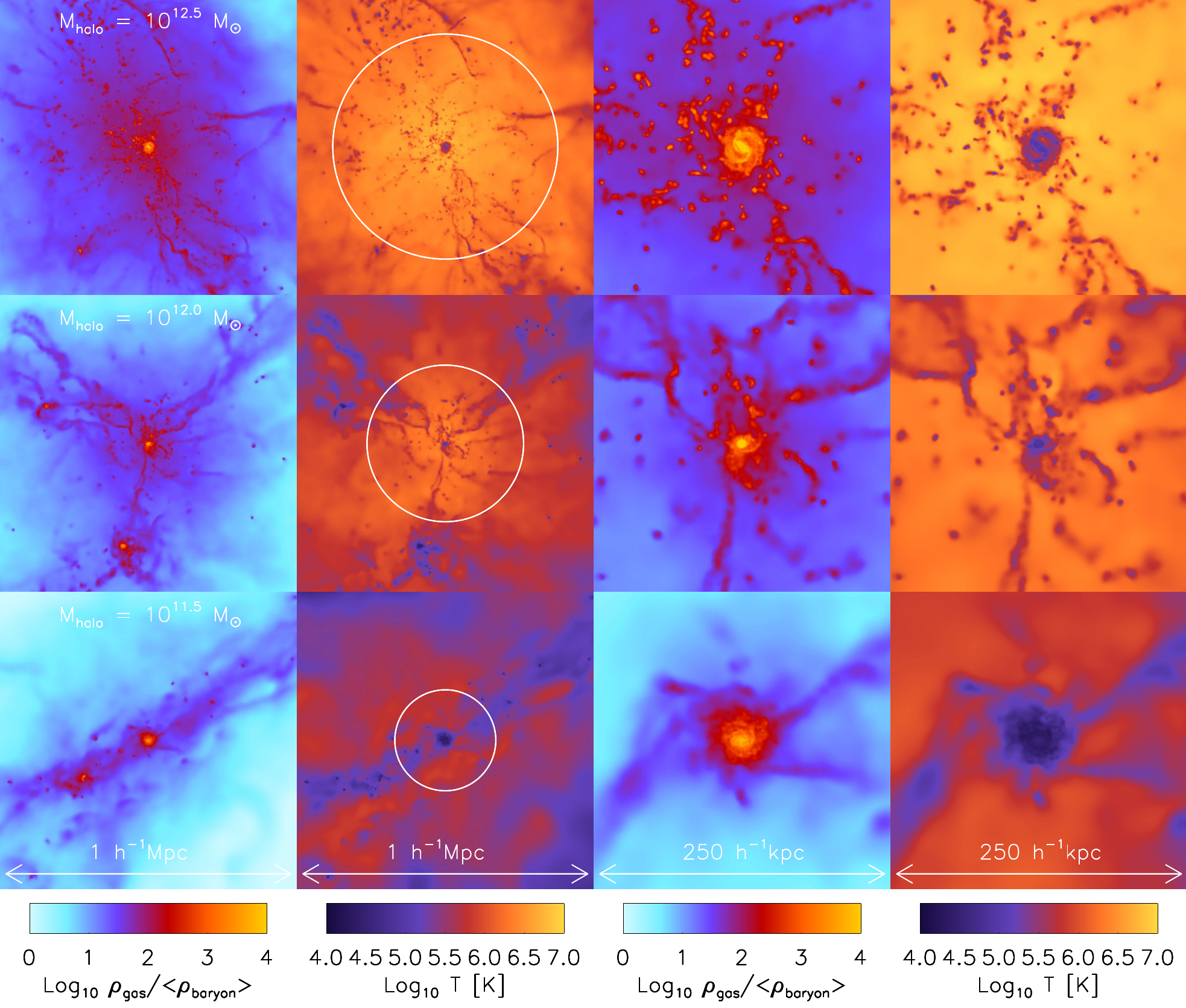}%
\caption{Density and temperature images of galaxies of different masses at $z\approx2$. Cool streams are clearly visible in temperature images as blue blobs and filaments (adopted from \protect\citet{vsb11}). \label{fig:coolimg}}
\end{figure}

Examples of cool steams in cosmological simulations from Overwhelmingly Large Simulation project \citep[OWLS, ][]{vsb11}) are shown in figure \ref{fig:coolimg}. As in a weird monster movie, the blue "tentacles" of cool gas try to reach the central galaxy; they break up into individual blobs for a massive one ($M=10^{12.5}\Msun$), remain as thin streams for a $M=10^{12}\Msun$ one, and completely swamp gas accretion for a Milky-Way type galaxy ($M=10^{11.5}\Msun$ at $z=2$). Images like that can be made from almost any cosmological simulation, and from any modern simulation code, be it an SPH code, an AMR, or a moving mesh code like AREPO\footnote{For these and other curious abbreviations check out Volker Springel's lectures in this volume.}\citep{springel2010}. All simulations agree that the cool flows dominate the gas accretion for halos above about  $M=10^{11.5}\Msun$ , with this mass being only weakly (if at all) redshift dependent.

Since most of the gas accretion occurs in low mass galaxies at all times, most of gas that ends up in galactic disks enters the halo as "cool" - significantly below the virial temperature, but it may still be well above the "ISM warm" of $10^4\dim{K}$ - at all cosmic times up to the present epoch. The contribution of cool streams is, however, diminishing with time, so by $z=0$ they, on average, only deliver about half of the accreting gas onto galactic disks.

\begin{figure}[t]
\sidecaption[t]
\includegraphics[width=0.64\hsize]{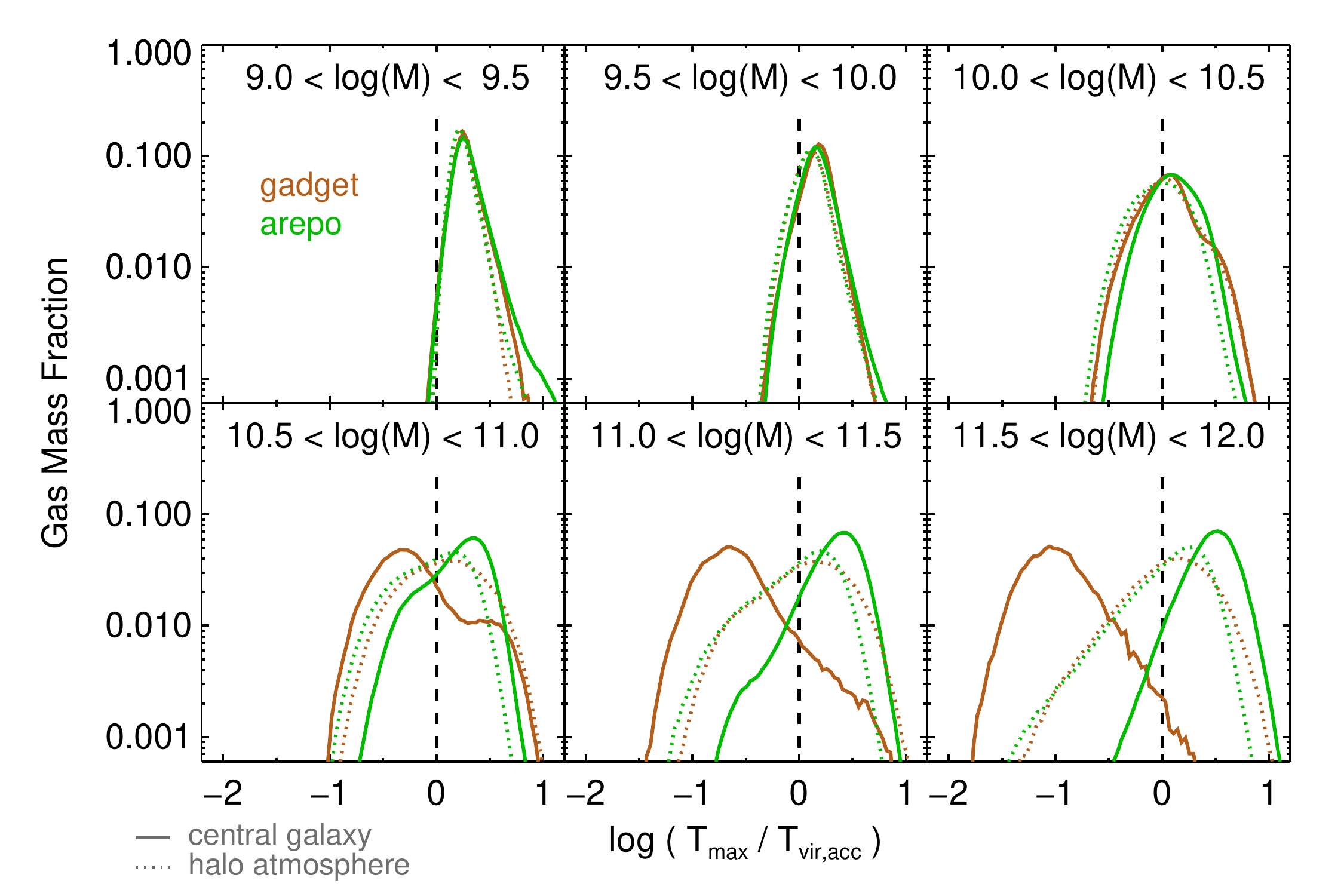}%
\caption{Temperature distribution function for galaxies in different mass bins simulated with GADGET and AREPO. The two codes predict significantly different distributions for high mass galaxies (adopted from \protect\citet{nvg13}). \label{fig:gadar}}
\end{figure}

A happy concordance is broken, however, when the fate of cool streams inside the halo is explored further. In a recent study, a carefully designed comparison between GADGET \citep{springel2005} and AREPO \citep{springel2010} codes found some disturbing differences \citep[][shown in figure \ref{fig:gadar}]{nvg13}. The two codes have the same gravity and dark matter solvers, but differ in the way gas dynamics is treated (for details, check Volker Springel's lectures in this volume). While in the SPH GADGET simulation the cool streams remain cool inside the halo and reach all the way to the galactic disk, in the mesh-based simulation with AREPO the cools streams heat up as they approach the disk. This discrepancy reflects the well-known dichotomy between SPH and mesh codes - the former do not have enough diffusion (without special fixes), while the latter may have too much numerical diffusion, especially in the poorly resolved regions. Which of the two codes is closer to reality is not yet clear; the progress in this field, though, happens at a relativistic speed, so as you are reading these lectures, the ambiguity may have been already resolved.

\subsection{Galactic Halos}

Few sane people doubt the existence of dark matter halos. Whether galaxies have gaseous halos is an entirely different matter. 

Cosmological simulations generically predict that galaxies like the Milky Way (MW) should be surrounded by hot gaseous halos in quasi-virial equilibrium. For two decades the actual existence of these hot halos was an even more hotly debated topic. The point of contention was the simple fact that hot gas emits X-rays, hence hot halos must be detectable in X-rays. The cruel reality is that the halo gas is rather tenuous, and for galaxies like the Milky Way it is expected to have temperatures that are very hard to detect observationally. 

How much gas one expects to reside in the Milky Way halo actually depends on the halo mass, which has been notoriously difficult to estimate. Proposed values range from $\sim 7\times10^{11}\Msun$ to over $2\times10^{12}\Msun$ (values outside this range are considered extremist and will provoke a French military intervention or an American bombing campaign). For the fiducial value of $10^{12}\Msun$ the cosmic share of baryons in the MW is $1.6\times10^{11}\Msun$. The stellar mass of the MW is about $6\times10^{10}\Msun$ (although, values up to $8\times10^{10}\Msun$ are sometimes used) and the disk gas mass is $\la1\times10^{10}\Msun$. Hence, the gaseous halo may contain up to $10^{11}\Msun$ (it may, of course, be much less if some of the gas is expelled from the Galaxy by stellar feedback and other energetic processes).

The contention about the existence of the hot halo finally has been resolved by \emph{Chandra} - not the brilliant man who resolved so many other contentions, but the remarkably successful space mission named after him. In a ground-breaking observation the Chandra team finally detected the X-ray emission from the hot gas around the Milky Way \citep{gmk12}. While measuring the total mass of the halo from Chandra observations is very challenging (try measuring the mass of a giant monster that swallowed you), the limits that the Chandra team has been able to place on the gas mass in the halo are consistent with our estimate of $10^{11}\Msun$.

\begin{figure}[b]
\sidecaption[t]
\includegraphics[width=0.64\hsize]{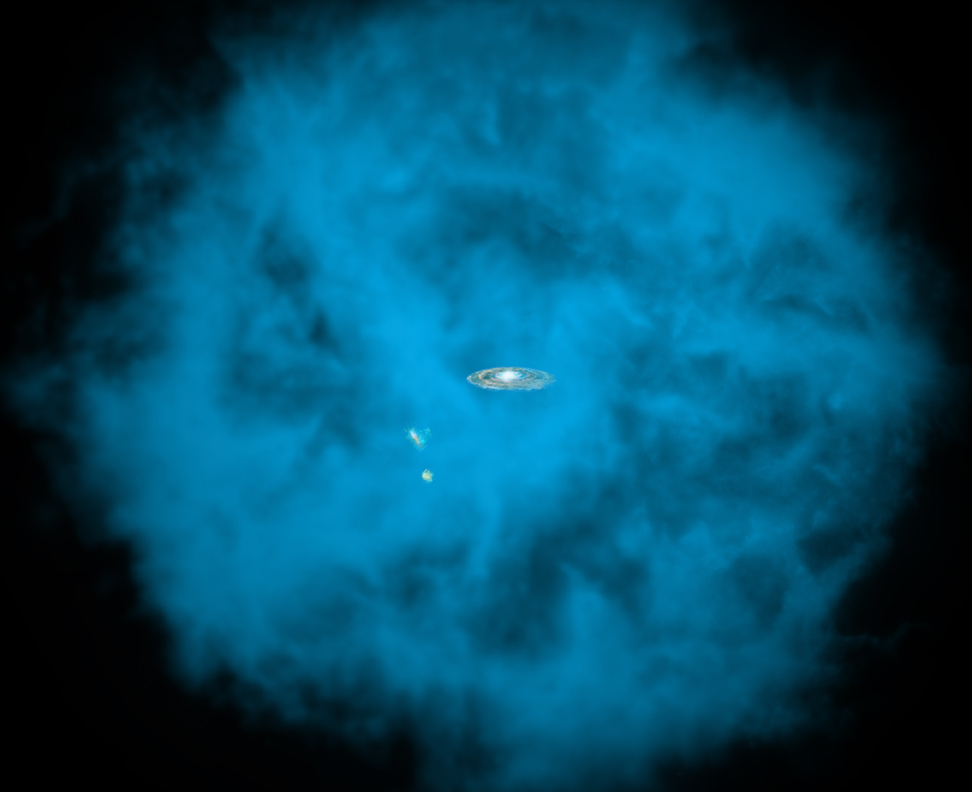}%
\caption{Radial density profile at $z=0$ of the same galaxy shown in figure \ref{fig:galz2}. Blue and red solid lines shows the actual simulated profiles of dark matter and gas, while dashed lines give the best-fit NFW (for dark matter) and rescaled by $0.05f_b$ NFW (for gas) profiles respectively. The red dotted line is the best-fit beta profile for the gaseous halo. Filled and open symbols are pre-Chandra  observational constraints \protect\citep{ww96,br00,gp09,qm01,sdk02,ab10}.\label{fig:halo}}
\end{figure}

X-ray detection of the halo is important, because it is a \emph{direct} evidence for the existence of a massive (from the point of view of the disk) gaseous halo. Historically, however, a large number of indirect constraints existed that all pointed out towards the same conclusion. In figure \ref{fig:halo} I show the $z=0$ dark matter and gas profiles for the same galaxy we met in figure \ref{fig:galz2}. The hot halo (solid red line) in that simulation is consistent with the existing pre-Chandra observational constraints as well as with the actual Chandra measurement. What is remarkable is that in the simulation all stellar feedback processes were switched off \citep[see][for details about the actual simulation]{ng:g12}. The galactic disk in the simulation is overly massive and has incorrect density profile, but the halo seems to be ok (at least within the precision of observational constraints). There is, actually a simple reason for it - the main physical process that matters for the gas in the halo is radiative cooling, it is cooling that determines which gas can rain on the disk and which remains in the halo in the hot phase. 

Hence, the physics of radiative cooling is our next stop.

\subsection{Diversion: Cooling of rarefied gases}

Before we proceed further along our yellow brick road, let's step aside for a short while and consider how  cosmic gas cools - the process we have already met in the IGM segment of our journey, and which we will be meeting over and over again in the future.

\emph{Radiative cooling} is an "umbrella" name for diverse physical processes through which gas transforms its thermal energy into the radiation that leaves the system. At low enough density, three processes dominate, and all three of them involve a collision of a free electron or an atom/ion with a neutral atom or a partially neutral ion. These three processes are
\begin{description}
\item[{\bf line excitation}:] a collision excites the neutral atom into a higher energy state, the state decays and the resultant photon leaves the system;
\item[{\bf collisional ionization}:] a collision ionizes the neutral atom and the binding energy of the freed electron is charged against the thermal energy account;
\item[{\bf recombination}:] an ion captures a free electron and the sum of the kinetic energy of the electron and the binding energy of the neutral atom is emitted as a photon.
\end{description}
All these collisional processes depend on the square of the density, so it is convenient (and customary) to factor our that density dependence explicitly in the cooling rate of the gas,
\[
  \left.\frac{dU}{dt}\right|_{\rm cool} = -n_b^2 \Lambda(T,...),
\]
where $n_b$ is the number density of baryons (I prefer it to another commonly used parametrization that factors out the hydrogen nucleus number density $n_\Ht$, because $n_b$ is directly proportional to the gas mass density for any value of helium abundance or gas metallicity) and $\Lambda$ is commonly called a \emph{Cooling Function}.

\begin{figure}[t]
\sidecaption[t]
\includegraphics[width=0.64\hsize]{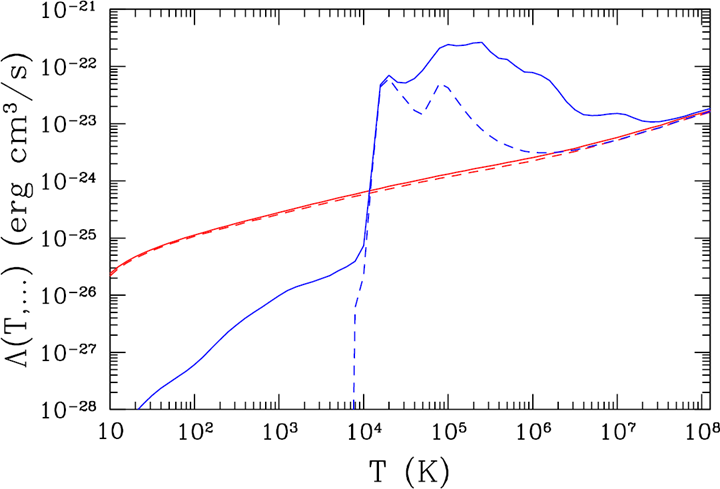}%
\caption{Cooling functions for the primordial gas ($Z=0$, dashed lines) and the gas at solar metallicity (solid lines). Blue lines show the "standard" CIE cooling functions, while red lines show the cooling functions for the fully ionized gas (the only cooling process is Bremsstrahlung).\label{fig:cfcie}}
\end{figure}

In the simplest case of gas in pure collisional equilibrium (no external or internal radiation of any kind - the so-called \emph{collisional ionization equilibrium}, or CIE) the cooling function is called "standard". If the relative abundance of various chemical elements is fixed and small variations in the helium abundance are neglected, the the cooling function only depends on the gas temperature $T$ and the total metallicity $Z$,
\[
  \Lambda_{\rm CIE} = \Lambda_{\rm CIE}(T,Z).
\]
Examples of this function for $Z=0$ and $Z=\Zsun$\footnote{Throughout these lectures I define "solar metallicity" as the metallicity of our galactic neighborhood, $\Zsun=0.199$ in absolute units, rather than  metallicity of an average-looking single star somewhere in the outskirts of the Galaxy.} are plotted in figure \ref{fig:cfcie}. 

\begin{figure}[b]
\sidecaption[t]
\includegraphics[width=0.64\hsize]{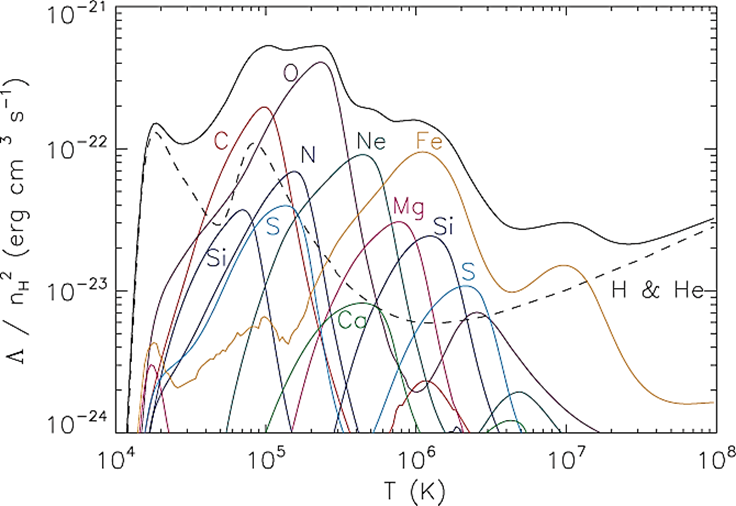}%
\caption{Contributions of individual chemical elements to the "standard" CIE cooling function (adopted from \protect\citet{sims:wss09}).\label{fig:cfcomps}}
\end{figure}

The specific shape of the CIE cooling function, with its "bumps and wiggles", is determined by the interplay between contributions of over a dozen various chemical elements. A good recent review is given by \citet{sims:wss09}, an illustration from which is reproduced here in figure \ref{fig:cfcomps}. In particular, one has to be aware that many of the atomic cooling rates used to construct the cooling function are know rather poorly, not better than a factor of 2, and that uncertainty propagates into the actual value of the cooling function. In realistic galactic and cosmological simulations this uncertainty is often, however, unimportant: the cooling time-scale is so much shorter than any other physical time-scale in the problem that it does not need to be known very precisely (all gas that can cool will indeed cool rapidly).

\citet{sims:wss09} paper offers another, much more important lesson, though. As they show, the actual cooling function in the IGM, CGM, and even ISM of galaxies at low and high redshifts may deviate from the "standard" one quite substantially. In other words, the "standard" CIE cooling function is actually highly non-standard and is almost never realized in nature. The reason for that is that low density cosmic gas is always affected by external radiation field.

\begin{figure}[t]
\sidecaption[t]
\includegraphics[width=0.64\hsize]{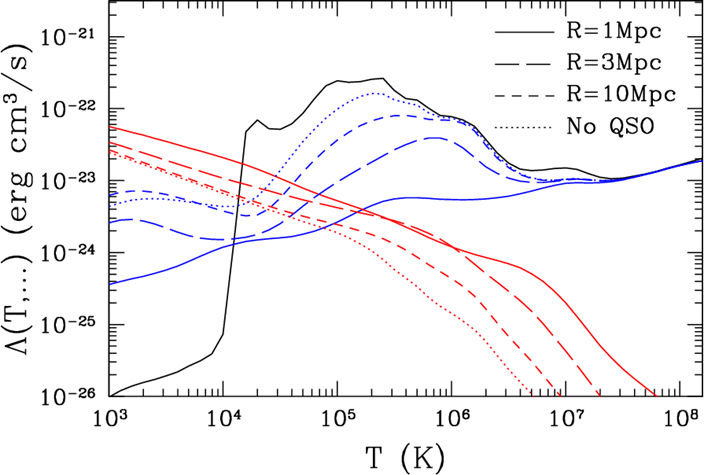}%
\caption{Illustration for the role of radiation field in suppressing cooling (blue lines) and enhancing heating (red lines). A gas in the galactic halo (at density 340 over the cosmic mean) is shined upon by the $10^{12}L_\odot$ quasar. Sufficiently close, the quasar radiation modifies the cooling and heating functions in a major way.\label{fig:cfpie}}
\end{figure}

Figure \ref{fig:cfpie} is a simple illustration of this process. Cooling (and heating) processes in a gaseous halo can be modified in a major way if it straddles too close to a strong source of ionizing radiation, such a bright quasar. Within $1\dim{Mpc}$ from the quasar, the equilibrium temperature in the halo goes all the way up to $200{,}000\dim{K}$, twenty times above our "canonical" $10{,}000\dim{K}$.

So, let us review the cooling function from the very beginning, this time being careful. In a most general case in addition to cooling there is also radiative heating by the radiation field. Hence, the change of the gas internal energy due to radiative processes has two terms with opposite signs,
\[
  \left.\frac{dU}{dt}\right|_{\rm rad} = n_b^2\left(\Gamma - \Lambda\right),
\]
where $\Lambda$ is our old acquaintance the cooling function and $\Gamma$ is the \emph{heating function}. Both of them depend on a multitude of parameters,
\begin{equation}
   [\Gamma,\Lambda] = {\cal F}\left(T,n_b,X_{ijl},J_\nu,\tau_{ijl}\right),
  \label{eq:chfex}
\end{equation}
where the density dependence reappears because not all processes are two-body, $X_{ijl}$ is the abundance of the chemical element $i=\Ht,\mbox{He},...$ in the ionization state state $j=\mbox{neutral},\mbox{single ionized},...$ in the quantum state $l$, $J_\nu$ is the spectrum of the incident radiation field that shines on a given (formally infinitesimally small) parcel of gas, and $\tau_{ijl}$ are opacities in each radiative transition (gas may be optically thick to some of its own cooling radiation if our parcel is embedded deep inside a huge cloud). For the sake of brevity in notation, we will use ${\cal F}$ to label either $\Gamma$ or $\Lambda$, since both functions always depend on the same set of arguments.

In order to compute the cooling and heating functions in such a detail one needs a highly sophisticated computer code that, in its complexity, rivals modern cosmological simulation codes. Fortunately, such codes exist, and the most famous and widely used of them is \emph{Cloudy}\footnote{Notice the convention, \emph{Cloudy} is a name, not an abbreviation.}. Conceived by Gary Ferland from the University of Kentucky and contributed to by many people, \emph{Cloudy} is freely available from its website, {\tt nublado.org}, and is well-documented for a fast start-up curve. 

There is one problem only with \emph{Cloudy} - it is way too complex to be used in modern simulation codes for computing cooling functions "on the fly". Perhaps in the future, in the era of exa-scale computing, it will be possible to run \emph{Cloudy} as a "sub-grid" model in real simulations, but for now we need to seek approximate short-cuts.

So, what one can do? Unless densities are very high (hence our focus on \emph{low density gas}), the gas will be optically thin to its own cooling radiation, so the dependence of cooling and heating functions on $\tau_{ijl}$ disappears - for this to be exactly true, we also should exclude all cooling and heating processes due to molecules, since those always require radiative transfer to be followed properly. Thus, if you need to follow molecular cooling/heating as well, you will have to add them "manually" to the cooling and heating functions that we discuss below.

Second, in almost all galactic and cosmological simulations the assumption of the \emph{ionization and excitation equilibrium} is not a bad one. In the ionization equilibrium the distribution of a given chemical element over various ionization states is uniquely determined by density, temperature, and the radiation field. The same is true about various quantum levels in the local thermodynamic equilibrium. If, in addition, we assume that relative abundances of chemical elements are fixed (say, to the solar abundance pattern), then the dependence on $X_{ijl}$ reduces to the simple dependence on the overall gas metallicity $Z$,
\begin{equation}
  {\cal F} = {\cal F}(T,n_b,Z,J_\nu).
  \label{eq:chfred}
\end{equation}
Very often this latter expression is what actually called a "cooling/heating function". But even the latter form is unusable in modern simulations codes, because it includes an explicit dependence on the radiation spectrum, which is an arbitrary function of frequency (in a strict mathematical sense ${\cal F}$ in equation (\ref{eq:chfred}) is actually an \emph{operator}, not a function). Hence, we still need to account for that dependence in an approximate manner.

One particular short-cut has been used in many cosmological codes for over a decade. \citet{sims:wss09} paper again serves as a good reference, although the first known (to me) example of such approach is used by \citet{sims:k03}. In the most of the volume of the universe the dominant source of external radiation is the cosmic background that we already met in the previous chapter. The cosmic background is uniform in space and is a function of the cosmic redshift only, hence in the limit when $J_\nu$ can be approximated by the cosmic background, cooling and heating functions become functions of 4 arguments, temperature, density, gas metallicity, and cosmic redshift, and hence can be easily tabulated and used in simulation codes efficiently via a simple table look-up.

Unfortunately, most of the volume in the universe contains only a modest fraction of the mass, and even smaller fraction of action. The radiation field in the ISM (and, in at least part of the CGM) of galaxies is dominated by local radiation sources (for example, the UV radiation field in the solar neighborhood is 500 times higher than the cosmic background; at the center of the galaxy that ratio jumps to 5{,}000). Since stars form in the ISM, any galactic or cosmological simulation that attempts to model star formation cannot use cooling and heating functions which only account for the cosmic background.

How one can attempt to construct a more accurate short-cut? After all, the effect of external radiation is in ionizing some of the chemical elements and/or exciting particular levels, and ionization and excitation rates are all integrals over the radiation spectrum with some cross-sections, which are broad and relatively slowly varying functions. Let's imagine the following thought experiment: we take a given spectrum and increase the radiation intensity in a narrow frequency bin between some $\nu_0$ and $\nu_0+\Delta\nu$. If the increase is large, the cooling and heating functions will be affected. Now shift the frequency bin to $\nu_0-\Delta\nu$  to $\nu_0$. Most of ionization and excitation rates will be barely affected (unless we choose $\nu_0$ very carefully to correspond exactly to the ionization/excitation threshold of an important cooling channel), since cross-sections of most physical processes will not change significantly between the two narrow bins. Hence, in order to compute the cooling and heating functions accurately, we do not need to know the radiation spectrum in excessive detail (say, in hundreds of frequency bins), but it may be sufficient to describe it by several "broadband filters".

There can be infinitely many choices for these filters. In a specific implementation of this idea, Nick Hollon and I decided to use photoionization rates of several chemical elements as "broadband filters". After all, the ionization balance is controlled by photoionization rates, so it makes sense from the atomic physics perspective. We have explored over 20 various chemical elements and their ionization states, and the best approximation that we have been able to come up depends on just 4 ionization rates \citep{ng:gh12}.

Specifically, we adopt the approximation in which the metallicity dependence of the cooling and heating functions is expanded into the Taylor series in gas metallicity,
\begin{eqnarray}
  {\cal F}(T,n_b,Z,J_\nu) = \sum_{i=0}^n \left(\frac{Z}{\Zsun}\right)^i {\cal F}_i(T,n_b,J_\nu),
  \label{eq:chfgh12a}
\end{eqnarray}
with $n=2$ providing a highly accurate approximation for $Z<5\Zsun$. Each of the expansion coefficients is approximated as
\begin{equation}
  {\cal F}_i(T,n_b,J_\nu) \approx {\cal F}_i(T,\left\{r_j\right\},n_b),
  \label{eq:chfgh12b}
\end{equation}
with several parameters $r_j$ encapsulating the full dependence of the cooling and heating functions on the external radiation field.

A parameter set that we found to work well is defined as follows:
\begin{eqnarray}
  r_1 & = &  \frac{P_{\rm LW}}{n_b}, \nonumber\\
  r_2 & = & \left(\frac{P_\HI}{P_{\rm LW}}\right)^{0.353}
            \left(\frac{P_\HeI}{P_{\rm LW}}\right)^{0.923}
            \left(\frac{P_\CVI}{P_{\rm LW}}\right)^{0.263}, \nonumber\\
  r_3 & = & \left(\frac{P_\HI}{P_{\rm LW}}\right)^{-0.103}
            \left(\frac{P_\HeI}{P_{\rm LW}}\right)^{-0.375}
            \left(\frac{P_\CVI}{P_{\rm LW}}\right)^{0.976},
  \label{eq:chfgh12c}
\end{eqnarray}
where $P_{\rm LW}$ is the rate of photo-destruction of molecular hydrogen (molecules are excluded from the cooling and heating functions, since they cannot be treated without radiative transfer, so we use $P_{\rm LW}$ just as a convenient "broadband filter" for the radiation below the hydrogen ionization threshold) and $P_\HI$, $P_\HeI$ and $P_\CVI$ are photoionization rates of $\HI$ (ionization edge of $1\dim{Ry}$), $\HeI$  (ionization edge of $1.8\dim{Ry}$), and $\CVI$  (ionization edge of $36\dim{Ry}$). These rates sample a large range of photon energies, and serve as a good set of more-or-less independent "broadband filters".\footnote{They are not fully independent, of course - a photon ionizing $\CVI$ can also ionize neutral hydrogen, but it is convenient to use photoionization rates rather that some other, arbitrary filter shapes, since the same rates can be useful in the simulation code for other purposes - for example, for computing the ionization balance of hydrogen, helium, or other chemical elements.} 

\begin{figure}[t]
\sidecaption[t]
\includegraphics[width=0.64\hsize]{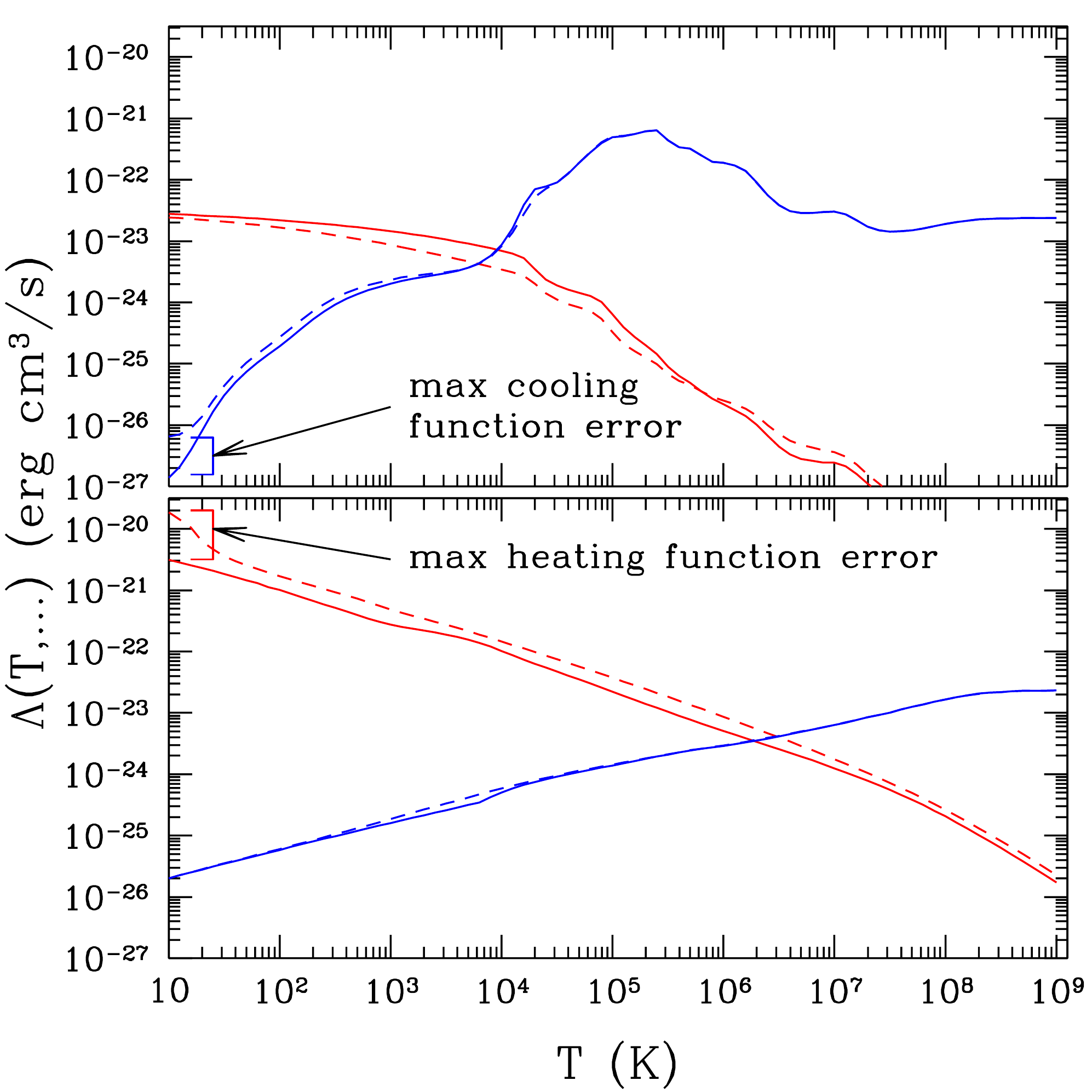}%
\caption{Cooling (blue lines) and heating (red lines) functions for our test
models that maximize the error in the cooling function (top panel) and the
heating function (bottom panel). Approximate functions from equations (\protect\ref{eq:chfgh12a}-\ref{eq:chfgh12c}) are shown as dashed lines, while exact calculations from Cloudy are shown with solid lines.\label{fig:cferr}}
\end{figure}

The main problem with approximation (\ref{eq:chfgh12a}-\ref{eq:chfgh12c}) is that it occasionally results in "catastrophic errors" - for example, if you choose the radiation field, gas temperature, density, and metallicity at random, in about 1 case out of the million the approximate cooling or heating function will deviate from the actual Cloudy calculation by a factor of several (that is a consequence of not being able to fully represent all possible variations in the radiation field by just 3 coefficients $r_j$, $j=1,2,3$). Figure \ref{fig:cferr} demonstrates the worst-case catastrophic error of the approximation. 

The good news is that these catastrophic errors occur for either highly implausible or completely irrelevant values of parameters - for  example, the large error in the heating function at $T\sim10\dim{K}$ in the bottom panel of figure \ref{fig:cferr} is not very important because the heating function there is much larger than the cooling function, and the equilibrium temperature (blue and red lines cross) is $T_{\rm eq}\approx 2\times10^6\dim{K}$. If the gas at $10\dim{K}$ finds itself suddenly in such conditions, it will be heated to above million Kelvins rapidly, quickly leaving the parameter space where the approximation is inaccurate. 

Similarly, the large error in the cooling function at $T\sim10\dim{K}$ in the top panel of figure \ref{fig:cferr} is irrelevant, because the heating function in those conditions is more than 3 orders of magnitude larger than the cooling function, hence it is not important to know the cooling function at all.

Undoubtedly, a better approximation for the cooling and heating functions is possible, but in the absence of such, equations (\ref{eq:chfgh12a}-\ref{eq:chfgh12c}) provide a practical way to fully account for the effect of the radiation field in modern cosmological and galactic simulations.

\subsection{Back to Galactic Halos}

Armed with the understanding of the cooling and heating functions, we can now return to the fate of gas in galactic halos. As gaseous halos are expected to become denser at the center, the cooling time will decrease towards the center. Hence, there must exist a \emph{cooling radius} $R_C$ at which the cooling time is equal to the age of the halo. Gas inside $R_C$ is able to cool efficiently and condense towards the halo center, while the gas outside $R_C$ cools too slowly and will remain in the (quasi-) hydrostatic equilibrium.

\begin{figure}[t]
\sidecaption[t]
\includegraphics[width=0.64\hsize]{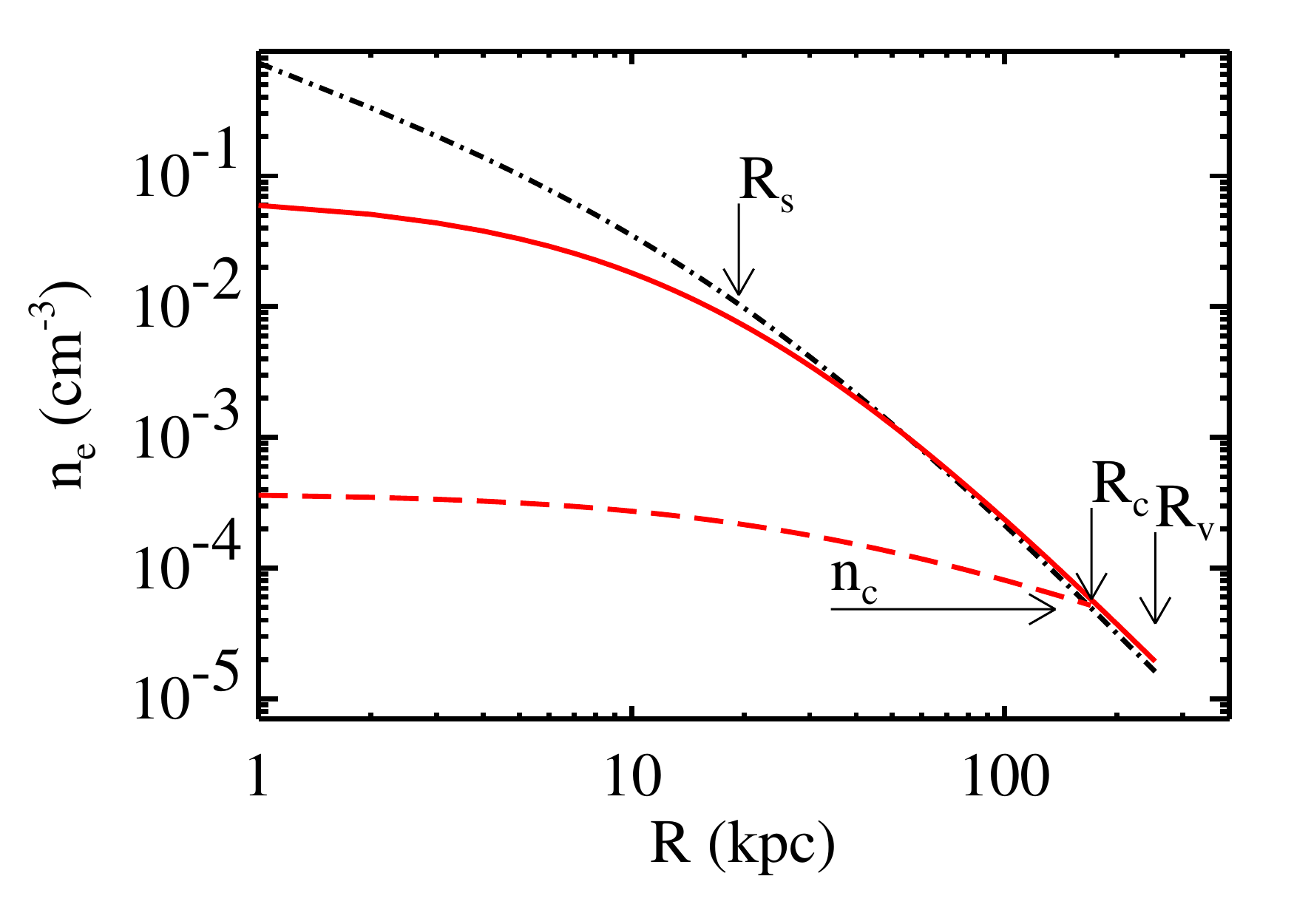}%
\caption{Density profiles of the hot phase of halo gas in \protect\citet{mb04} model in the absence of cooling (solid red line) and with cooling properly accounted for (dashed red line) for a Milky Way like galaxy at $z=0$. The dot-dashed black line shows the NFW profile (adopted from \protect\citet{mb04}).\label{fig:mb04}}
\end{figure}

A detailed analysis of the cooling process is well presented in \citet{mb04}, although they were not the first group who considered that process. In figure \ref{fig:mb04}, adopted from that paper, the final profile of the hot gas is shown with the dashed red line. The density profile is cored - all the gas above some threshold density is able to cool, and the core density is set by the requirement that the cooling time in the remnant of the core gas is longer than the age of the halo.

The gas that is able to cool will stream towards the center and will settle into a galactic disk. It can do that, however, in two distinct ways: it can either develop a “cooling flow” and smoothly flow in a quasi-spherical way all the way to the center, or it can experience thermal instability, split into individual dense clouds, which then fall onto the disk along parabolic orbits like rain drops fall on the ground. Which of these two ways dominates is still a completely open question, with the observational evidence being sparse and inconclusive.

Clouds of neutral hydrogen (hence dense and cool) are indeed detected in the halo of the Milky Way, they are commonly known as "high velocity clouds" (HVC), since they are detected in radio observations as neutral hydrogen at velocities significantly offset from the gas in the galactic disk (clouds that are not offset in the velocity would not be distinguishable from the disk itself).

\begin{figure}[b]
\includegraphics[width=1\hsize]{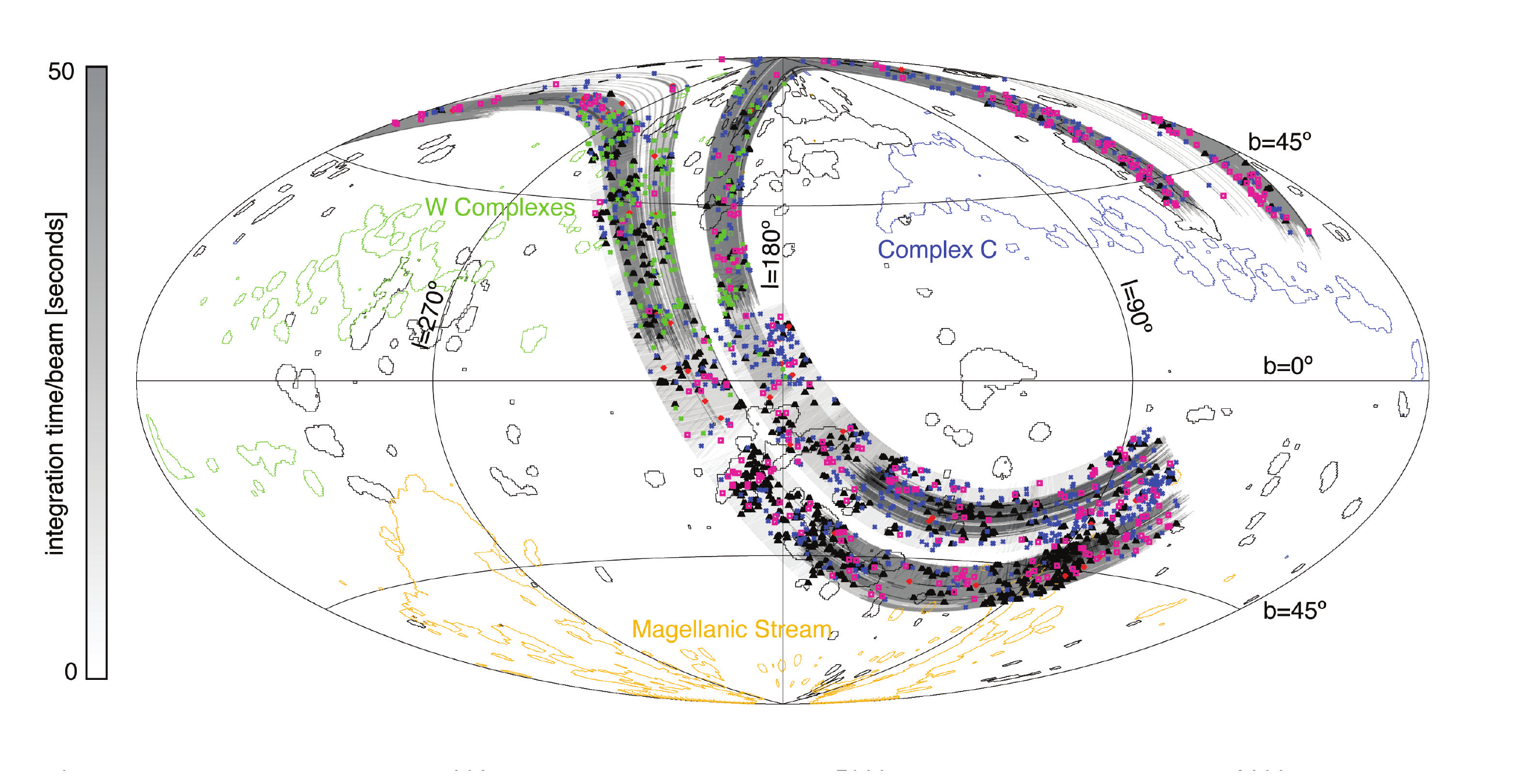}%
\caption{Neutral hydrogen clouds in the halo of the Milky Way discovered by the GALPHA-HI survey. Various colors mark the cloud type, with HVC plotted in black (adopted from \protect\citet{sea12}).\label{fig:hvc}}
\end{figure}

For example, the recent GALPHA-HI survey by Arecibo telescope uncovered a large number of new clouds \citep[][as shown in figure \ref{fig:hvc}]{sea12}. Unfortunately, from the radio observations alone it is very hard to determine the distances to those clouds. Perhaps, they are not located in the halo but form the so-called "galactic fountain", with the gas being thrown up by stellar feedback.

One way to resolve the ambiguity is to search for high velocity clouds in external galaxies. Alas, even in our neighbor Andromeda galaxies none have been found. Andromeda is sufficiently far away for sufficiently small clouds to remain undetected, so the jury is still out on whether HVCs are indeed the halo gas raining onto the galactic disks or the disk gas pushed (temporarily) into the halo.

One part of the problem is that the 21 cm line that is used to detect neural hydrogen in radio observations is one of the weakest lines in this universe. Neutral hydrogen also has one of the strongest lines - Lyman-$\alpha$. However, it is not easy to excite $n=2$ level in the hydrogen atom, hence Lyman-$\alpha$ is usually seen in absorption.

\begin{figure}[b]
\sidecaption[t]
\includegraphics[width=0.64\hsize]{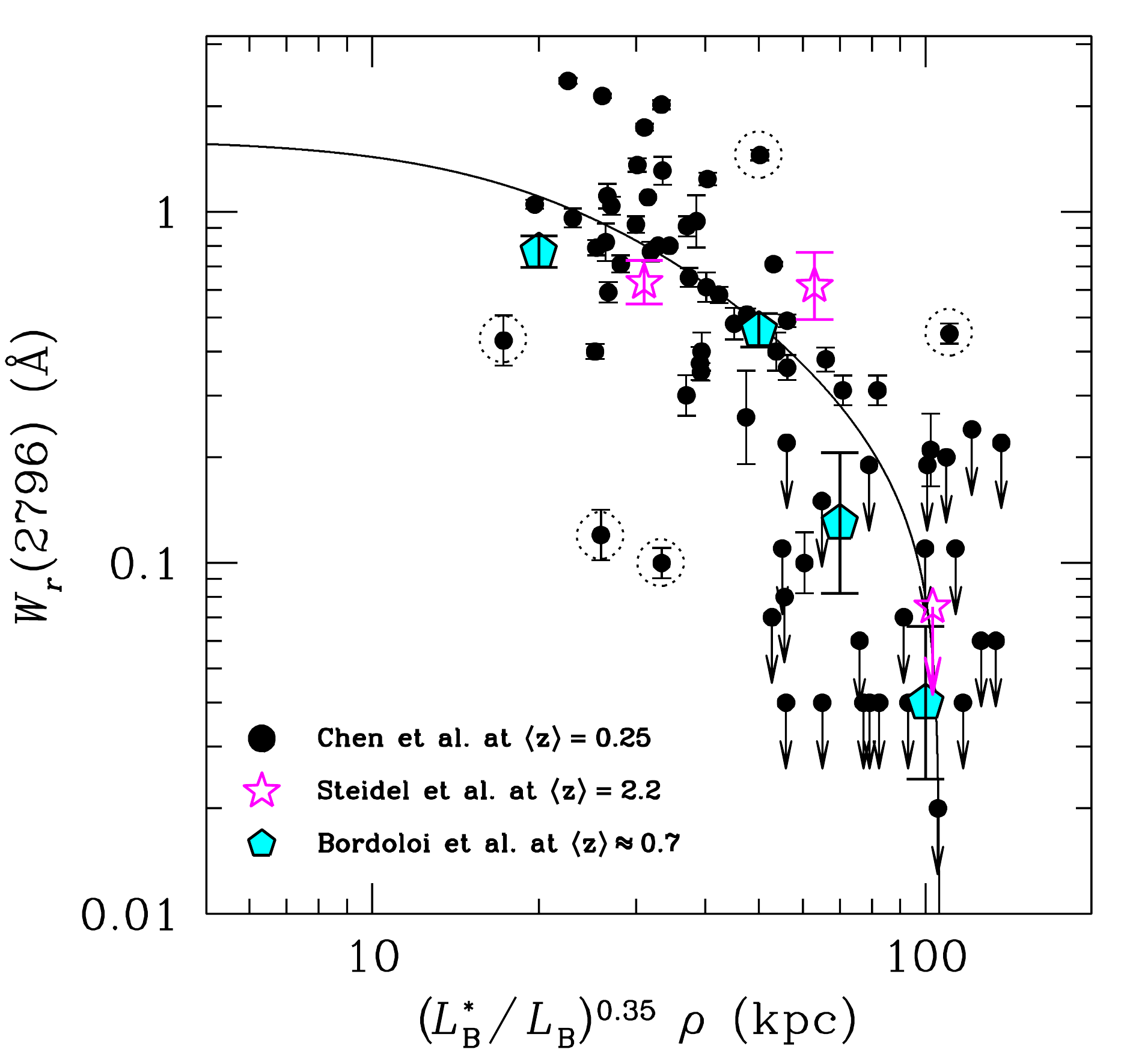}%
\caption{Strength of $\ion{Mg}{II}$ absorption as a function of distance from he host galaxy. There is a sharp drop in absorption for distances in excess of about $100\dim{kpc}$, probably indicating the cooling radius for halos (adopted from \protect\citet{c12}).\label{fig:mg2}}
\end{figure}

That is where other chemical elements come to rescue. Even while we are primarily after hydrogen, a trace amount of heavy elements may produce enough absorption in some of their, more easily excitable and observable lines. One such element is Magnesium, the ionization threshold of its singly ionized state is just $15\dim{eV}$, very close to the hydrogen ionization threshold of $13.6\dim{eV}$. Because of that, $\ion{Mg}{II}$ has been used as a proxy for neutral hydrogen in absorption studies of galaxies for several decades. Figure \ref{fig:mg2} shows a plot from a recent compilation of observational constraints in several ions by \citet{c12}. A general feature of all observations is that $\ion{Mg}{II}$ drops precipitously further away that about $100\dim{kpc}$ from a galaxy (with a mild dependence on the galaxy luminosity). It is highly tempting to associate this drop with the cooling radius for the halo, and $\ion{Mg}{II}$ with the cool clouds formed by thermal instability, but in the absence of additional evidence such a proposition will remain no more than a plausible conjecture.

One way or the other the gas from the halo (and beyond) ends up in the galactic disk, making up the Interstellar Medium (ISM) of galaxies. This is where our yellow brick road leads us next.

\def\D{D_{\rm MW}}
\def\U{U_{\rm MW}}

\abstract*{We are now arriving into the galactic ISM. After a brief refresher on galaxy formation, we explore the structure and stability of galactic disks, going well beyond standard Toomre criterion. We then focus our attention to gas, and after a brief overview of atomic and ionized gas dive even deeper into the molecular ISM. We will explore the atomic-to-molecular transition in exquisite detail, before committing a grave mistake of opening a cosmic Pandora box of the $X$ factor, inside which, as in a Russian Matrioshka doll, we find another Pandora box, and then another, until in total desperation we give up and ready ourselves to jumping into the domain of stars.}

\section{ISM: Gas In Galaxies}
\label{ism}

The field of Interstellar Medium takes easily a quarter of all of Astronomy. Any attempt to review it at any reasonable level will result in me still writing these lectures on my deathbed. Hence, our journey through the ISM realm will be brief and highly focused - we will be mainly concerned with "gas in galaxies", i.e.\ gas as a medium (forget about chemistry, except for one very specific topic), and gas as a galaxy component (i.e.\ not small-scales behavior of gas, but rather the role of gas as a citizen of a galaxy). Even with these restrictions, the journey that lays ahead is extremely biased towards my own research interests and topics I find fascinating.

\subsection{Galaxy Formation Lite}

Galaxies are rather complex creatures; understanding galaxy formation and evolution is the current frontier of extragalactic astronomy and cosmology. Never-the-less, the basic sketch of how galaxies form and evolve has been developed - it is captured by the \citet{gals:mmw98} model (hereafter MMW98).

The cornerstone assumption of MMW98 model is that the cool ($\sim10^4\dim{K}$) gas is delivered to the bottom of the potential well of a dark matter halo - either by radiative cooling in the halo or by inflow along cool flows. The specific way by which gas is delivered is unimportant; what matters is that the angular momentum is conserved, and hence the cool gas settles into a rotationally-supported disk.

It is convenient to parametrize the mass of the disk $M_d$ as a fraction $m_d$ of the halo mass $M_h$,
\[
  M_d = m_d M_h,
\]
and the disk angular momentum $J_d$ as a fraction $j_d$ of the halo angular momentum $J_h$,
\[
  J_d = j_d J_h.
\]
For an exponential disk with constant circular velocity $V_c$ and the surface density profile
\[
  \Sigma(R) = \Sigma_0\exp(-R/R_d),
\]
$M_d = 2\pi\Sigma_0R_d^2$ and $J_d=4\pi\Sigma_0R_d^3V_c$. From these two equations the disk density profile (parameters $\Sigma_0$ and $R_d$) can be expressed as functions of $m_d$, $j_d$, and $V_c$.

The distribution of angular momenta for dark matter halos is usually quantified by the \emph{spin parameter}
\[
  \lambda = \frac{J_h|E_h|^{1/2}}{GM_h^{5/2}},
\]
where $E_h$ is the binding energy of the halo (which depends on the actual adopted density profile). In hierarchically clustered universe spins of dark matter halos are induced by tidal torques from the surrounding material \citep{hp88}. The distribution of spin parameters of halos of various masses turns out to be surprisingly independent of anything else (halo mass, shape of the matter power spectrum, cosmological parameters, redshift, etc) and is approximately lognormal,
\[
  p(\lambda)d\lambda = \frac{1}{\sqrt{2\pi}\sigma_\lambda}\exp\left(-\frac{\ln^2(\lambda/\bar\lambda)}{2\sigma_\lambda^2}\right) \frac{d\lambda}{\lambda},
\]
with $\bar\lambda\approx0.05$ and $\sigma_\lambda\approx0.5$ - that result remains unchanged from the first N-body simulations \citep{be87} to the present day \citep{tlb13}.

The final step in the MMW98 model is the connection between the disk circular velocity $V_c$ and the virial velocity of the halo,
\[
  V_{\rm vir} = \left(\frac{GM_h}{r_{\rm vir}}\right)^{1/2}.
\]
In the original MMW98 model the coefficient of proportionality between $V_c$ and $V_{\rm vir}$ was assumed to be 1, but it does not have to be. For example, for the NFW profile
\[
  \left(\frac{V_c(r)}{V_{\rm vir}}\right)^2 = \frac{1}{x} \frac{\ln(1+cx)-cx/(1+cx)}{\ln(1+x)-c/(1+c)}
\]
where $x\equiv r/r_{\rm vir}$ and $c$ is the concentration of the halo. In this case, however, $V_c$ is a function of radius and is not constant, so which one should we use? One solution is to consider the ``maximal'' disk, i.e.\ take the largest value of $V_c$ for any radius, commonly referred to as $V_{\rm max}$, as the disk circular velocity. That value is mildly dependent on the halo concentration $c$,
\begin{eqnarray}
V_{\rm max} & = & 1.0\, V_{\rm vir}\mbox{ for }c=3,\nonumber\\
V_{\rm max} & = & 1.2\, V_{\rm vir}\mbox{ for }c=10,\nonumber\\
V_{\rm max} & = & 1.6\, V_{\rm vir}\mbox{ for }c=30.\nonumber
\end{eqnarray}

The MMW98 model is controlled by two main parameters, $m_d$ and $j_d$. In principle, they can be arbitrary. However, recently an interesting property of real galaxies has been noticed by \citet{k13}: disk sizes (both for stellar disks and gaseous disks) are linearly proportional to the virial radii, with the scatter in the relation entirely consistent with the distribution of $\lambda$ parameters for halos of a given mass. In other words, parameters $m_d$ and $r_d$ must be such that for stellar disks $R_d\approx 0.01 R_{\rm vir}$ and for gaseous disks it is about a factor of 2.5 larger.

\subsection{Galactic Disks}

We now descend into the actual galactic disks. The common lore is that disks are exponential, rotationally supported, and have flat rotation curves. While all these statements are kind of true, they are very far from being exact. 

\begin{figure}[t]
\includegraphics[width=0.5\hsize]{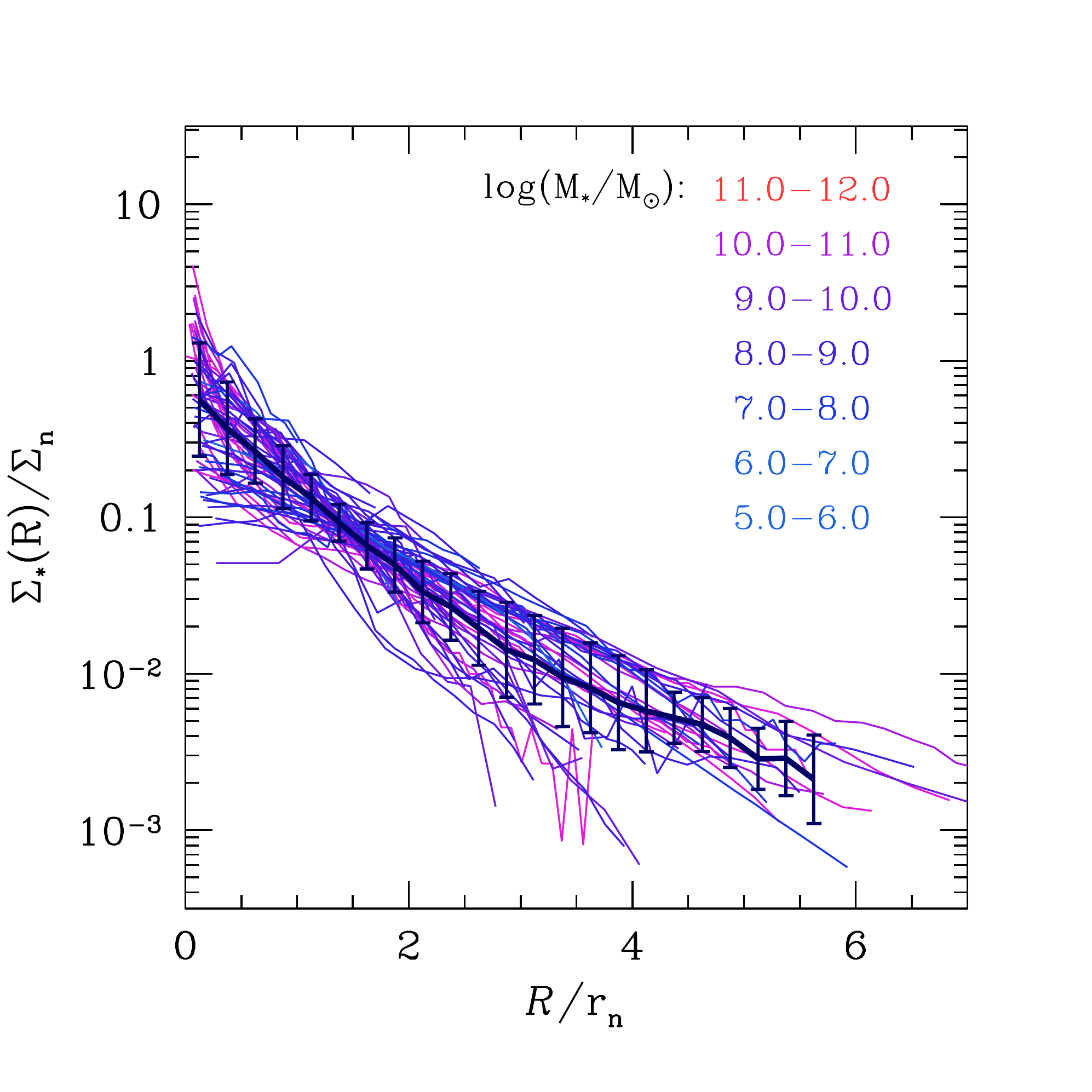}%
\includegraphics[width=0.5\hsize]{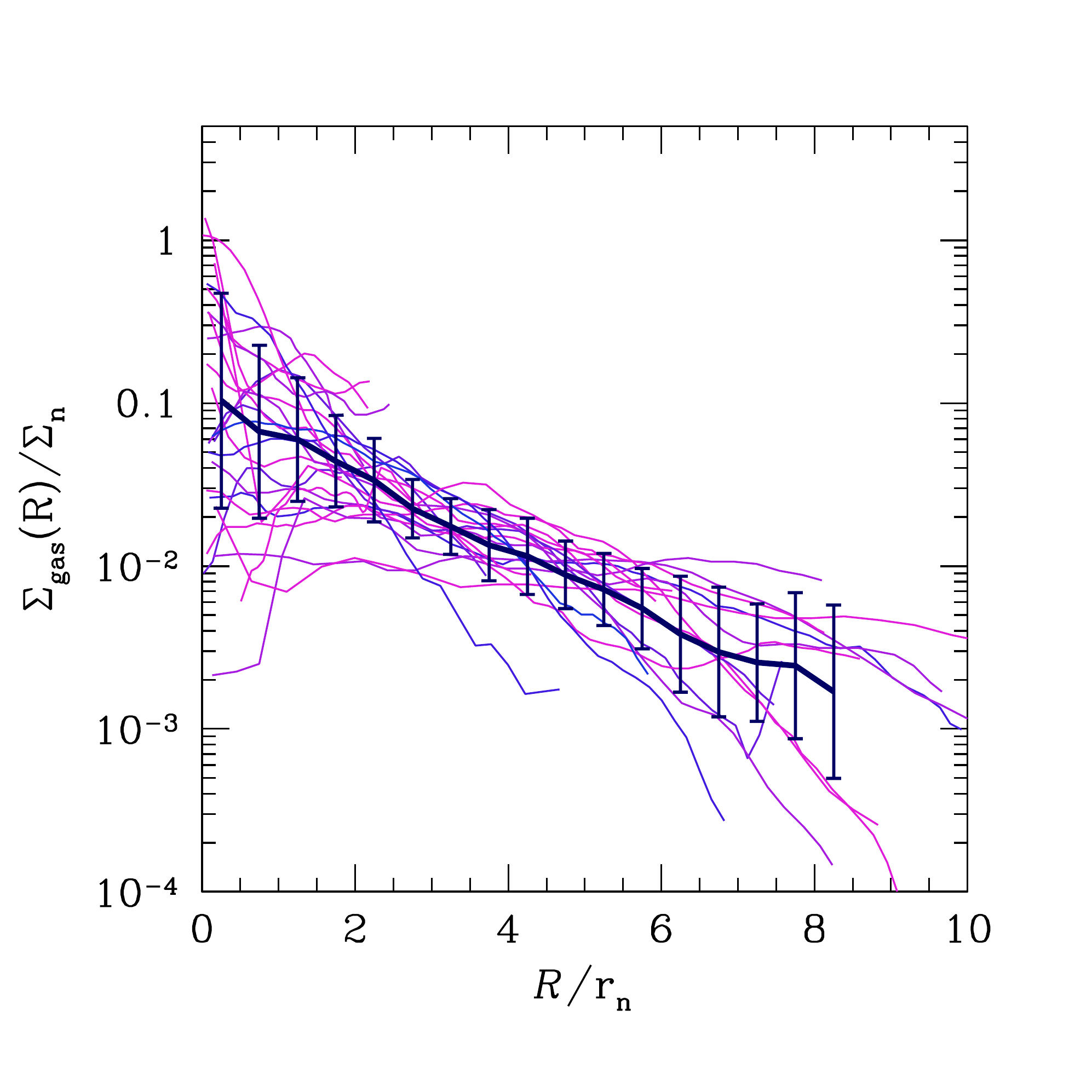}%
\caption{Normalized surface density profiles of stars and neutral gas for late-type galaxies \protect\citep[adopted from][]{k13}).\label{fig:spro}}
\end{figure}

Disks come with a variety of density profiles and a variety of rotation curves. For example, figure \ref{fig:spro} shows surface density profiles for stars and gas for several samples of disk galaxies \citep{k13}. On average profiles are indeed exponential, but deviations of individual galaxies from the mean can easily reach a factor of several.

\begin{figure}[b]
\sidecaption[t]
\includegraphics[width=0.64\hsize]{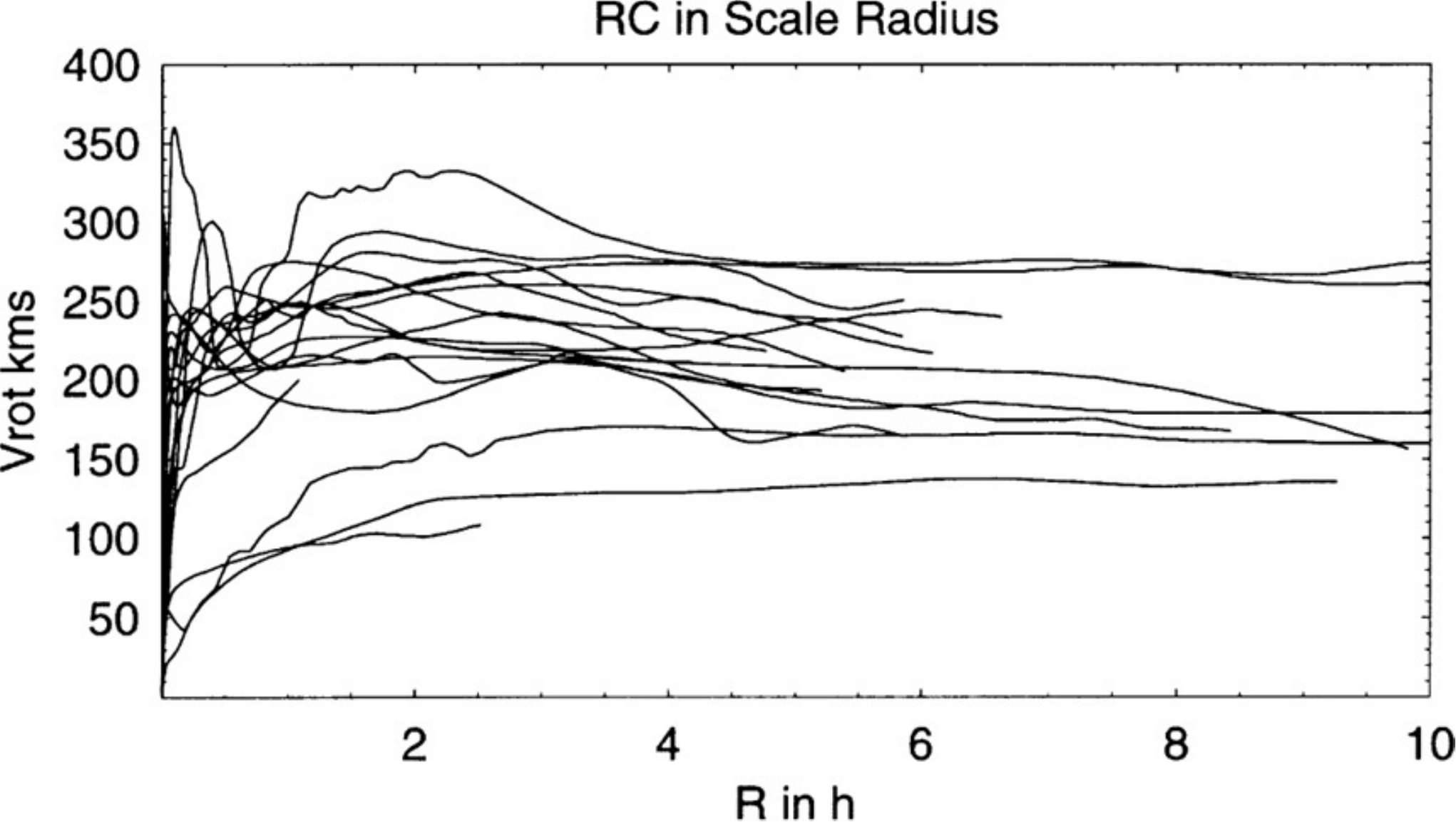}%
\caption{Rotation curves of several spiral galaxies from \protect\citet{sth99}.\label{fig:rotcur}}
\end{figure}

Similarly, rotation curves of individual galaxies (figure \ref{fig:rotcur}) show large deviations from the canonical flat shape - some rotation curves are rising, some are falling, some remain truly flat all the way to the outer edge of the disk.

Disk dynamics in general is a very complex affair. A large number of various disturbances and waves can propagate over the disks - in addition to spiral arms, there exit bending modes, bars, warps, etc. All these perturbations cause orbits of stars and gas to deviate from spherical symmetry. For example, spiral arms are shock waves, gas changes its velocity abruptly by a large factor (up to several times its sound speed) as it crosses the shock, and hence the gas in front of and behind the spiral arm shock cannot remain on the same circular orbit - one of the sides has to deviate substantially. For example, in the classical example of the grand design spiral, M51, the deviations of the gas rotational velocity from the circular velocity reach $20\dim{km/s}$ \emph{almost everywhere in the disk} \citep{hks09}.

Such deviations, in fact, may be responsible, at least partially, for the notorious cusp-core controversy. Some of the ``observed'' cusps may, in fact, be just an erroneous consequence of the incorrect assumption that the rotational velocity is equal to the circular velocity for gas \citep{gals:vrkg07}.

\subsubsection{Disk Stability}

How one would investigate such waves and features? Nonlinear treatment would require numerical simulations, but some widely known (and not so widely known) results can be obtained analytically for the linear stability of disk systems. A standard approach to studying linear stability of any system is to impose small fluctuations on the system and derive their dispersion relation. For an infinitely thin disk one can represent the radially perturbed (i.e.\ a perturbation remains azimuthally symmetric) surface density $\Sigma(t,R)$ as
\[
  \Sigma(t,R) = \bar\Sigma(R) + \Delta\Sigma(t,R),
\]
where the perturbation $\Delta\Sigma(t,R)$ is assumed to be a collection of linear waves, each wave characterized by the frequency $\omega$ and the wavevector $\vec{k}=(k_R,k_\phi)$. Let's first focus on purely radial perturbations, $k_\phi=0$. In that case the dispersion relation for the gaseous disk becomes \citep{gals:bt87}
\begin{equation}
  \omega^2 = \kappa^2 - 2\pi G\bar\Sigma|k_R| + c_s^2k_R^2,
  \label{eq:diskdr}
\end{equation}
where $\kappa^2\equiv R(d\Omega^2/dR)+4\Omega^2$ is the so-called \emph{epicyclic frequency} and $\Omega(R)$ is the disk angular velocity, $V_c(R) = R\Omega$.
 
The disk is stable when the right hand side is always positive, which is achieved if and only if
\begin{equation}
  Q \equiv \frac{c_s\kappa}{\pi G\bar\Sigma} > 1.
  \label{eq:toomre}
\end{equation}
This condition is universally known as \emph{Toomre stability criterion}, although for gaseous disks it has been obtained earlier by \citet{s60}, while Alan Toomre derived a similar relation for stellar disks \citep{t64}, a much more difficult exercise.

When $Q<1$, some of the radial modes in the disk become unstable,
\[
  \frac{\kappa}{Qc_s}\left(1-\sqrt{1-Q^2}\right) < k_{\rm unstable} < \frac{\kappa}{Qc_s}\left(1+\sqrt{1-Q^2}\right).
\]
An interesting property of this relation is that only a limited range of wavenumbers become unstable, the disk remains stable at very large ($k\rightarrow0$) and very small ($k\rightarrow\infty$) scales.

\subsubsection{Beyond Toomre}

Toomre stability criterion is often used in galactic and extragalactic studies. However, it is, unfortunately, often forgotten that it is incomplete. No disk is infinitely thin, and no perturbation is perfectly radial.

A case of arbitrary, not necessarily radial, perturbations was considered by \citet{pp97}, who found that the critical value for the $Q$ parameter is actually larger than 1. This is not surprising - at $Q=1$ radial perturbations go unstable; however, for the disk to become unstable it is only enough for \emph{some} waves to become unstable, and these first unstable waves do not have to be radial. Thus, some of the non-radial (i.e.\ non-axially-symmetric) perturbations may become unstable when all radial perturbations remain stable with $Q>1$.

The critical value of the $Q$ parameter turns out to depend on the disk density profile,
\[
  Q^2_{\rm crit} = \frac{3\alpha^2-3}{2\alpha^2-3} > 1,
\]
where
\[
  \alpha^2 = \frac{2\Omega(R)}{R|d\Omega/dR|}.
\]
For example, for a flat rotation curve ($\Omega\propto R^{-1}$) $\alpha^2=2$ and 
\[
  Q_{\rm crit} = \sqrt{3}.
\]
This is the reason why most actively star-forming (and, thus, instability-developing) disk galaxies have $Q$ parameters above unity but not significantly greater than 2 \citep{ism:lwbb08}.

Another generalization of the Toomre stability criterion is obtained when the finite thickness of a disk is taken into account. In that case the dispersion relation has been introduced by \citet{bs09}, although in a highly convoluted form it has been derived earlier by \citet{s60},
\begin{equation}
  \omega^2 = \kappa^2 - 2\pi \frac{G\bar\Sigma|k_R|}{1+|k_R|h} + c_s^2k_R^2,
  \label{eq:diskdrh}
\end{equation}
where $h$ is the disk \emph{scale height}, $\bar\Sigma(z) \propto \exp(-z/h)$. For a non-exponential vertical profile the dispersion relation becomes more complex and is not presentable analytically in a closed form.

Relation (\ref{eq:diskdrh}) is remarkable in that in the limit of very small scales, well below the disk scale height, $kh\gg1$ (in which case the disk cannot be considered as a flattened system any more), it reduces to
\[
  \omega^2 = - 4\pi G\bar\rho + c_s^2k_R^2,
\]
(with $\bar\Sigma=2\bar\rho h$), which is nothing else as a usual Jeans stability dispersion relation, familiar to any astrophysicist since kindergarten.

\subsubsection{Modeling Disks}

Modeling disks numerically is a subject of itself, and cannot be covered in these lectures. However, a word of caution is in order here. Let's imagine one is trying to model a galactic disk (or, for that matter, a disk around a supermassive black hole, or any other self-gravitating disk). A natural setup is to start with an axially-symmetric disk and let the instabilities develop. 

So, you prepared your symmetric disk as the initial condition for your powerful numerical code that includes all kind of important physical processes (cooling, star formation, feedback, etc). To be specific, let's say you set the gas temperature to $10^4\dim{K}$ in the disk with the circular velocity of $200\dim{km/s}$.

You press the magic button, simulation starts, and in an instant your disk cools off to the lowest temperatures your cooling module allows (indeed, cooling times in astrophysical environments are often very short), the $Q$ parameters plunges to very small values, and your disk fragments into tiny clumps of size comparable to the wavelength of fastest growing instability mode $\lambda_{\rm fast}$,
\[
  R \sim \lambda_{\rm fast} = 2\pi Q\frac{c_s}{\kappa}.
\]
Such a state, however - cold homogeneous disk - is \emph{unphysical}, there is no plausible physical process that can create such a system: after all, you started with an artificial initial condition; try running it backward in time, the disk is still cooling, so shortly before your initial moment it should have been blazingly hot, at $10^7-10^8\dim{K}$, and how would you propose to keep $10^8\dim{K}$ plasma in a disk with $200\dim{km/s}$ circular velocity?

Ok, that does not work. Let's now start with an initially stable disk ($Q\gg1$) and let it become unstable gradually (either by artificially introducing cooling gradually, or disabling cooling below $10^4\dim{K}$, or, even better, gradually adding mass to the disk). As $Q$ decreases gradually, at some moment it will reach a critical value $Q_{\rm crit}>1$. At that moment some non-radial perturbations become unstable and start growing, turning into non-linear waves; any non-linear wave in the gas steepens to a shock; any shock in a differentially-rotating disk becomes an oblique spiral wave; oblique shocks are known to generate an energy cascade a-la turbulence (although it may not be turbulence in the exact meaning of that word). Turbulence will provide extra support to the gas, replacing the sound speed $c_s$ in equation \ref{eq:diskdr} with $\sqrt{c_s^2+\sigma_t^2}$ and will limit the fragmentation scales to $R\sim2\pi\sqrt{c_s^2+\sigma_t^2}/\kappa$.

In other words, in the latter scenario the $Q$ parameter never had a chance to become much lower than the critical value, but must linger at around it, maintaining the disk in the just-unstable-enough state to generate enough turbulence. Hence the conclusion that the author arrived at himself after much suffering and erring: if your disk simulation has $Q\ll1$, you are doing something wrong...

\subsection{Ionized, Atomic, and Molecular Gas in Galaxies}

Everyone knows that ISM consists of several gas phases. The ionized gas comes in two flavors, as hot ($\sim10^6\dim{K}$) coronal gas and warm/cool ($\sim10^4\dim{K}$) ionized gas (known under many names: warm ionized medium (WIM), diffuse ionized gas (DIG), Reynolds Layer); atomic gas exists as warm/cool ($\sim10^4\dim{K}$) and cold ($\sim10^2\dim{K}$) neutral media (WNM and CNM respectively); finally, molecular gas is almost always cold ($<10^2\dim{K}$).

\subsubsection{Ionized Gas}

A story of coronal gas is misty and messy - it is not even clear how much of it there is in the Milky Way ISM, or what fraction of it comes from stellar feedback processes and what fraction is merely halo gas intermixed into the ISM due to various disk instabilities. Warm ionized medium is understood better because it is primarily located at the outer edges of the disk.

What causes WIM? We can get a hint on its origin from its temperature - gas at $10^4\dim{K}$ is likely to be photo-ionized. If we recall that only gods have the power to switch off the Cosmic Ionizing Background, the ionizing source is there too - plus whatever ionizing radiation escapes from star-forming regions inside the Milky Way disk.

\begin{figure}[t]
\sidecaption[t]
\includegraphics[width=0.64\hsize]{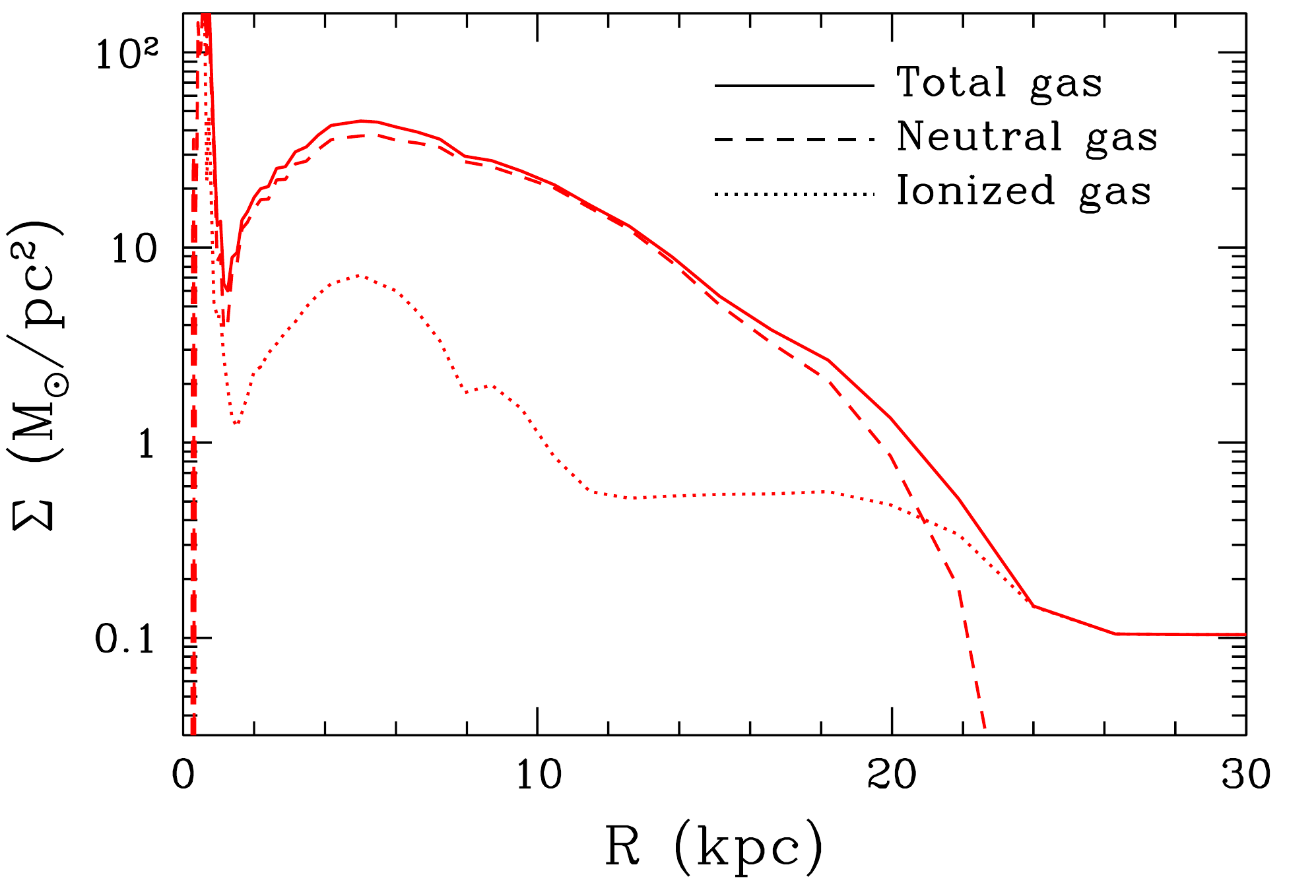}%
\caption{Surface density profiles for the total, ionized, and neutral (atomic and molecular) gas for a model Milky-Way-like galaxy (from the simulation described in \protect\citet{ng:g12}).\label{fig:wim}}
\end{figure}

An example of how the relative distribution of neutral and ionized gas may look like in the Milky Way galaxy (or other similar galaxies) is shown in figure \ref{fig:wim}. The WIM contribution stays more-or-less constant at about $0.5\Msun/\dim{pc}^2$ (column density $N_\Ht = 6\times10^{19}\dim{cm}^{-2}$) in the outer disk, but increases to several $\Msun/\dim{pc}^2$ inside the solar radius because of the increased radiation field and a contribution of coronal gas. Broadly, such behavior is consistent with actual observations of the ionized gas in the Milky Way and other galaxies. For example, in the Milky Way the contribution of ionized gas at the solar radius is about $1\Msun/\dim{pc}^2$.

The outer parts of the disk are consistent with being ionized by Cosmic Ionizing Background, and the transition from neutral to ionized gas is often very sharp. However, consistency does not imply causality. There could be other ionizing sources, such as stellar radiation escaping from star-forming regions or cosmic rays. Since stars do not form in the ionized gas (as far as we can tell), we leave the WIM-land on our way to denser and colder domains; interested readers should check an excellent recent review by \citet{hdb09}.

\subsubsection{From Atomic To Molecular Gas}

Stars (at least most of them) form from molecular gas. Few astronomers would question this conjecture. While a minority of all stars may form in the atomic gas (at least Pop III stars certainly form in gas that is 99\% atomic), on this journey we are chasing the bulk of star formation. Hence, the transition from atomic to molecular gas is a necessary condition for (the bulk of) star formation.

Chemistry of molecular hydrogen is not particularly complex; $\H2$ forms through two physically distinct channels: in numerous reactions in the gaseous phase, from rare ions $\Ht-$ and $\H2^+$ (the best reference for these processes is \citet{ism:ga08}), and on the surface of cosmic dust, which serves as a catalyst. The gas processes are slow exactly because  $\Ht-$ and $\H2^+$ are rare; fraction of molecular hydrogen forming in the gas phase saturates at $10^{-3}-10^{-2}$ and only jumps to close to 1 when 3-body reactions become sufficiently efficient (which only happens at densities above about $10^{12}\dim{cm}^{-3}$). This channel of $\H2$ formation does not require any metals and can proceed in the primordial gas (indeed, this is how Pop III stars form).

Formation of $\H2$ on dust grains is not fully understood. It is usually assumed that atomic hydrogen accumulates on grains where two atoms can find each other much more easily (young couples tend to live in cities). The formation rate $R_D$, defined as
\[
  \left.\frac{dn_\H2}{dt}\right|_{\mbox{dust}} = R_D n_\Ht n_\HI,
\]
has been modeled (somewhat inconclusively) theoretically and measured observationally by Wolfire et al. (2008):
\[
  R_D = \D R_0,
\]
with $R_0\approx 3.5\times10^{-17}\dim{cm}^3/\dim{s}$,
where from now on I will use a convenient parameters $\D$ that measures the abundance of dust relative to the solar neighborhood; i.e.\ $\D=1$ implies the same abundance of dust per unit mass of gas as in the Milky Way ISM around us.

It is not, however, enough to know the formation rate to determine the abundance of molecular hydrogen - like predator and prey, ultraviolet radiation plays with $\H2$ the game of life and death. Particularly deadly for molecular hydrogen is radiation in the so-called Lyman and Werner bands, at energies between $11.3$ and $13.6\dim{eV}$ (actually, the bands extends further, but hydrogen ionizing radiation is often well shielded by neutral atomic ISM). In addition, molecular hydrogen is destroyed by collisions with atoms and other molecules when gas temperatures raise above about $5{,}000\dim{K}$. Hence, in order to predict the abundance of molecular hydrogen in specific conditions, we need to know the Interstellar Radiation Field (ISRF).

ISRF is not measured directly, but rather modeled based on the observations of various line ratios in the ISM. Two canonical references to such modes are \citet{ism:d78} and \citet{ism:mmp83}, which are perfectly consistent with each other. In the solar neighborhood $J_0\approx 10^6\mbox{phot}/\mbox{cm}^2/\mbox{s}/\mbox{eV}/\mbox{rad}$, but in the Galaxy the radiation field changes with the distance from the center. At the center it is up to 10 times higher than around the Sun.

Just like masses and luminosities are convenient to measure in solar units, in galactic studies it is convenient to measure the radiation field and other quantities (like dust abundance) in the Milky Way units. Hence, hereafter we will also use $\U\equiv J_{\rm LW}/J_0$ (where $J_{\rm LW}$ is the average radiation field in the Lyman and Werner bands). By definition, $\U=1$ in the solar neighborhood, but in high redshift galaxies it can be large, $\U=30-300$ at $z\sim2$ \citep{gals:cppp09}.

Even the Milky Way radiation field is extremely strong from the molecular hydrogen point of view - if it could shine on typical molecular clouds unimpeded, the molecular fraction would only be $10^{-6}-10^{-5}$. The only reason molecular clouds exist in the universe is because all that radiation is \emph{shielded}.

There are two distinct shielding processes: dust shielding and molecular self-shielding. Dust absorbs radiation over a very large range of wavelengths, from infra-red to X-rays. Dust opacity is a smooth function of wavelengths, and in the first approximation it can be considered constant over a narrow Lyman and Werner bands \citep[for detailed plots of dust opacity see][]{misc:wd01}. In different galaxies the dust opacity is different, but in the three galaxies it was studied best - Milky Way and two Magellanic clouds - it is roughly proportional to the dust-to-gas ratio,
\[
  \sigma_{\rm LW} = D_{\rm MW} \sigma_0
\]
with $\sigma_0=1.7\times10^{-21}\mbox{cm}^2$ for the Milky Way ($D_{\rm MW}=1$),$\sigma_0=1.6\times10^{-21}\mbox{cm}^2$ for the LMC ($D_{\rm MW}\approx0.5$), and $\sigma_0=2.2\times10^{-21}\mbox{cm}^2$ for the SMC ($D_{\rm MW}\approx0.2$). Thus, it is possible to simply take $\sigma_0$ as a universal constant, 
\[
  \sigma_0\approx2\times10^{-21}\mbox{cm}^2.
\]

Accounting for continuum shielding over a narrow band is easy; the molecular hydrogen photo-destruction rate $\Gamma$ is then simply
\[
  \Gamma = c \sum_j \int_{\nu_1}^{\nu_2}\sigma_j(\nu)\underbrace{e^{-\sigma_{d}(\nu) N_\Ht} n_\nu}_{\mbox{radiation field}} d\nu \approx e^{-\bar\tau_{d}} \Gamma_{\rm LW},
\]
where $N_\Ht$ is the total hydrogen column density, $\bar\tau_d\equiv\bar\sigma_dN_\Ht$ is the average dust opacity in the Lyman and Werner bands, $\Gamma_{\rm LW}$ is the so-called "free space" photo-destruction rate (i.e. photo-destruction rate in the absence of any shielding), and the sum is taken over all $\H2$ lines in the Lyman and Werner bands. It is convenient to define a \emph{shielding factor} $S_D$ that parametrizes the suppression of the free space field by dust shielding, $\Gamma = S_D\Gamma_{\rm LW}$, with
\[
  S_D(\D,N_\Ht) = e^{\displaystyle-\D\sigma_0N_\Ht}.
\]

\begin{figure}[b]
\sidecaption[t]
\includegraphics[width=0.64\hsize]{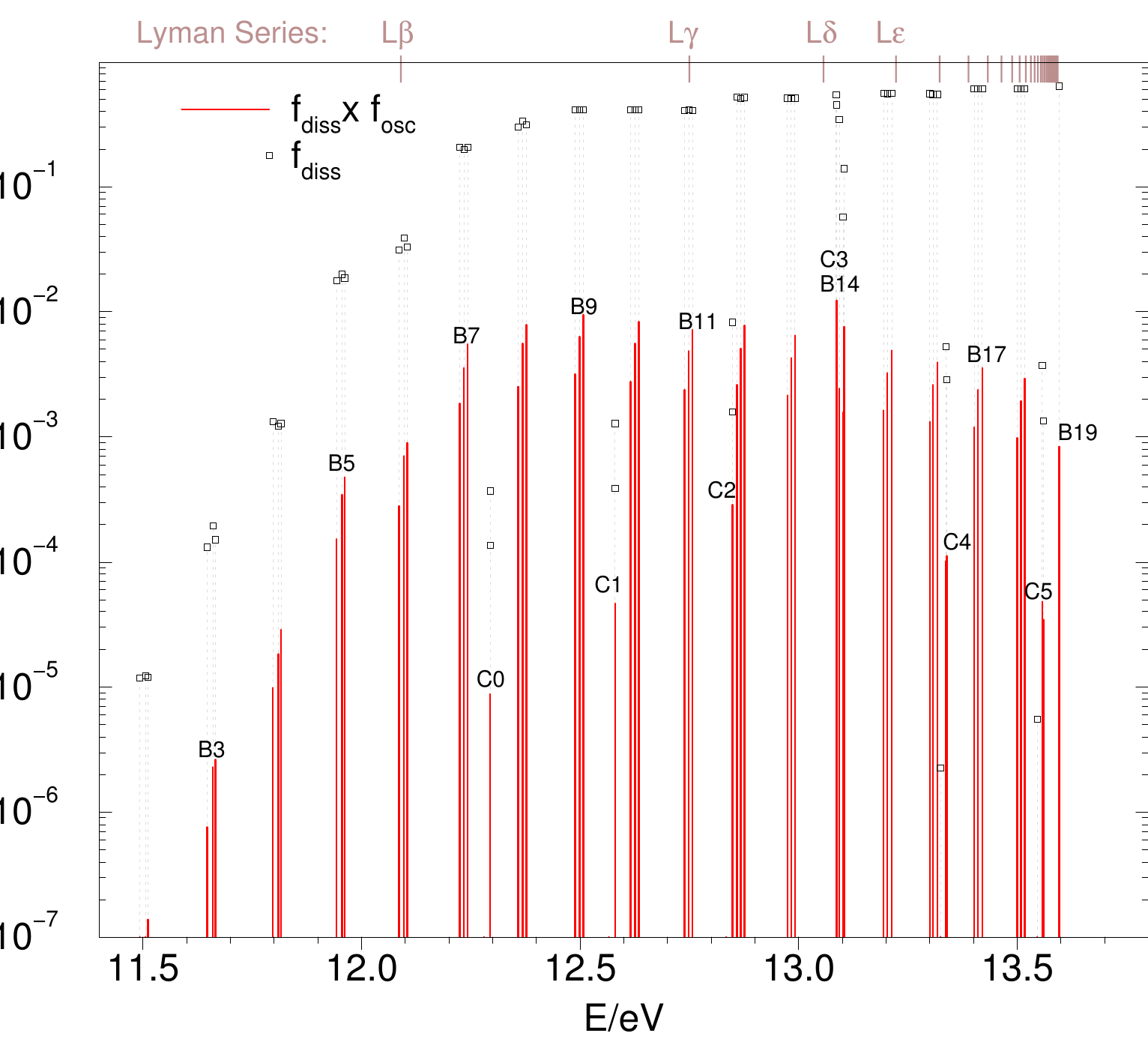}%
\caption{Molecular lines in the Lyman and Werner bands. A hydrogen molecule has a non-zero probability to be photo-dissociated $f_{\mbox{diss}}$ when it is excited into any of these states (adopted from \protect\citet{har00}).\label{fig:lwlines}}
\end{figure}

Self-shielding of molecular hydrogen is much more complicated. Lyman and Werner bands consists of numerous lines of various strengths (figure \ref{fig:lwlines}). Absorbing a photon in one of those lines may or may not lead to the destruction of the hydrogen molecule, and the probability of dissociation varies significantly for different lines.

Hence, the shielded photo-destruction rate can be represented as a sum over individual lines, each with its own cross section $\sigma_{j}(\nu)$,
\begin{equation}
  \Gamma = c \sum_j \int_{\nu_1}^{\nu_2}\sigma_j(\nu)\underbrace{e^{-\sigma_{j}(\nu) N_\H2} n_\nu}_{\mbox{radiation field}} d\nu \approx \sum_j e^{-\bar\tau_{j}} \Gamma_{{\rm LW}, j} = S_\H2(N_\H2)\Gamma_{\rm LW}.
  \label{eq:sh2}
\end{equation}

The self-shielding factor $S_\H2(N_\H2)$ is much harder to compute, but its general behavior may be guessed. As individual Lyman and Werner bands lines become optically thick, some of the terms in the sum in equation (\ref{eq:sh2}) become small, but weaker lines will remain optically thin and un-shielded for much higher column densities than the stronger lines, thus allowing the destructing radiation to sneak deeper into a molecular cloud. Hence, as the column density of molecular hydrogen increases, the self-shielding factor will fall at a rate, which is much slower than the exponential decline of an individual line.

The self-shielding of molecular hydrogen has been modeled extensively; a specific approximation for the self-shielding factor that is most commonly used is due to \citet{ism:db96},
\begin{equation}
S_{\H2} = \frac{0.965}{(1+x/b_5)^\alpha} + \frac{0.035}{\sqrt{1+x}}\exp\left(-\frac{\sqrt{1+x}}{1180}\right),
  \label{eq:db96}
\end{equation}
where $x \equiv N_\H2/5\times10^{14}\dim{cm}^{-2}$, $b_5 \equiv b/\dim{km/s}$, and in the original approximation $\alpha=2$. \citet{whb11} suggested that at higher temperatures a better fit is $\alpha=1.1$, but the first term in equation (\ref{eq:db96}) is not important anyway.

\begin{figure}[t]
\sidecaption[t]
\includegraphics[width=0.64\hsize]{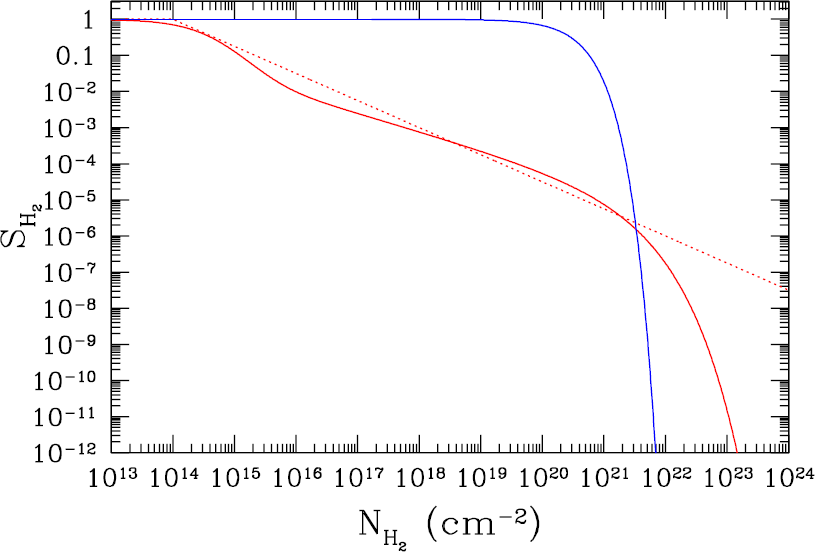}%
\caption{\protect\citet{ism:db96} molecular self-shielding factor as a function of $\H2$ column density (solid red line). For comparison, exponentially falling off shielding factor (dust shielding with Milky Way dust and fully molecular gas, $N_\H2=2N_\H2$)is shown as a blue line. Red dotted line is a power-law approximation for the self-shielding factor, $S_{\H2}\propto N_\H2^{-0.75}$ that has been also used in the past.\label{fig:db96}}
\end{figure}

Figure \ref{fig:db96} shows the \citet{ism:db96} approximation as a function of the molecular column density. A gradual decline of the self-shielding factor ($S_{\H2}$ going approximately as $N_\H2^{-0.75}$) is apparent for almost 8 orders of magnitude. However, at very high column densities, $N_\H2>10^{22}\dim{cm}^{-2}$, the fall-off becomes steeper, with the last factor in equation (\ref{eq:db96}) dominating. What could cause such a steep decline?

Our deduction above that the weaker lines remain optically thin and serve as avenues for the radiation to sneak into a molecular cloud remain correct for as long as each absorption line can be treated as independent. However, just like in human society neighbors sooner or later will put a stop on a weak person misbehaving, so in the society of Lyman and Werner bands stronger lines begin to interfere in the affairs of weaker one at sufficiently high column densities. Since each excited state in an atom or molecule lives for a finite time, lines have non-trivial \emph{natural width} (see section 2.2.1). In the high column density limit the natural width dominates, and the equivalent width of a line (the area of the spectrum the line takes out) grows as $N_\H2^{1/2}$. As the strongest lines begin to overlap, the nature of self-shielding changes - instead of individual lines absorbing UV radiation each by itself, the absorption cross-section now becomes a continuous function of frequency, with cross-sections of individual lines all blending together into a single, continuum-like absorption. Hence, self-shielding becomes much stronger, and that is manifested in the drop-off in the \citet{ism:db96} formula at $N_\H2>10^{22}\dim{cm}^{-2}$.

Finally, we need to figure out what $N_\H2$ actually is. Let's imagine that we have a line-of-sight through a molecular cloud with the total hydrogen column density $N_\Ht$. The first inclination is to simply use $N_\H2=0.5N_\Ht$ (let's assume the cloud is fully molecular), but that is actually \emph{wrong}!

Equation (\ref{eq:db96}) is suitable for the idealized case of a slab of gas with no internal motions. Real molecular clouds are, however, supersonically turbulent on scales above the \emph{sonic length}, $l_s\la 1\dim{pc}$. In other words, if you take two parcels of gas inside a molecular cloud separated by a distance $l$, the rms velocity dispersion between them satisfies what is known as Larson's law,
\[
  \delta v(l) \approx c_s\left(\frac{l}{l_s}\right)^{0.5}
\] 
with $c_s$ being the gas sound speed (in fact, the definition of the sonic length is that $\delta v(l_s) = c_s$). For $l\gg l_s$, the velocity difference between them would be much larger than the width of each Lyman and Werner bands line $b \sim c_s$. Hence, these two fluid elements would shield each other only if they happen accidentally to fall at the same line-of-sight velocity, which would occur with the probability $b/\delta v$.

 \begin{figure}[t]
\sidecaption[t]
\includegraphics[width=0.64\hsize]{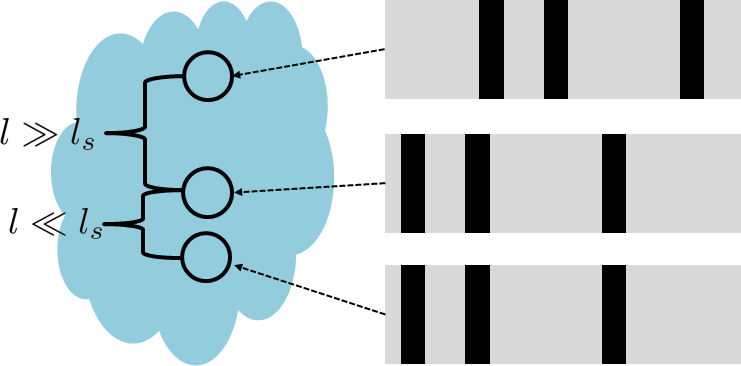}%
\caption{A cartoon illustrating the role of ISM turbulence in suppressing self-shielding of molecular hydrogen on scales above the sonic length.\label{fig:H2cartoon}}
\end{figure}

\def\mc{{\rm MC}}

This is illustrated in a cartoon fashion in figure \ref{fig:H2cartoon}. Hence, a fluid element inside a molecular cloud sees a column density of about $N_\H2 \sim \langle n_\H2\rangle_s L_\mc b/\delta v \approx N_\H2 \sim \langle n_\H2\rangle_s (l_s L_\mc)^{1/2}$, where $\langle n_\H2\rangle_s$ is the average molecular hydrogen density on a sonic scale at the location of interest and  $L_\mc$ the width of the whole molecular cloud. This is valid, however, only until individual lines do not overlap. With line overlap relative velocity shifts between different fluid elements become unimportant (lines overlap anyway). In other words, at sufficiently large column densities line radiative transfer in the Lyman and Werner bands effectively behaves as continuum radiative transfer, and the effective length over which the column density is accumulated approaches  $L_\mc$.

In equation (\ref{eq:db96}) the line overlap is described by the last exponential factor. To account for the supersonic turbulence inside the molecular cloud, equation (\ref{eq:db96}) can be modified as
\begin{equation}
S_{\H2} = \frac{0.965}{(1+x_1/b_5)^2} + \frac{0.035}{\sqrt{1+x_1}}\exp\left(-\frac{\sqrt{1+x_2}}{1180}\right),
\label{eq:lofac}
\end{equation}
where $x_1\equiv  \langle n_\H2\rangle_s (l_s L_\mc)^{1/2}/5\times10^{14}\dim{cm}^{-2}$ is proportional to the $\H2$ column density over the sonic length, while $x_2\equiv  \langle n_\H2\rangle_\mc L_\mc/5\times10^{14}\dim{cm}^{-2}$ accounts for the column density of the whole molecular cloud. Obviously, $x_2 \gg x_1$.

Armed with understanding of dust and self-shielding, we can consider some interesting limiting cases. In the kinetic equilibrium the rates of photo-destruction and molecular hydrogen formation balance, hence
\[
   \Gamma_{\rm LW} S_{\H2} e^{-\sigma_{\rm LW}N_{\Ht}}n_{\H2} = R_Dn_{\Ht} n_{\HI}.
\]
The free-space radiation field is parametrized by the introduced above $\U$ parameter, $\U\equiv\Gamma_{\rm LW}/\Gamma_0$. Hence,
\begin{equation}
   \frac{ f_{\H2}}{(1-f_{\H2})} = \frac{\D}{\U} \frac{R_0}{S_{\H2}\Gamma_0} e^{D_{\rm MW}\sigma_0N_{\Ht}} n_{\Ht}.
  \label{eq:h2bal}
\end{equation}

As we already know, in low metallicity environments self-shielding is expected to dominate over dust shielding,
\[
  \frac{ f_{\H2}}{(1-f_{\H2})} = \frac{\D}{\U} \frac{R_0}{S_{\H2}\Gamma_0}  n_{\Ht}.
\]
Let's say we are interested in densities at which the gas becomes 50\% molecular ($f_{\H2} = 0.5$). In that case 
\[
  S_\H2 \propto \frac{\D}{\U},
\]
and for high enough column density, when
\[
	S_\H2 \sim e^{-\mbox{const}\times N_\H2^{1/2}},
\]
we find
\[
  N_{1/2} \equiv N_\Ht(f_\H2=1/2) \propto \ln^2\left(\frac{\U}{\D}\times\mbox{const}\right),
\]
i.e.\ the column density of the atomic-to-molecular transition depends only weakly on the dust abundance or the interstellar radiation field.
 
In the opposite extreme, in high radiation fields the dust shielding dominates, 
\[
  \frac{ f_{\H2}}{(1-f_{\H2})} = \frac{\D}{\U} \frac{R_0}{\Gamma_0} e^{\D\sigma_0N_{\Ht}} n_{\Ht},
\]
hence
\[
  N_{1/2} \propto \frac{1}{\D} \ln\left(\frac{\U}{\D}\times\mbox{const}\right).
\]
As could have been easily guessed, higher dust abundance pushes the atomic-to-molecular transition towards lower (column) densities.

How should we go now from shielding factors for individual parcels of gas to the factors that should be used in actual numerical simulations? Modern cosmological or galactic scale simulation may not resolve molecular clouds at all or may resolve them down to parsec scales. Hence, in the most general case we can imagine whole space being tessellated into regions (say, simulation cells) of size $L$ some of which include pieces of molecular clouds. Each such piece has a distribution of column density inside it, $\phi_j(N_\H2)$, where $j$ refers to a given piece. Hence, the average shielding factor is 
\[
  \langle S_\H2\rangle_j = \int S_\H2(N_\H2) \phi_j(N_\H2) dN_\H2 = S_\H2(N_{{\rm eff}, j}) \int \phi_j(N_\H2) dN_\H2 = S_\H2(N_{{\rm eff}, j})
\]
since $\int \phi_j(N_\H2) dN_\H2 = 1$ by definition. If the distribution $\phi_j(N_\H2)$ was known, one can also compute $N_{{\rm eff}, j}$, but at present there are no models that attempt to determine $\phi_j$. Hence, we need to come up with an ansatz for $N_{{\rm eff}, j}$.  For example, in the absence of a better alternative, we can simply take equation (\ref{eq:lofac}) with the sonic length $l_s$ being fixed to some small value ($0.1-1\dim{pc}$) and a model for the size of molecular cloud $L_\mc$.

\begin{figure}[t]
\sidecaption[t]
\includegraphics[width=0.64\hsize]{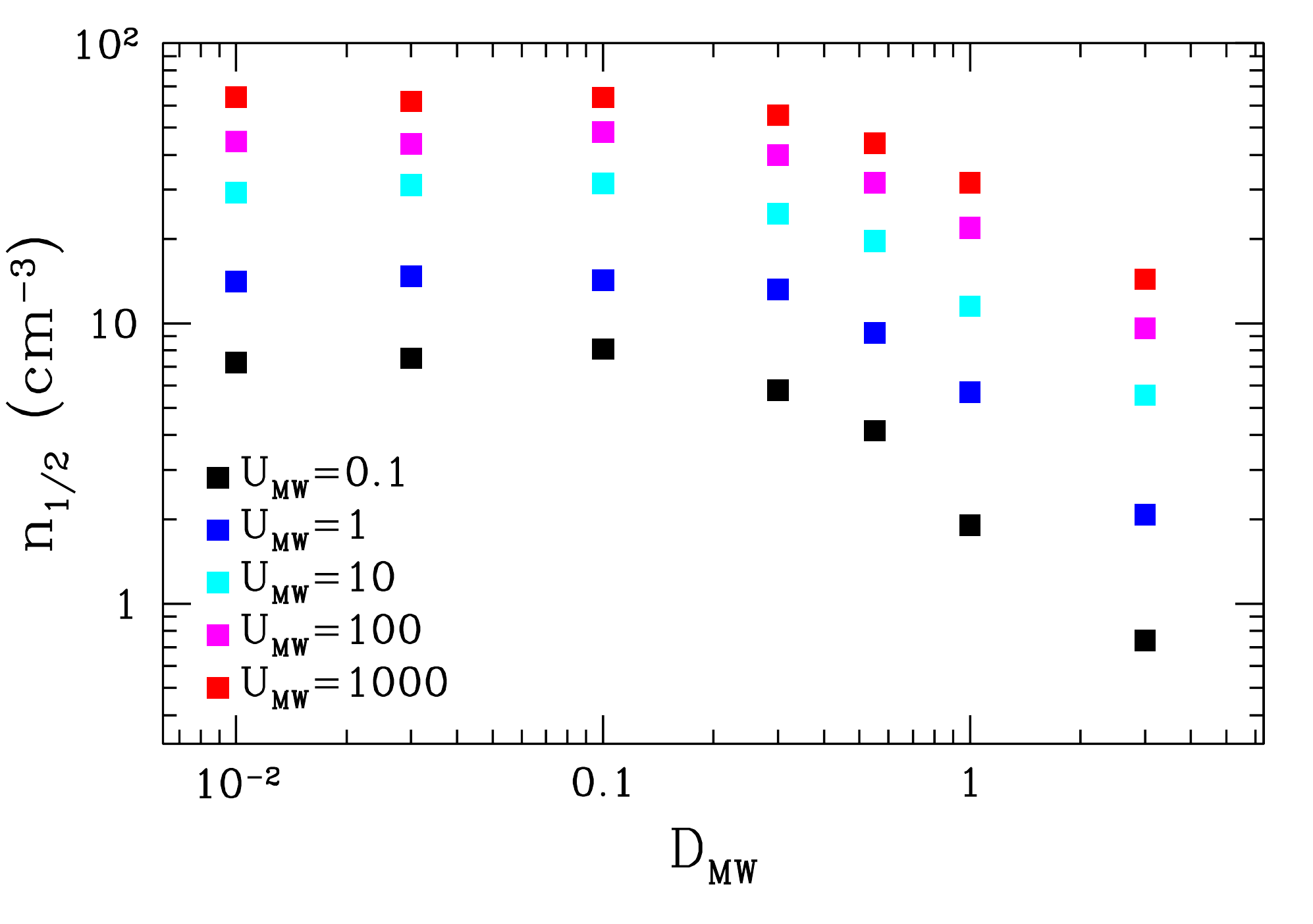}%
\caption{Average total hydrogen number density of atomic-to-molecular  gas transition (defined as $f_\H2=1/2$) as a function of the  dust-to-gas ratio $\D$ and the interstellar radiation field $\U$ on scales $L=65\dim{pc}$.\label{fig:nthd2g}}
\end{figure}

Perhaps the simplest such model is a "Sobolev-like" approximation that Andrey Kravtsov and I introduced a few years ago \citep{ng:gk11},
\[
  L_\mc \equiv \frac{\rho}{2|\nabla\rho|}.
\]
With such an approximation the complete set of equations is obtained. The dependence of the characteristic density of the transition on the environmental parameters on the particular spatial scale $L=65\dim{pc}$ (read "resolution of your simulation") is shown in figure \ref{fig:nthd2g} - the two limiting regimes are easily noticeable in the figure. 

\begin{figure}[t]
\sidecaption[t]
\includegraphics[width=0.64\hsize]{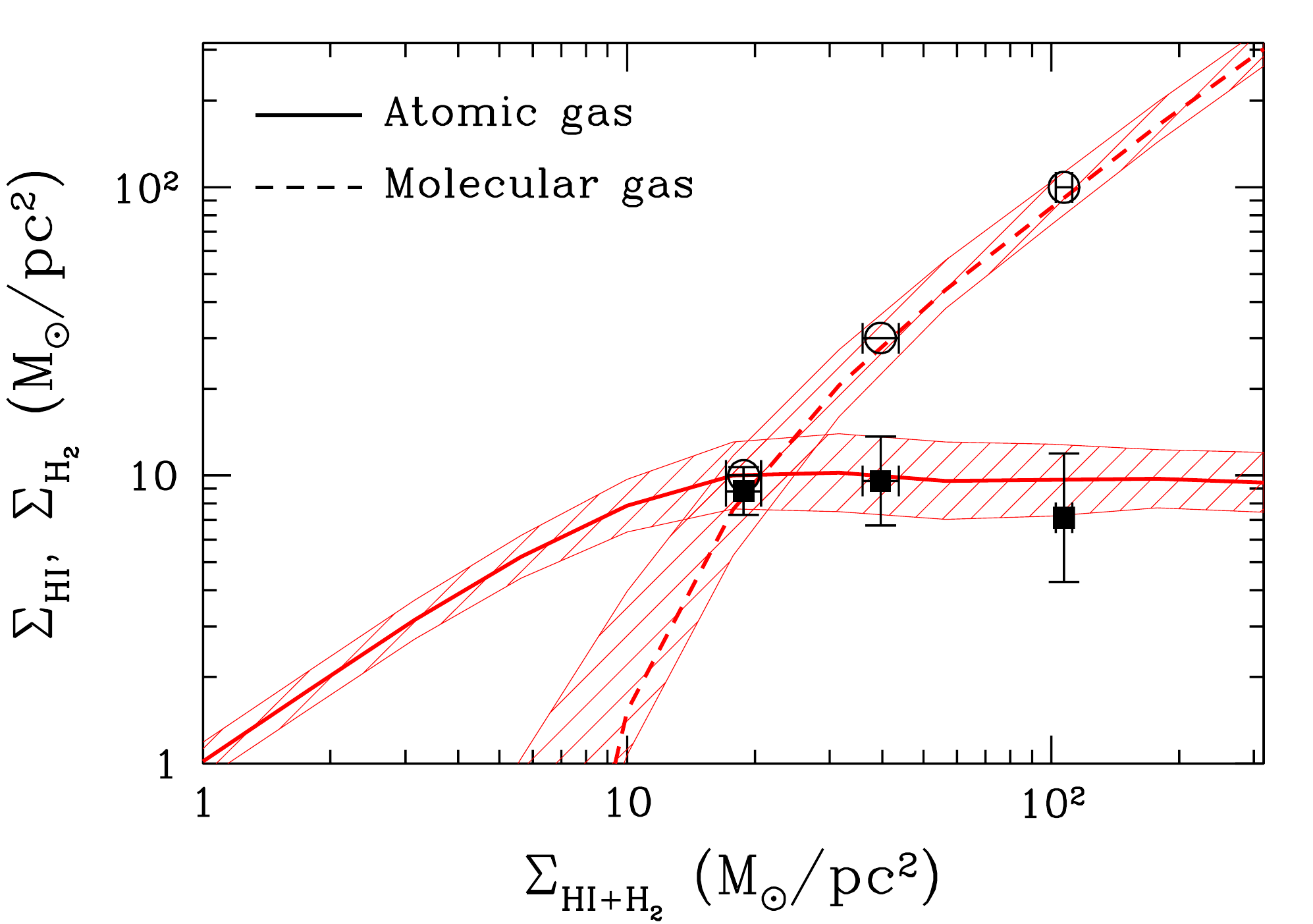}%
\caption{Average atomic and molecular gas surface densities as
  functions of the total (neutral) hydrogen gas surface density 
  averaged over $500\dim{pc}$ scale for the $(\D=1,\U=1)$
  simulation case (red lines/bands for mean/rms). Filled squares and open circles
  with error bars mark the observed average and rms atomic and molecular
  hydrogen surface densities from \protect\citet{misc:wb02}.\label{fig:wong}}
\end{figure}

In figure \ref{fig:wong} I show how such a model fares in matching the observed surface densities of atomic and molecular gas on larger scales, where they are actually measured. The main achievement of models like this one is that they capture the observed saturation of the atomic surface density at about $10\Msun/\dim{pc}^2$ (for $\D=\U=1$ case; the saturation level does depend on the environment, just like $n_{1/2}$). A detailed description of the latest edition of \citet{ng:gk11} model is presented in Gnedin, Kravtsov, \& Draine (2013, in preparation). 

An alternative model for the atomic-to-molecular transition is due to \citet{sfr:kmt09a} - that model is simpler to implement, but does not account for line overlap, and, hence, breaks down for metallicities (or, rather, dust-to-gas ratios) below about 20\% of the Milky Way value.

\subsection{Molecular ISM}

Ok, we arrived into the molecular ISM. Now what? Why do we even care about the molecular gas? After all, many experts in star formation will tell you that molecules are \emph{not} required for star formation. Now we know this is not quite true - line overlap makes $\H2$ self-shielding important at low dust abundances, and, hence, in that regime molecules \emph{are required} for star formation. 

A second answer to that question is offered by \citet{sfr:klm11} (and by the nature herself, but that story is still well ahead). Shielding in molecular gas actually performs two functions at once - it protects hydrogen molecules from photo-destruction by Lyman and Werner bands photons, but it also allows gas to cool to the state that is properly called \emph{cold} ($100\dim{K}$ and below) - without shielding, UV and optical photons can eject energetic electrons from dust grains by photoelectric effect (the one Einstein got the Nobel prize for); these energetic electrons thermalize in the gas, effectively transferring the energy of radiation into the gas thermal energy. With shielding, this process becomes much less efficient and the gas can cool to low temperatures - and, hence, fragment into small clumps from which stars can form.

Thus, even in the regime when molecular self-shielding is not important, molecular gas plays a role of a "paint" someone poured into the ISM - the "painted" (i.e.\ molecular) gas is cold and can form stars, while gas without "paint" is too hot for star formation to take place there. Think of this as a lucky "coincidence", if you like.

\subsubsection{Thermodynamics of $\H2$}

Before we move further down the yellow brick road towards star forming regions, let us pause for a short while and refresh what we know about the hydrogen molecule. After all, it is the simplest molecule one can imagine, containing just two atoms (hence \emph{diatomic}), and its thermodynamics can be solved (almost) exactly.

If you recall some college thermodynamics, you may remember that the diatomic gas has a polytropic index of $7/5$ (or, equivalently, specific heat $c_V=5/2$). If you have forgotten that, it should not be hard to re-derive that result! After all, the partition function for the diatomic molecule is simply 
\[
  Z = e^{-E_n/(k_BT)}Z_{\rm rot}Z_{\rm vib},
\]
where the vibrational part is
\[
  Z_{\rm vib} = \sum_{v=0}^\infty e^{\displaystyle -\hbar\omega(v+1/2)/(k_BT)}.
\]
The rotational part is a bit tricky, but really just a bit - since $\H2$ is a symmetric molecule and two protons are indistinguishable, two nuclear states (with the spins aligned, total nuclear spin is 1 and the spins anti-aligned, total nuclear spin is 0) behave almost like two different molecules (transitions between the two states are possible, but highly suppressed and only occur at high enough densities). The state with the nuclear spin of 1 is called an \emph{ortho}-hydrogen molecule, and only allows odd values for the total angular momentum $J$, while the state with the 0 nuclear spin is a \emph{para}-hydrogen molecule and only has even values of the angular momentum. Ortho-$\H2$ has a higher statistical weight than the para-state, hence
\[
  Z_{\rm rot} = \frac{3}{4}Z_{\rm ortho} + \frac{1}{4}Z_{\rm para},
\]
where
\begin{eqnarray}
  Z_{\rm ortho} & = & \sum_{J=1,3,...} (2J+1) e^{\displaystyle-\hbar^2J(J+1)/(2Ik_BT)},\nonumber\\
  Z_{\rm para} & = & \sum_{J=0,2,...} (2J+1) e^{\displaystyle-\hbar^2J(J+1)/(2Ik_BT)}.\nonumber\\
 \end{eqnarray}

The partition function $Z$ is a magic wand of thermodynamics, all other quantities are derived from it: free energy
\[
  F = -k_BT\ln\left[\frac{V}{N!}\left(\frac{mk_BT}{2\pi\hbar^2}\right)^{3/2}Z\right],
\]
internal energy
\[
  E = F - T\left.\frac{\partial F}{\partial T}\right|_V,
\]
specific heat
\[
  c_V = \frac{1}{k_BN} \left.\frac{\partial E}{\partial T}\right|_V,
\]
etc.

\begin{figure}[t]
\sidecaption[t]
\includegraphics[width=0.64\hsize]{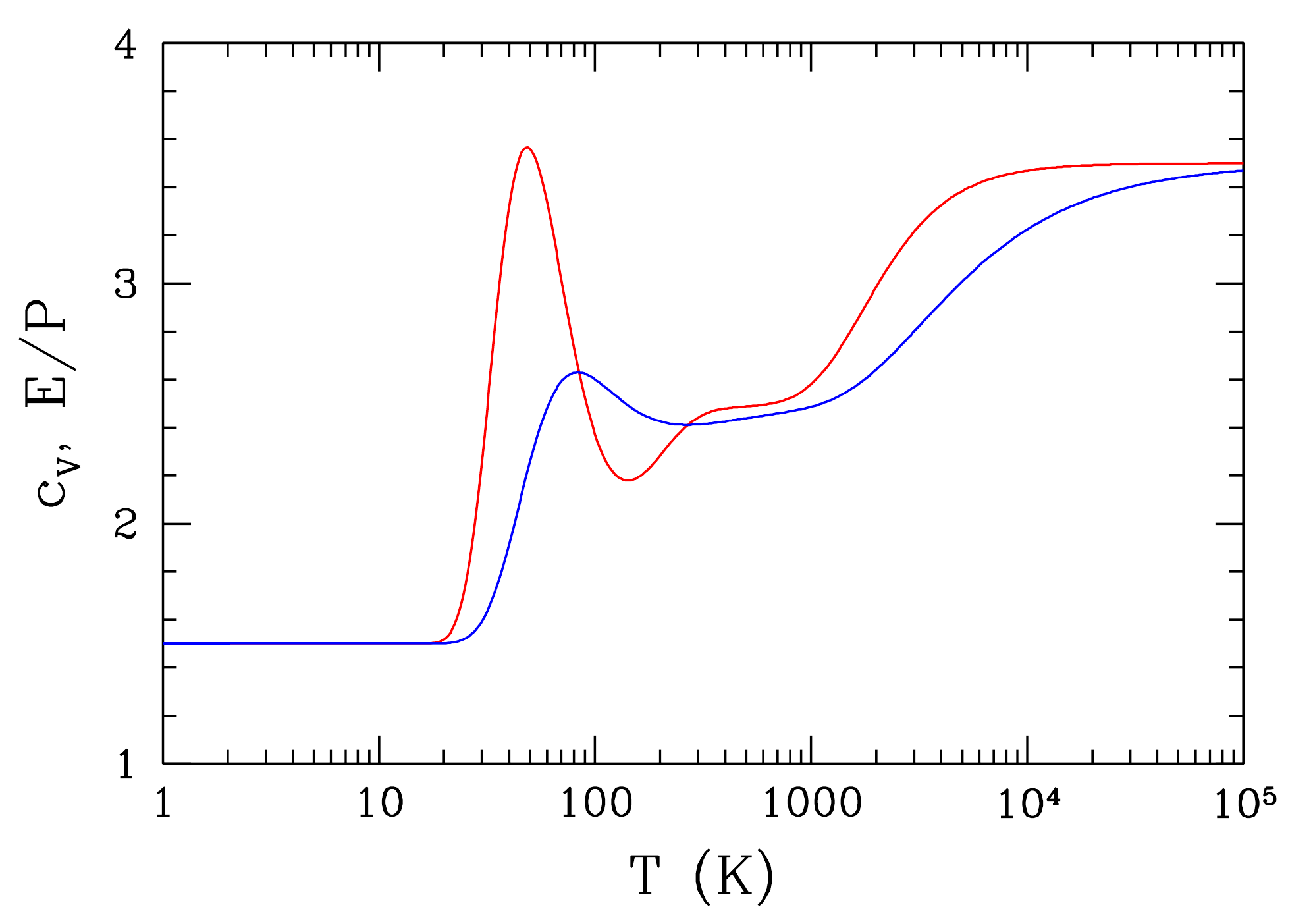}%
\caption{Specific heat $c_V$ (red) and the internal energy over pressure (blue) for molecular hydrogen gas as a function of temperature. Notice that $\H2$ never behaves as classic diatomic gas ($c_V=E/P=5/2$).\label{fig:cvet}}
\end{figure}

For example, figure \ref{fig:cvet} shows $c_v$ and $E/(k_bT)$ for $\H2$ gas with $3:1$ ratio of ortho-to-para molecules. If that plot does not surprise you, then you are a true expert in quantum thermodynamics - {\bf molecular hydrogen actually never behaves as classic diatomic gas with $c_V=5/2$ (or, equivalently, $\gamma=7/5$)}. More than that, it does not even behave as \emph{polytropic} gas with $E=P/(\gamma-1)$ except for very low temperatures ($T<20\dim{K}$) where it behaves as \emph{monoatomic} gas with $c_V=3/2$! If you did not know that, you can be excused - some of highly distinguished astrophysicists made that error too...

\subsubsection{Cosmic Pandora Box: The X-Factor}

We are now approaching the most confused, abused, and misused subject in the studies of molecular ISM -  $\CO$ emission and the $X_{\CO}$ factor. 

Molecular hydrogen is a great example of a classic catch-22 - $\H2$ has to be shielded from the outside to exist, hence the outside (i.e.\ us observing it) cannot actually see its emission in the Lyman and Werner bands. And to add insult to injury, the same dust obscures background sources, making absorption spectroscopy extremely difficult. Historically, by far the most common method to observe molecular gas was via its $\CO$ emission.

Rotational transitions of the $\CO$ molecule are equally spaced in frequency, $\nu_J=\hbar J/(2\pi I)$ (the molecule is asymmetric, so we do not need to worry about ortho/para mess). For the most common $^{12}\ion{C}{}^{16}\ion{O}{}$ isotope the first ($1\rightarrow0$) transition is located at $\nu_{1\rightarrow0}=115\dim{GHz}$ (or $\lambda_{1\rightarrow0}=0.26\dim{cm}$). This is a major convenience, since $\CO$ emission lines are easy to identify (just look for a uniform fence in the millimeter wavelengths). On the other hand, there is no a priori reason why $\CO$ should be a good tracer of $\H2$: $\CO$ needs higher dust shielding to form and it gets saturated at too high column densities. Hence, $\CO$ emission comes from a narrow range of column densities, both cloud outskirts and cloud centers emit little.

\begin{figure}[t]
\sidecaption[t]
\includegraphics[width=0.64\hsize]{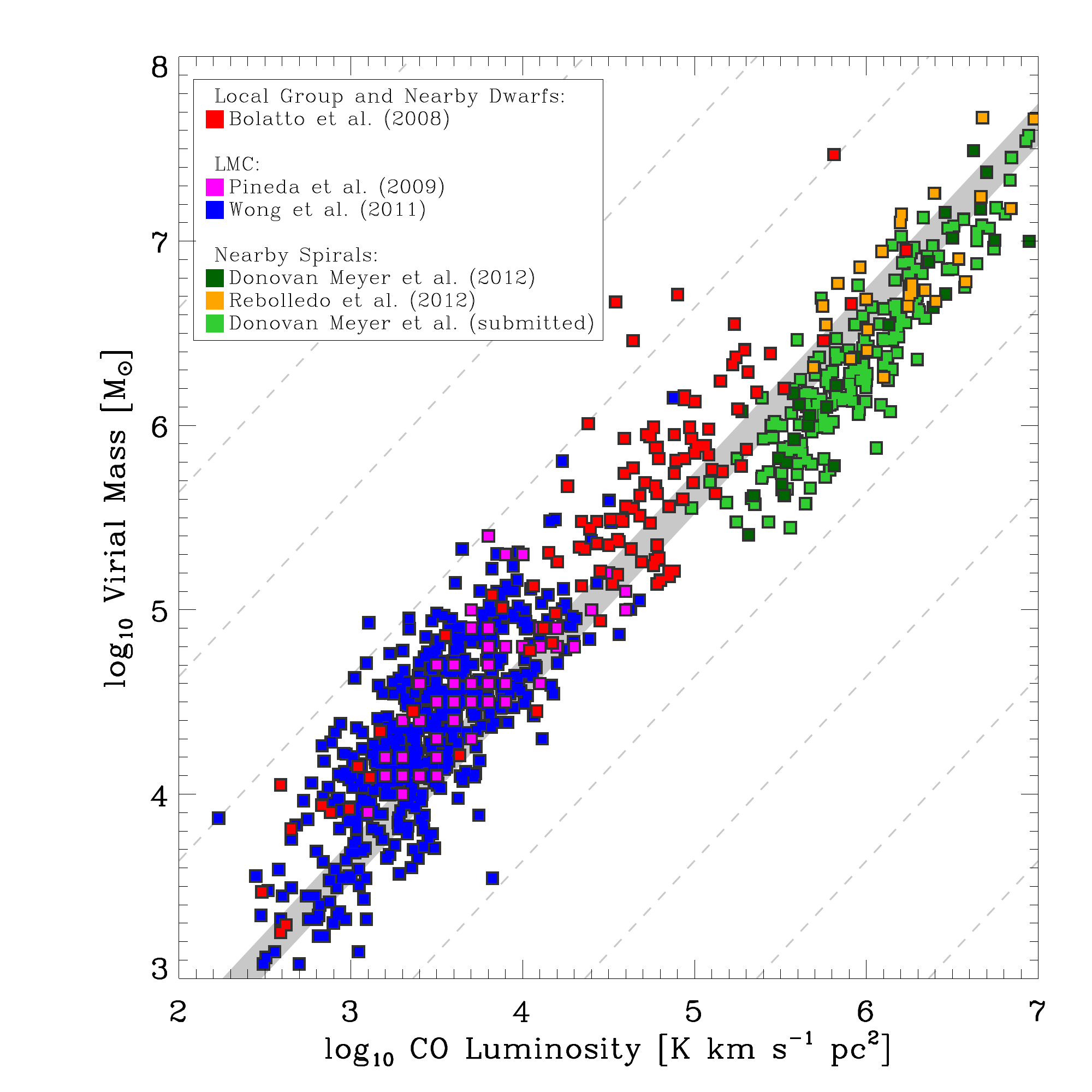}%
\caption{$\CO$ luminosity versus the virial mass for extragalactic molecular clouds. The gray band shows the average values in the Milky Way (adopted from \protect\citet{bwl13}).\label{fig:xfac}}
\end{figure}

Never-the-less, whenever a mass of molecular gas can be estimated by other means (usually the virial theorem), observations show a good correlation between the $\CO$ luminosity and the gas mass, albeit with substantial scatter from one cloud to another (figure \ref{fig:xfac}).

In galactic studies the relevant conversion factor between the molecular gas and $\CO$ luminosity is the infamous \emph{X-factor},
\[
  X_\CO \equiv \frac{N_\H2}{W_\CO},
\]
where $W_\CO$ is the equivalent width of a $\CO$ emission line (which will be different for different transitions),
\[
  W_\CO = \int T_A(v)dv
\]
with $T_A$ being the antenna temperature of the radio emission. The canonical Milky Way value for the X-factor is $X_\CO = 2\times10^{20}\dim{cm}^{-2}\dim{K}^{-1}\dim{(km/s)}^{-1}$ (enjoy the elegance of units!). The reason for this particular combination is that a measurement of the equivalent width in your telescope beam can be directly converted into the column density of molecular hydrogen along the line-of-sight.

In extra-galactic studies most of the time a galaxy is not spatially resolved (at least until the full ALMA comes online), so a single observation measures the total $\CO$ luminosity $L_\CO$ of a galaxy, and a convenient quantity is 
\[
  \alpha_\CO \equiv \frac{1.36 M_\H2}{L_\CO}
\]
(the factor of 1.36 is a contribution of Helium, and it really should be $1/(1-Y)$, since $Y$ does depend slightly on the metallicity). The Milky Way value is $\alpha_\CO = 4.3\Msun/\dim{pc}^2/\dim{K}/\dim{(km/s)}$. Notice, that the connection between $X_\CO$ and $\alpha_\CO$ is somewhat non-trivial; $\alpha_\CO$ can be re-written as
\[
  \alpha_\CO \propto \frac{\int N_\H2 dA}{\int W_\CO dA} = \frac{\langle N_\H2\rangle}{\langle W_\CO\rangle},
\]
where $A$ is the area on the sky. Hence, $\alpha_\CO$ is \emph{not} directly proportional to the average $X_\CO$ for a galaxy. Rather, it is proportional to the ratio of average $N_\H2$ to the average $W_\CO$. Alternatively, we can re-interpret the averaging procedure for $X_\CO$ in a highly non-trivial way,
\[
  \bar X_\CO^{-1} \equiv \frac{\langle W_\CO\rangle}{\langle N_\H2\rangle} = 
  \frac{\langle (W_\CO/N_\H2) N_\H2\rangle}{\langle N_\H2\rangle} = 
  \left\langle \frac{1}{X_\CO}\right\rangle_{N_\H2}.
\]
I.e., $X_\CO$ should be averaged harmonically and with the $\H2$ column density weighing.

So, how should we approach modeling $\CO$ emission in modern cosmological or galactic-scale simulations? Scales on which $\CO$ emission originates are not yet resolvable in modern simulations, hence, it needs to be followed with a sub-grid model. However, since $\CO$ emission is not important dynamically, it can be modeled in post-processing, after the simulation had been completed. There exist many approaches to constructing a sub-grid model, and the best (at least in principle) sub-grid model is a someone's else simulation!

The field of modeling internal structure of molecular clouds with sufficient physics is rather new, with only a few attempts made so far, but it certainly developing rapidly. One example of how $\CO$ emission can be modeled is the work that was led by Robert Feldmann in two series of paper in 2012 \citep{ng:fgk12a,ng:fgk12b}. This is just an illustration, one can follow a similar path with newer, better small-scale simulations for an undoubtedly better result.

\begin{figure}[t]
\sidecaption[t]
\includegraphics[width=0.64\hsize]{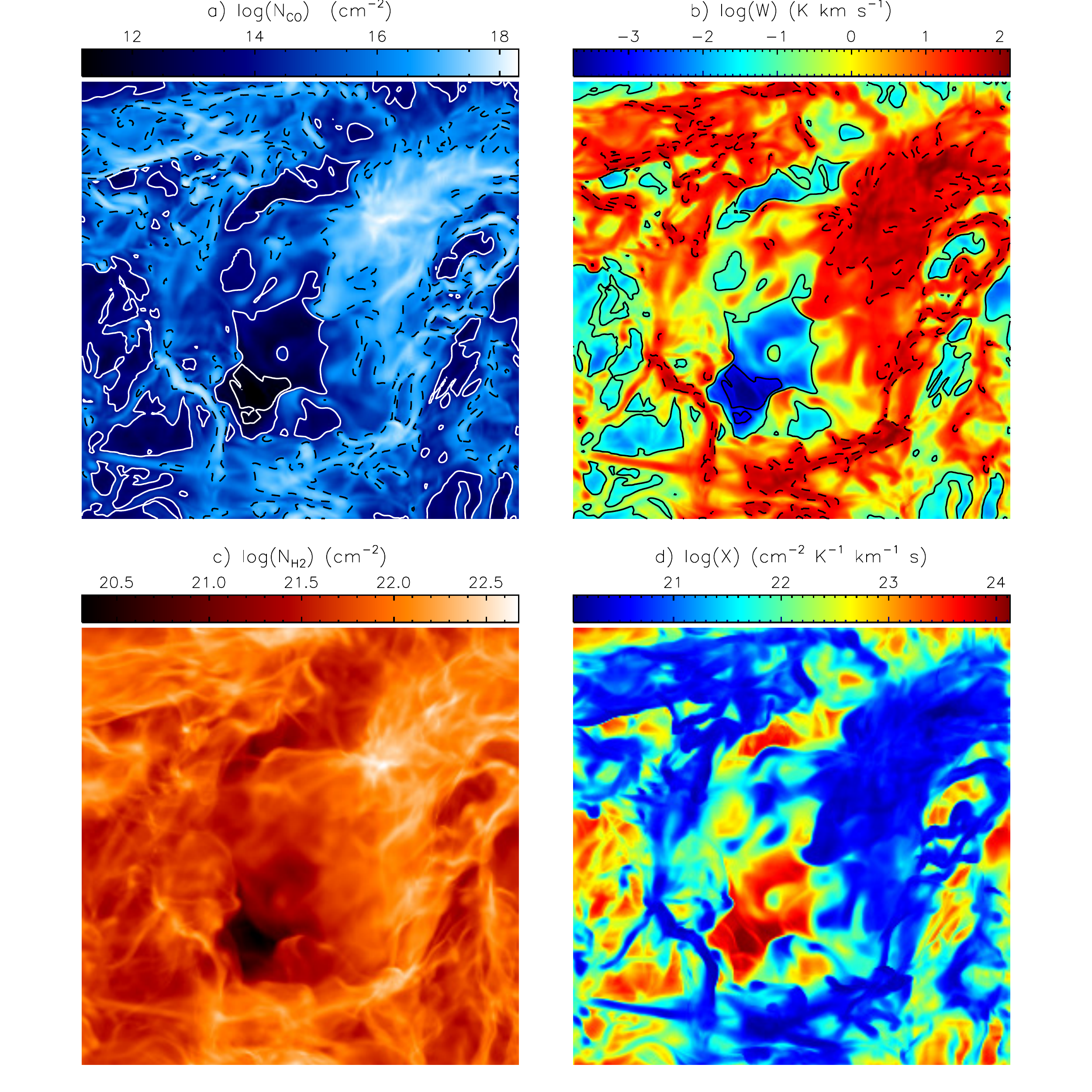}%
\caption{Images of $\H2$ and $\CO$ column densities, $\CO$ equivalent width $W_\CO$, the $X_\CO$ factor from \protect\citet{ism:sgd11a} simulations.\label{fig:gmcsim}}
\end{figure}

One of the very first attempt to model $\CO$ emission directly in GMC-scale simulations was done by Simon Glover and collaborators \citep{ism:gm11,ism:sgd11a,ism:sgd11b}. Images of $\H2$ and $\CO$ column densities, $\CO$ equivalent width $W_\CO$, the $X_\CO$ factor from these simulations are shown in figure \ref{fig:gmcsim}. As can be expected in a turbulent ISM, there are large variations in the $X_\CO$ factor on very small, sub-pc scales. Never-the-less, when averaged over the whole simulated region, $X_\CO$ dependence on the properties of the molecular cloud exhibits remarkable regularity - \citet{ism:gm11} found that the main parameter that controls the $X_\CO$ factor is (surprise!) the dust opacity (sometimes parametrized as the visual extinction $A_V \sim \tau_D$). 

\begin{figure}[b]
\includegraphics[width=0.48\hsize]{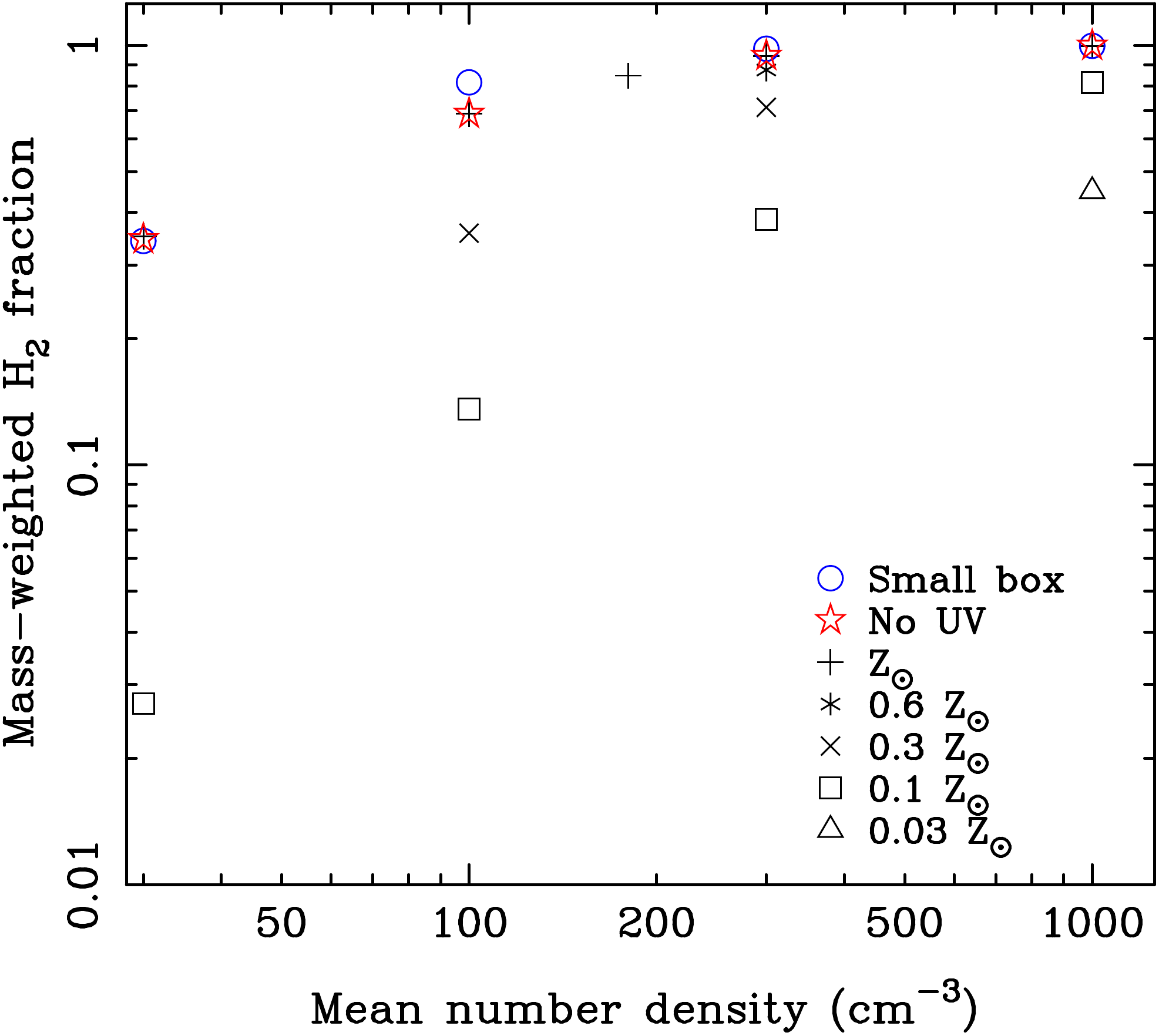}~~%
\includegraphics[width=0.48\hsize]{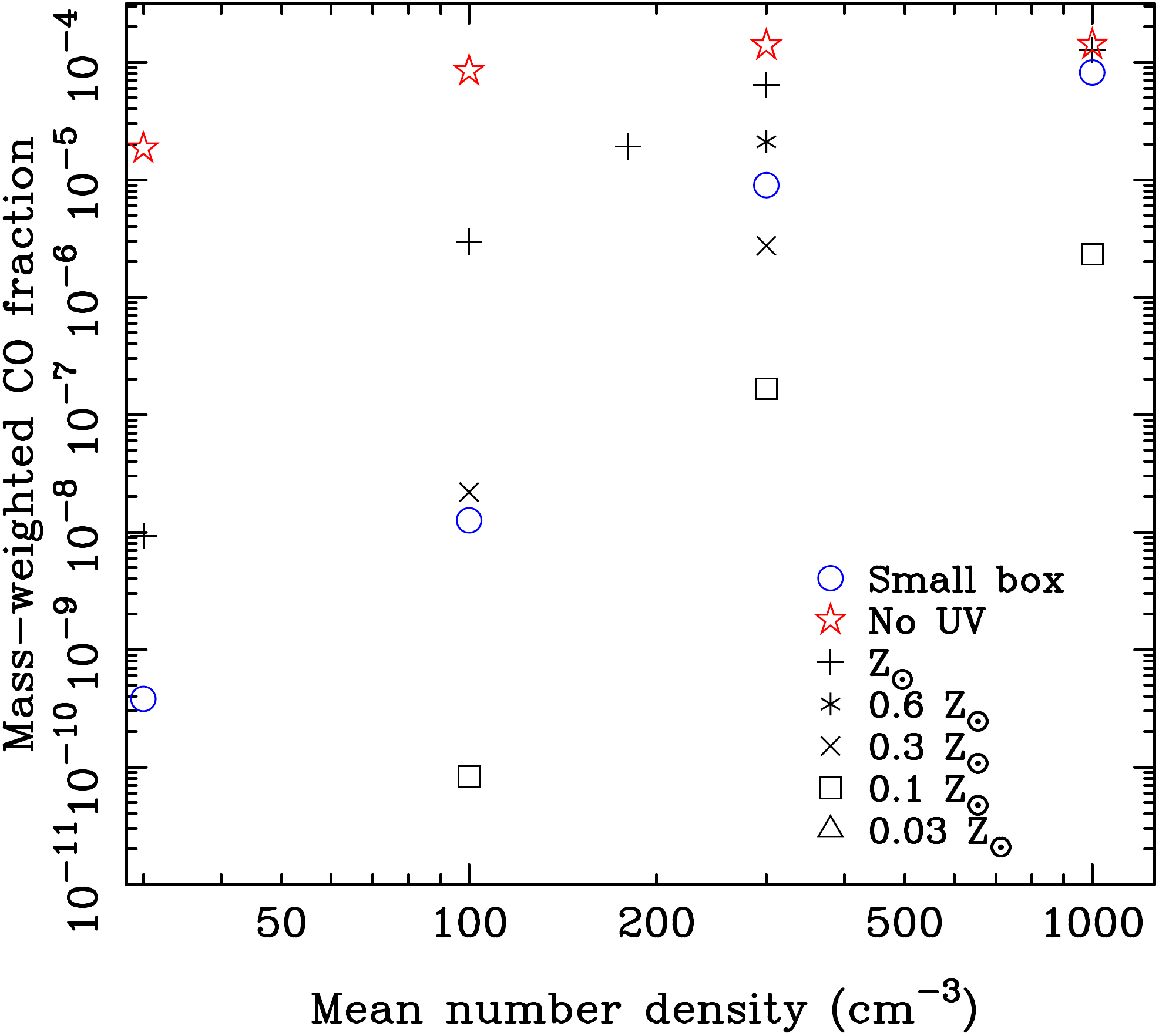}%
\caption{Mass-weighted fractions of $\H2$ and $\CO$ as a function of average gas density in \protect\citet{ism:gm11} simulations.\label{fig:gmcxco}}
\end{figure}

Figure \ref{fig:gmcxco} shows the mass-weighted molecular and $\CO$ fractions from \citet{ism:gm11} simulations. Using these tabulated values, \citet{ng:fgk12a,ng:fgk12b} developed a sub-grid model that can be used in cosmological and galactic-scale simulations for computing the $X_\CO$ factor in each simulation cell. Realistic simulated galaxies have complex ISM, with gas densities, metallicities, dust abundances, and interstellar radiation field varying from place to place. Hence, one can and \emph{should} expect the $X_\CO$ factor to vary significantly inside a given galaxy and from galaxy to galaxy.

\begin{figure}[t]
\sidecaption[t]
\includegraphics[width=0.64\hsize]{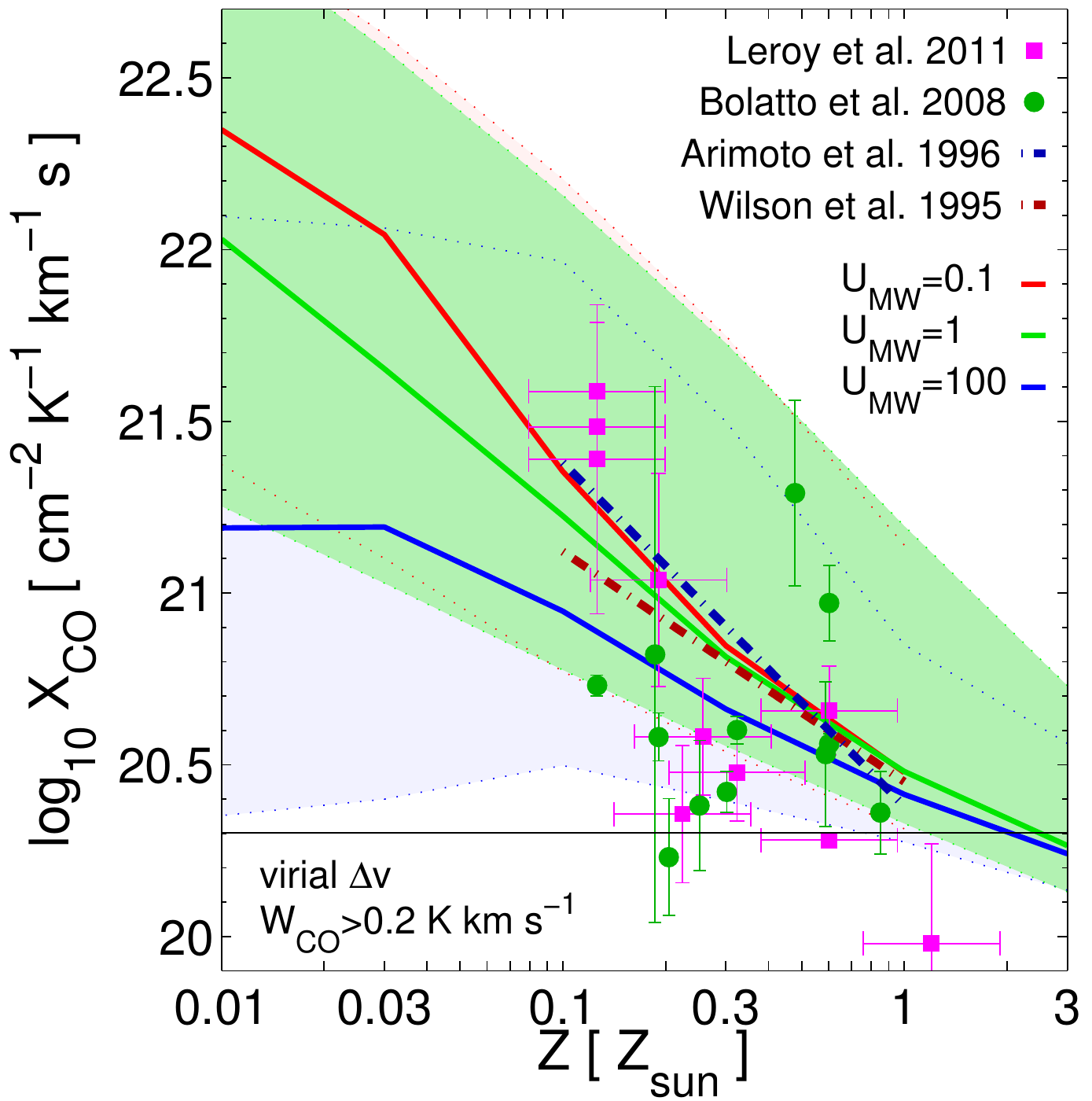}%
\caption{Dependence of the $X_\CO$ factor on the environmental parameters: gas metallicity $Z$ and the interstellar radiation field $\U$, on $\sim50\dim{pc}$ scales. Colored bands show the variation over different locations in a single simulated galaxy (adopted from \protect\citet{ng:fgk12a}).\label{fig:fgk}}
\end{figure}

As the result, the \citet{ng:fgk12a,ng:fgk12b} model predicts a range of values for $X_\CO$ even for a given metallicity and $\U$, not a single number, as is shown in figure \ref{fig:fgk}. Overall, the predictions of the model are within the existing observational measurements, although observations are still too imprecise to provide a serious constraint on the theoretical models.

\begin{figure}[t]
\includegraphics[width=0.48\hsize]{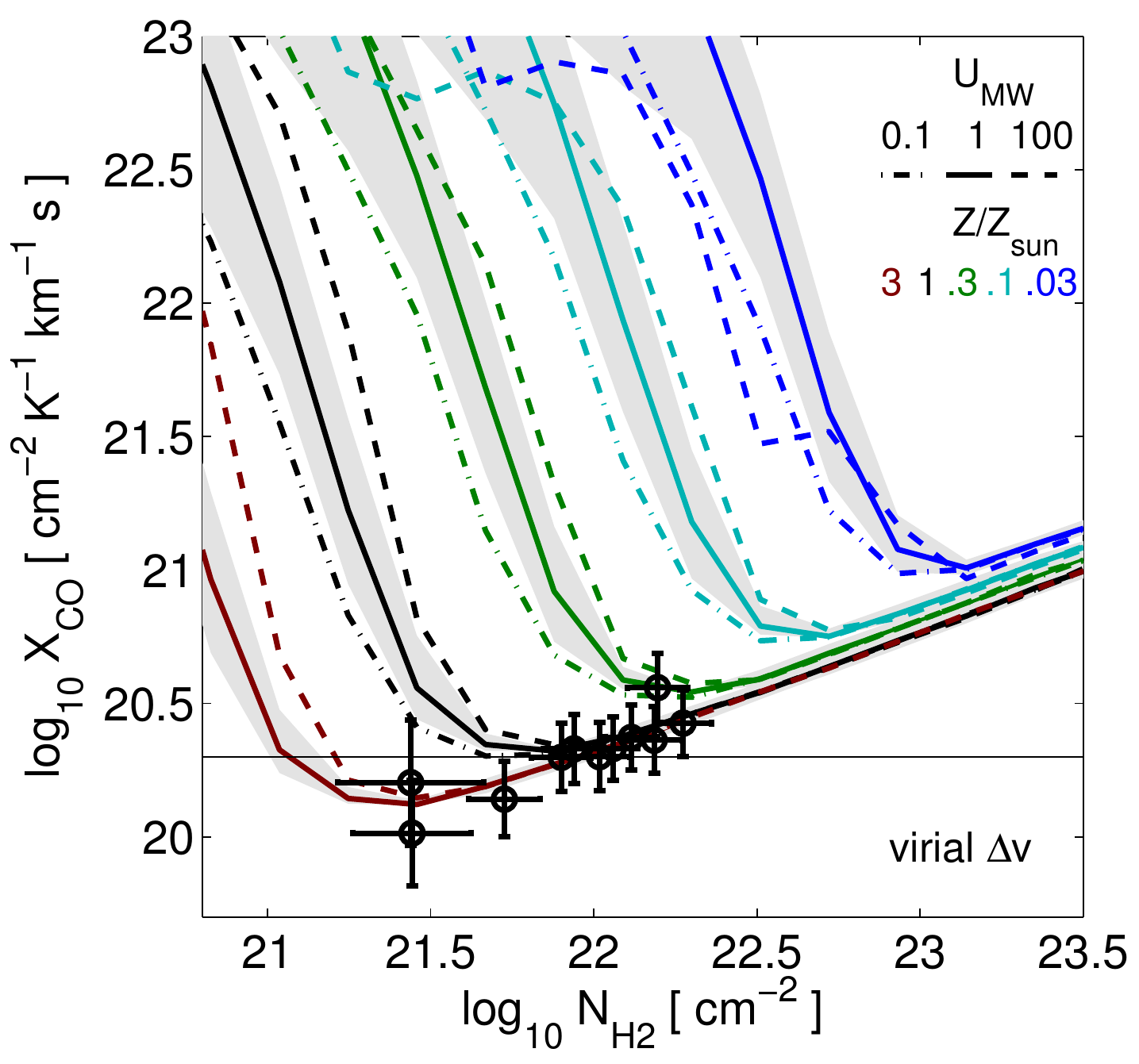}~~%
\includegraphics[width=0.48\hsize]{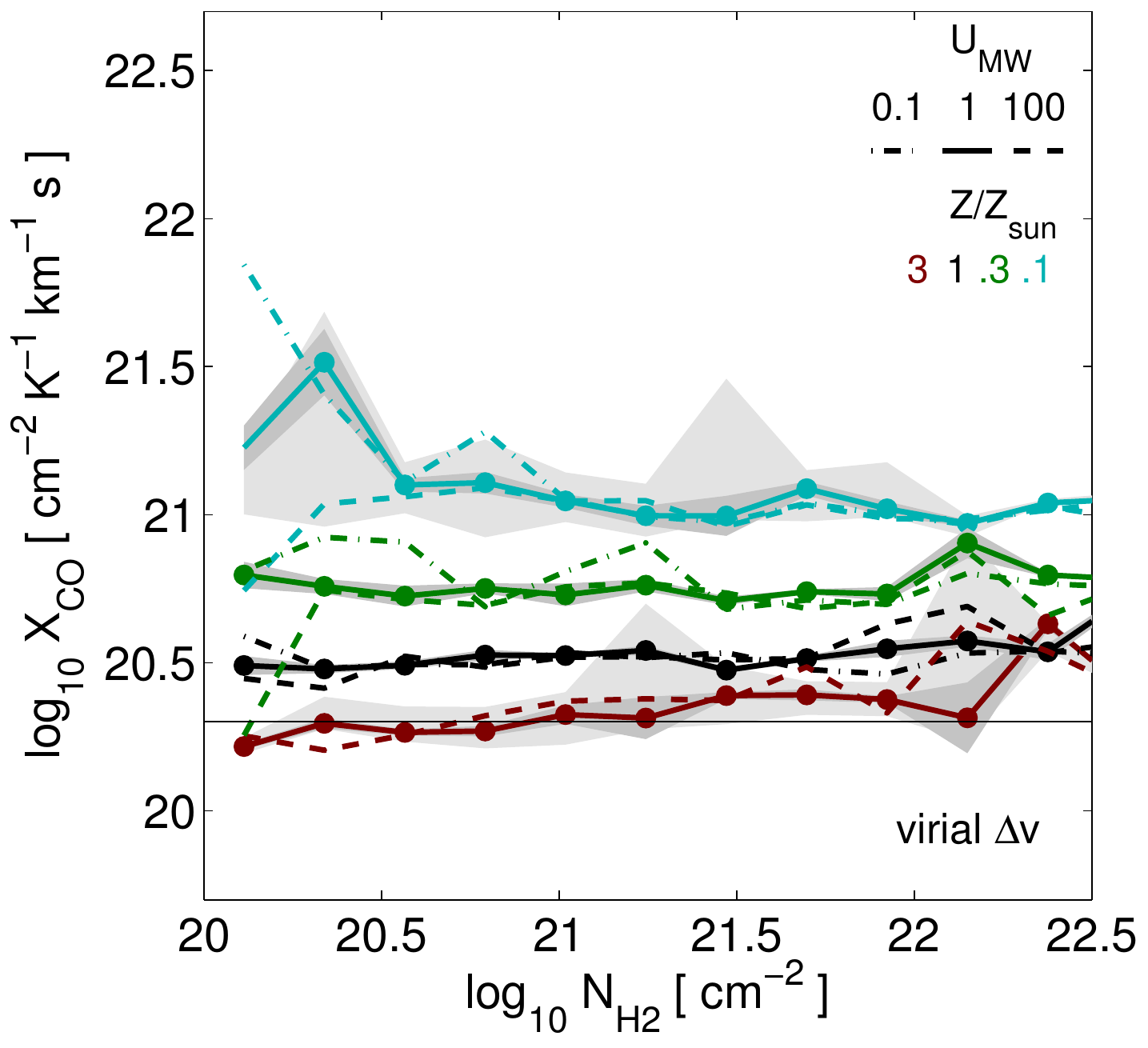}%
\caption{$X_\CO$ factor as a function of $\H2$ column density for a variety of values for the gas metallicity and the interstellar radiation field $\U$ on small, $~50\dim{pc}$ scale (left) and large, $\sim1\dim{kpc}$ scale (right). The data points on the left panel are from \protect\citet{hea10}. The role of large-scale averaging in making $X_\CO$ approximately constant is apparent (adopted from \protect\citet{ng:fgk12a}).\label{fig:xcoscale}}
\end{figure}

Using the \citet{ng:fgk12a,ng:fgk12b} model, we can explore why observers are often extremely stubborn in using a constant value for $X_\CO$ (or, alternatively, for $\alpha_\CO$). In figure \ref{fig:xcoscale} I show the dependence of the $X_\CO$ factor on the molecular hydrogen column density on small (GMC) scales and on large (galactic) scales. Averaging over large scales performs a miracle - almost all the complicated variations in the $X_\CO$ factor with various environmental parameters disappear (except for the mild residual dependence on the metallicity) and $\bar X_\CO\propto\alpha_\CO$ becomes a surprisingly robust conversion factor from the observed $\CO$ luminosity to the total mass of the molecular gas in a distant galaxy (this is indeed nothing short of a miracle).

Before we depart from the domain of sub-grid modeling of the $X_\CO$ factor, a word of caution is in order. Such modeling is, obviously, not unique. In addition, the existing observational constraints that can be used to calibrate such modeling are still in their infant stage. Hence, any sub-grid model for the $X_\CO$ factor will remain highly imprecise for some time. For example, an alternative model was proposed by \citet{ism:nko11} in which $X_\CO$ is a \emph{decreasing} function of $\H2$ column density - the dependence that has the \emph{opposite} sign to the left panel of figure \ref{fig:xcoscale}. That does seem somewhat inconsistent with the data from \citet{hea10}, but the measurements are not yet fully constraining. In any event it is clear that if two different models predict opposite signs, there is a large amount of work laying ahead...

\subsubsection{Cosmic Pandora Box, level 2: The X-Factor in ULIRGS}

Cosmic Pandora boxes are like Russian Matrioshka dolls, inside one there is always another one...

The remarkable property of the $X_\CO$ factor to average out on large scales has been used extensively in many extragalactic studies. From an observer's point of view, it is very convenient to be able to determine the molecular gas mass of a distant galaxy by a simple multiplication. There is, however, a complication. For an optically thick emission, like $\CO$, the equivalent width of the line $W_\CO = T_B \int \beta(v) dv$, where $T_B$ is the brightness temperature of the emitting gas and $\beta(v)$ is the escape probability from a parcel of gas with velocity $v$. The second factor can be thought of as an effective line width $\Delta v_{\rm eff}$, so that
\[
  X_\CO = \frac{N_\H2}{T_B \Delta v_{\rm eff}}.
\]
Variations in $N_\H2$ and $\Delta v_{\rm eff}$ do average out, so it is $T_B$ that we are concerned with now. In LTE brightness temperature is equal to the gas temperature. In normal galaxies molecular gas is very cold, $T_B\sim10\dim{K}$, but dust is usually warmer than the gas, $T_{\rm dust}=40-60\dim{K}$. Hence, if dust and gas couple thermally, $T_B$ can increase systematically in at least some molecular clouds, causing a systematic decrease in the $X_\CO$ factor that will not average out.

For dust and gas to couple, densities must be really high, significantly higher than is achieved in normal molecular clouds, so in normal galaxies coupling occurs only in a tiny fraction of the most dense molecular gas. The situation is different in Ultra-Luminous IR Galaxies (ULIRG), which are major merger of large galaxies. In mergers substantial fraction of the total gas in both galaxies gets channeled towards the center, where it gets extremely dense, piling up to many thousands of solar masses per square parsec (versus a few tens for galaxies like the Milky Way). At such high densities (and column densities) dusts starts coupling to (and, hence, heating) the gas.

In a classical study \citet{sdr97} explored that effect in several nearest ULIRG, and concluded that the $X_\CO$ factor (or, rather, $\alpha_\CO$, since we are talking about external galaxies) could be as low as $\alpha_{\CO,\rm min}=0.8\Msun/\dim{pc}^2/\dim{K}/\dim{(km/s)}$. That value, however, was only a strict lower limit, as their results depended on several assumptions that all added a factor of 2 factors on top of $\alpha_{\CO,\rm min}$. Alas, in an ironic mis-interpretation of the \citet{sdr97} paper many observers took that lower limit as the actual value, and for almost 2 decades it was quite common to hear a fairy tale of "two modes of star formation", each with its own value of $\alpha_\CO$ (0.8 and 4.3). 

Obviously, such a "bimodality" makes no physical sense - a miracle of nature may make $\alpha_\CO$ a universal constant, but if it is not, then there must be either a distribution of $\alpha_\CO$ for different galaxies or a systematic trend of the average $\alpha_\CO$ value with some of galaxy properties, like the mean surface density or IR luminosity.

Fortunately, the dust settled (or, more precisely, was observed directly), thanks to Herschel (again, not a somewhat eccentric, clever, and compassionate man but a space telescope). Measurements of dust emission over several bands between 100 and 1000 microns, when taken together with optical and sub-mm observations from the ground, allow to fit detailed models of dust spectral energy distribution and, hence, derive dust temperature and mass, in a substantial sample of ULIRG over a wide redshift range, all the way to $z\sim3$ \citep[c.f.][]{ism:mdns12}. 

These observations can then be combined with measurements of gas metallicities and $\CO$ luminosities in the same galaxies in  two different ways.
\begin{enumerate}
\item If one assumes dust-to-gas ratio as a function of metallicity $M_{\rm gas}/M_{\rm dust}(Z)$ (for example, by calibrating from the measurements of nearby galaxies), then from $M_{\rm dust}$ one gets $M_{\rm gas}$, and under the assumption that all gas is molecular, $M_{\rm gas}$ and $L_\CO$ give $\alpha_\CO$.
\item Alternatively, if one adopts a value for $\alpha_\CO$, the dust-to-gas ratio can be derived in the reverse order of steps.
\end{enumerate}

\begin{figure}[t]
\sidecaption[t]
\includegraphics[width=0.64\hsize]{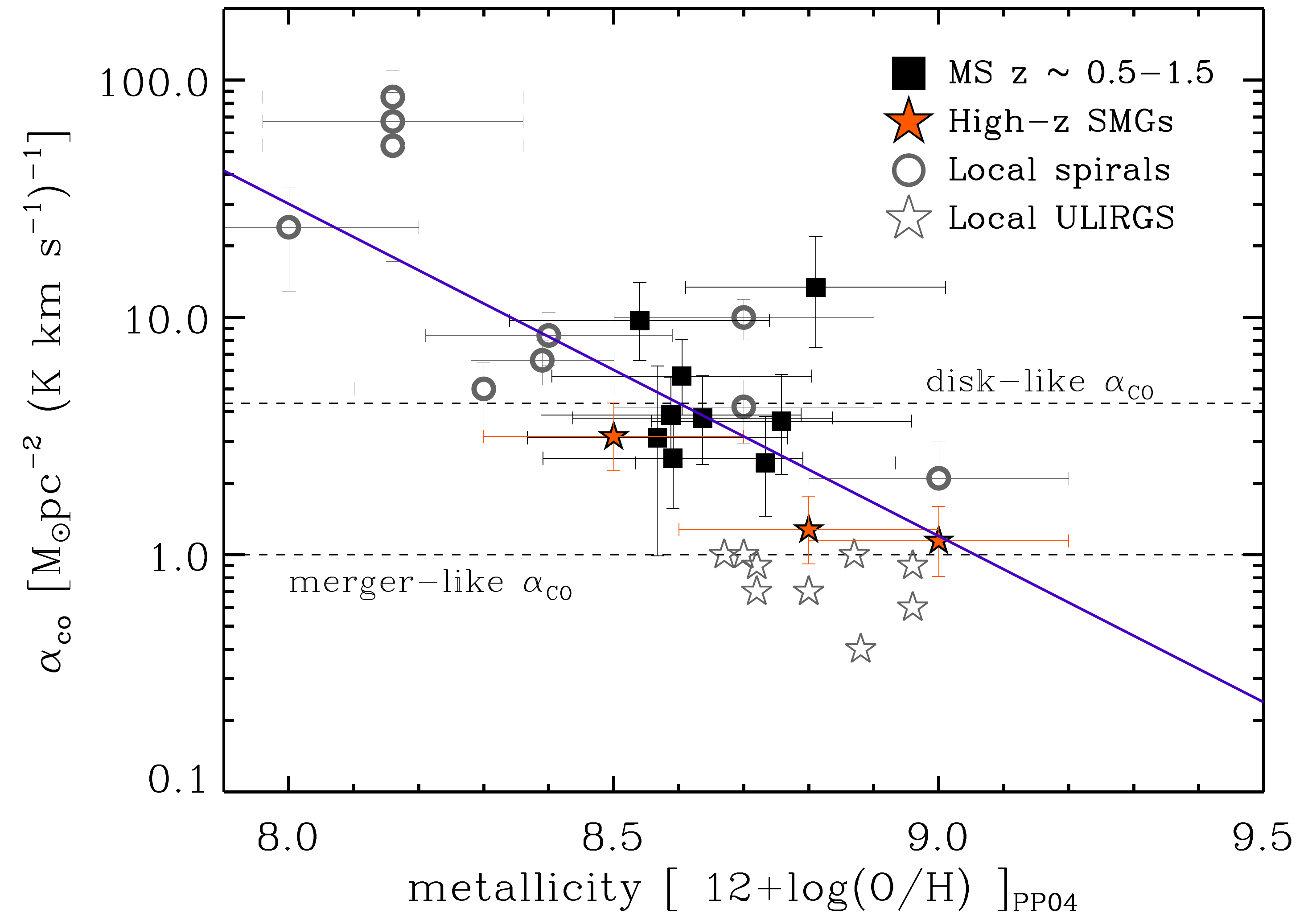}%
\caption{$\alpha_\CO$ as a function of gas metallicity for several samples of local and high redshift galaxies (adopted from \protect\citet{ism:mdns12}).\label{fig:aco}}
\end{figure}

The measurements of $\alpha_\CO$ vs $Z$ for \citet{ism:mdns12} sample and other available samples are shown in figure \ref{fig:aco}. The data are inconclusive - a trend with metallicity, a wide distribution, even bimodality cannot yet be excluded, but the main conclusion is clear - the $X_\CO$ \emph{is not universal}.

\subsubsection{Cosmic Pandora Box, level 3: Which Transition Dominates?}

\begin{figure}[b]
\sidecaption[t]
\includegraphics[width=0.64\hsize]{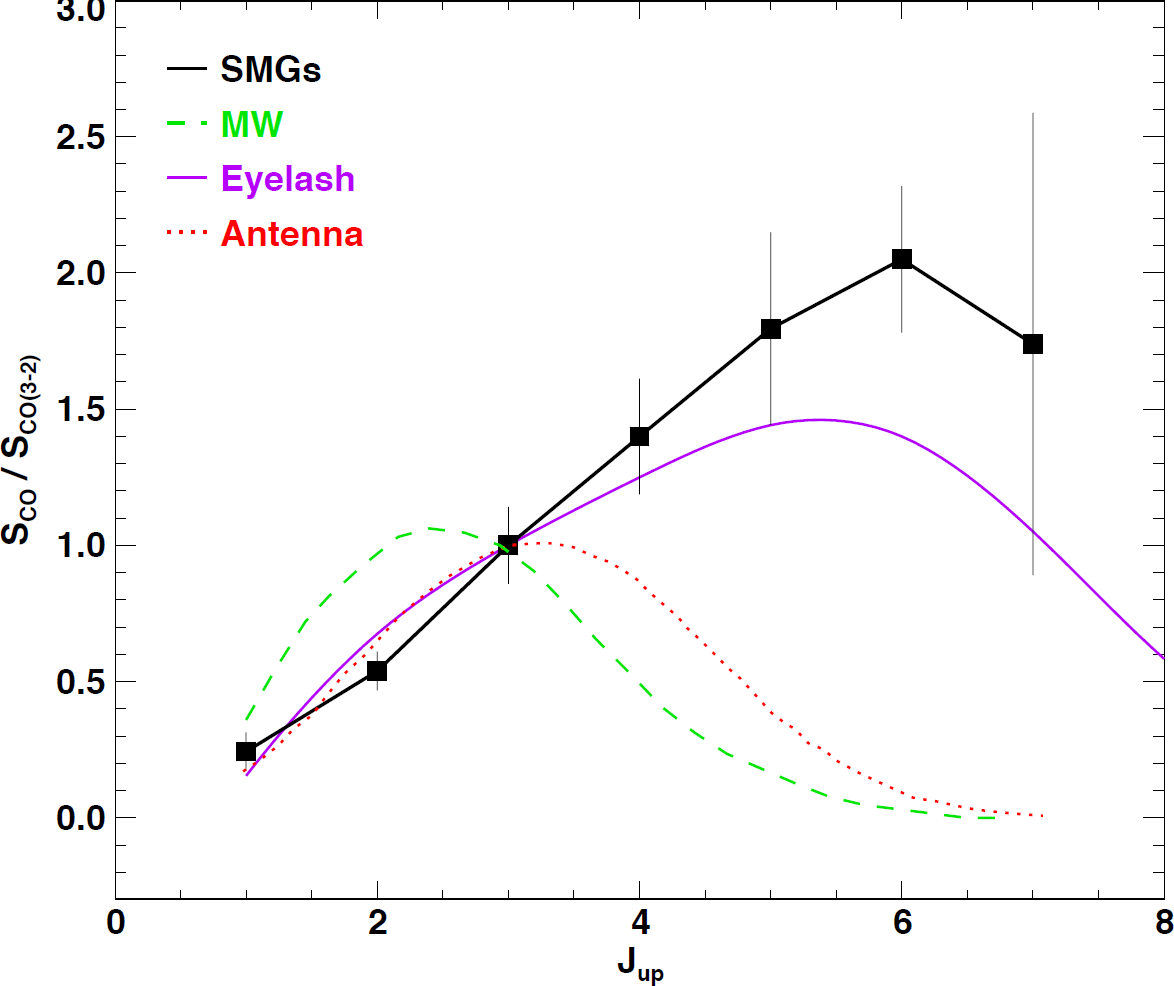}%
\caption{Distribution of $\CO$ emission over the rotational transition $J$ for several galaxies (colors) and the average distribution for high-redshift sub-millimeter galaxies (SMG). For most galaxies the $1\rightarrow0$ transition is not the dominant one (adopted from \protect\citet{bsc13}).\label{fig:coj}}
\end{figure}

If, by now, you are totally disenchanted with the $X_\CO$ factor, here is an insult to your injury - in figure \ref{fig:coj} I show a distribution of $\CO$ emission over the rotational transitions $J\rightarrow(J-1)$ for several galaxies. Even in our own Milky Way $\CO$ emits most of its energy in the $2\rightarrow1$ transition, in more active/merging galaxies the peak of the emission is shifted to even high transitions (i.e.\ higher gas temperatures). Hence, the $X_\CO$ factor is different for different $J\rightarrow(J-1)$ transitions, so to compare apples to apples, we need to convert different observed transitions to one baseline one (say, $1\rightarrow0$). These new conversions factors will also depend on the galactic environment, dust temperature, perhaps redshift, etc. A hierarchy of nested Pandora boxes never ends...

\abstract*{Now we are at the Holy Grail of all of astronomy - star formation. The subject is so huge, that we will only touch it gently, at the largest scales, and will not go much beyond the canonical Kennicutt-Schmidt relation. We will, though, think about in 2D, realizing that the spatial scale is as an important physical parameter as the density of the molecular gas itself. We will theorize about how the two-dimensional scale/density plane can be charted by star formation rate, but will not arrive at an answer. At the end we will briefly get familiar with a novel idea of using the Excursion Set formalism as a theory of star formation.}

\def\Ssfr{\Sigma_{\rm SFR}}
\def\Sgas{\Sigma_{\rm gas}}
\def\Sntr{\Sigma_{\HI+\H2}}
\def\Sh2{\Sigma_\H2}
\def\Shi{\Sigma_\HI}
\def\Shii{\Sigma_\HII}
\def\tsf{\tau_{\rm SF}}
\def\rhomol{\rho_{\rm mol}}

\section{Star Formation}
\label{sf}

If the field of ISM is large, what one can say about star formation - it is at least another quarter of all Astronomy research. So, we must thread very carefully, or we will be lost forever in the jungle of clouds, disks, and outflows. We will attempt to stay on largest scales, and will look only on the most generic relations between gas and stars. We are not even going to paint the broad picture, we will just look at the frame...

\subsection{Kennicutt-Schmidt and All, All, All}

For us, looking down on star formation from galactic scales and above, the story of star formation begins in March 1959, with the classical paper by Maarten Schmidt \citep{s59}, who noticed that the \emph{surface} density (and let us be precise here, we still have very little observational clues on what the \emph{volumetric} density of star formation is doing) of star formation correlates with the surface density of gas approximately as a power-law,
\[
  \Ssfr \propto \Sgas^n,
\]
with $n=1-2$.

This relationship was firmed up later by Rob Kennicutt \citep{sfr:k89,sfr:k98a}, resulting in what is nowadays commonly referred to as the Kennicutt-Schmidt (KS) relation\footnote{God save you from calling it a "law" in the presence of a devout physicist!},
\begin{equation}
  \Ssfr = (2.5\pm0.7)\times10^{-4}\frac{\Msun}{\dim{kpc}^2\dim{yr}}
  \left(\frac{\Sgas}{1\Msun/\dim{pc}^2}\right)^{1.4\pm0.15}.
  \label{eq:k98}
\end{equation}

In this form the KS relation survived for 10 years. But THINGS does matter (orthography is correct), The Nearby $\HI$ Galaxies Survey was an important step in shaping our modern understanding and interpretation of the KS relation, in large part because in addition to $\HI$, the THINGS team assembled a large amount of other data on their target galaxies, from $\CO$ emission to UV and  H-$\alpha$ measurements of star formation rates.

\begin{figure}[t]
\includegraphics[width=1\hsize]{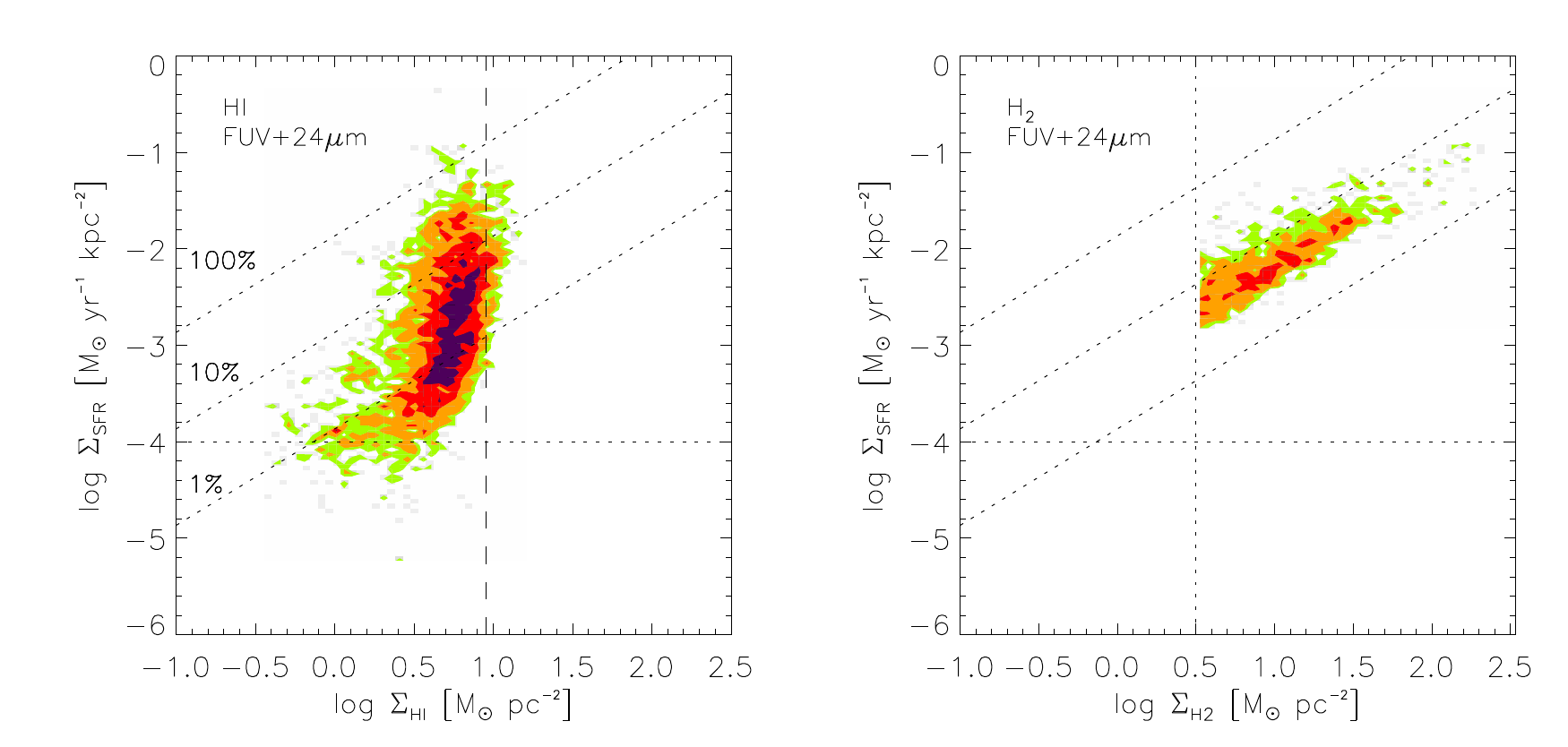}%
\caption{Star formation surface density as a function of $\HI$ (left) and $\H2$ surface densities from the THINGS survey (adopted from \protect\citet{sfr:blwb08}).\label{fig:things1}}
\end{figure}

The THINGS survey unambiguously proved what everyone knew in their hearts: stars form from molecular gas\footnote{At least, the vast majority of them - by itself, the THINGS result does not exclude a possibility of a small fraction of stars forming in the atomic gas.}. In figure \ref{fig:things1} there is a clear strong correlation between $\Ssfr$ and the surface density of the molecular gas, but there is almost no correlation with the atomic gas. Hence, we have not wasted our time discussing the atomic-to-molecular transition, it is one of the bottlenecks that controls star formation in galaxies.

\begin{figure}[b]
\includegraphics[width=0.48\hsize]{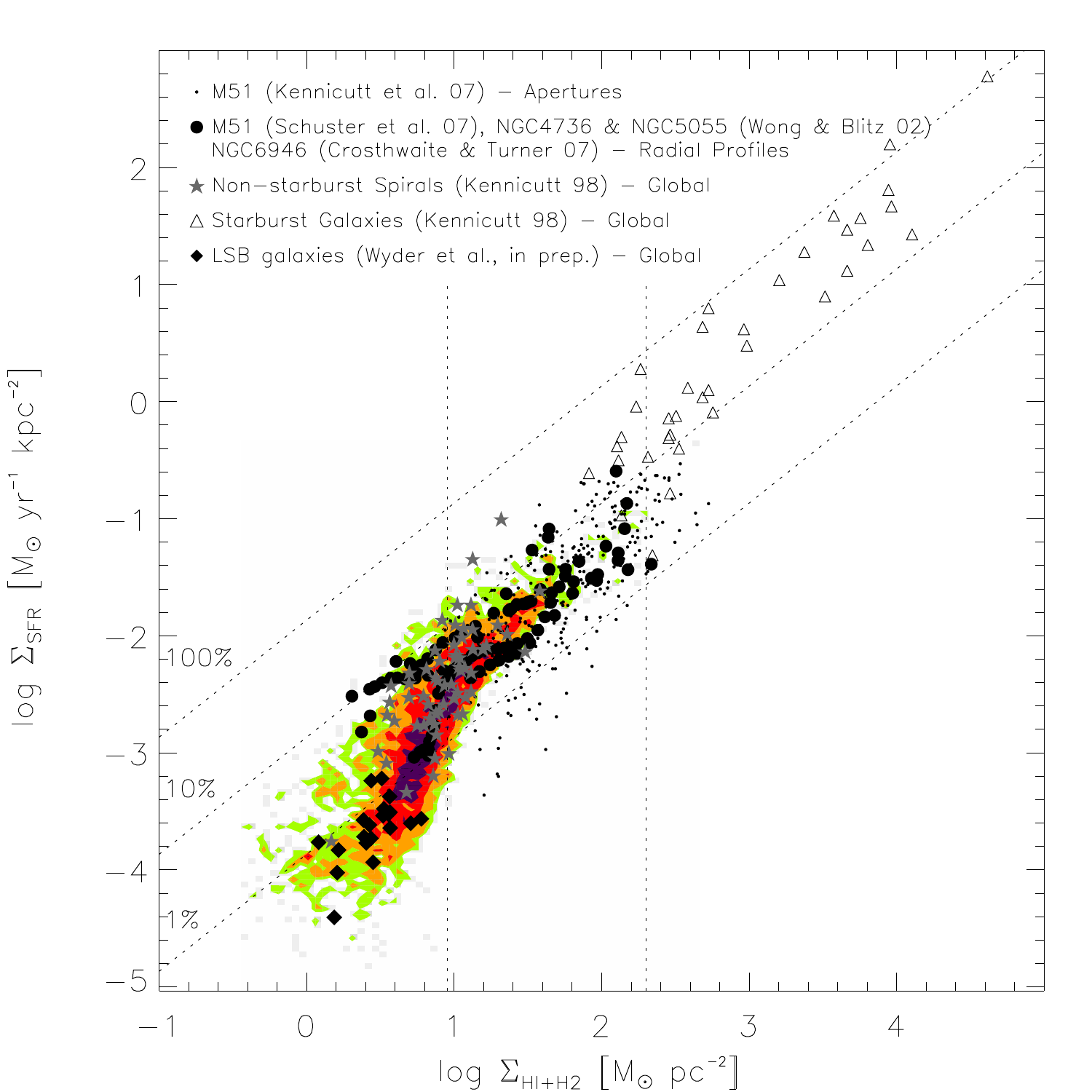}~~%
\includegraphics[width=0.48\hsize]{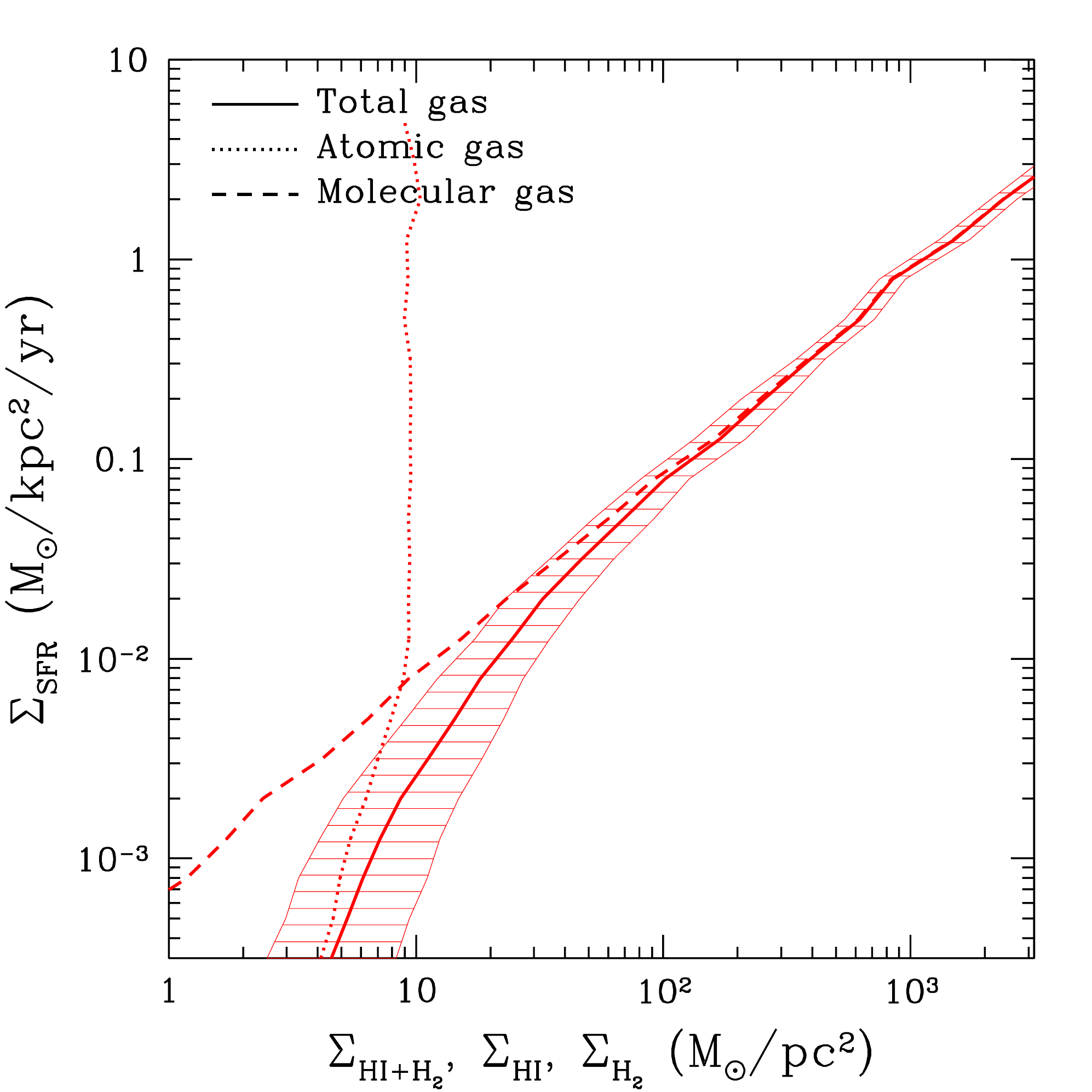}%
\caption{\emph{Left}: "classical" Kennicutt-Schmidt relation from THINGS (adopted from \protect\citet{sfr:blwb08}). \emph{Right}: separate average KS relations for the atomic (dotted), molecular (dashed), and neutral (solid) gas from the already familiar to us cosmological simulation of the Milky Way like galaxy. The solid line is the sum of the dotted and dashed along the horizontal direction.\label{fig:ksr}}
\end{figure}

Historically, it was common to represent the KS relation as the relation between the star formation surface density and the surface density of the "total" gas, which actually meant the sum of atomic and molecular (i.e.\ \emph{neutral}) gas. The left panel of figure \ref{fig:ksr} shows this "classical" form of KS relation from the THINGS data, together with the original measurements from \citet{sfr:k98a} (although the latter are tricky to interpret, since they use a different value of $\alpha_\CO$ to convert $\CO$ emission to the molecular gas surface density).

In order to illustrate how that particular shape appears, the right panel shows the KS relation from a numerical simulation of the Milky-Way-like galaxy that we already met several times in two previous chapters. In the simulation the star formation rate is postulated to be linearly proportional to the molecular gas surface density,
\begin{equation}
  \Ssfr = \frac{1.36\Sh2}{\tsf},
  \label{eq:tsf}
\end{equation}
where the factor 1.36 is, again, to account for Helium, and $\tsf$ is the gas \emph{depletion time}, assumed to be constant $\tsf=1.5\dim{Gyr}$ in the simulation (we will come back to that number shortly).

The "classical" KS relation then forms from the separate atomic $\Shi$ and molecular $\Sh2$ surface densities as
\[
  \begin{array}{lcrcr}
    \Ssfr & = & {\displaystyle\frac{1.36}{\tsf}}\Sh2 & + & 0\times\Shi,\\
    \Sntr & = & \Sh2 & + & \Shi.\\
  \end{array}
\]
The steepening of the KS relation at low surface densities is simply due to gas becoming predominantly atomic, and is fully explained by the physics of the atomic-to-molecular transition. Indeed, observations support this interpretation (but we won't dive into that question here, it is too wide and deep for us to linger in it, as we are rushing along our yellow brick road).

\subsubsection{How we should think about star formation}

If star formation correlates well with molecular gas, it is useful to think about equation (\ref{eq:tsf}) as our primary ansatz, and consider how $\tsf$ may depend on other properties (for example, density). That thinking, however, is \emph{totally wrong}!

A simple fact that we often forget is that density is \emph{not even defined} without a particular scale. After all, $\rho=M/V$, and if there is no $V$, there is no $\rho$. Hence, both theoretically and observationally, we need to explicitly consider the range of spatial scales that is relevant for our problem.

Let us take some spatial scale $L$. One can imagine the whole universe divided into boxes of size $L$, like in a super-huge numerical simulation, or the universe observed with a given telescope resolution. If we average gas densities on scale $L$, they become meaningfully defined. Thus, equation (\ref{eq:tsf}) should really be replaced with
\begin{equation}
  \langle\dot{\rho}_*\rangle_L = \frac{\langle\rhomol\rangle_L}{\tsf},
  \label{eq:tsfL}
\end{equation}
where $\rhomol=1.36\rho_\H2$ is the density of the molecular gas, and averaging is done over the spatial scale $L$. In that case depletion time becomes the function of other gas properties on scale $L$,
\[
  \tsf = \tsf(L,\langle\rhomol\rangle_L,...).
\]
In other words, we need to explicitly think of star formation relation as (at least) a two-dimensional relation on the plane $(L,\langle\rhomol\rangle_L)$, or, perhaps, even a higher-dimensional relation if chemistry, magnetic fields, properties of ISM turbulence, etc are also important.

Armed with this understanding, we can now reinterpret the existing observational constraints on various scales on a uniform basis. 

\begin{figure}[t]
\includegraphics[width=0.56\hsize]{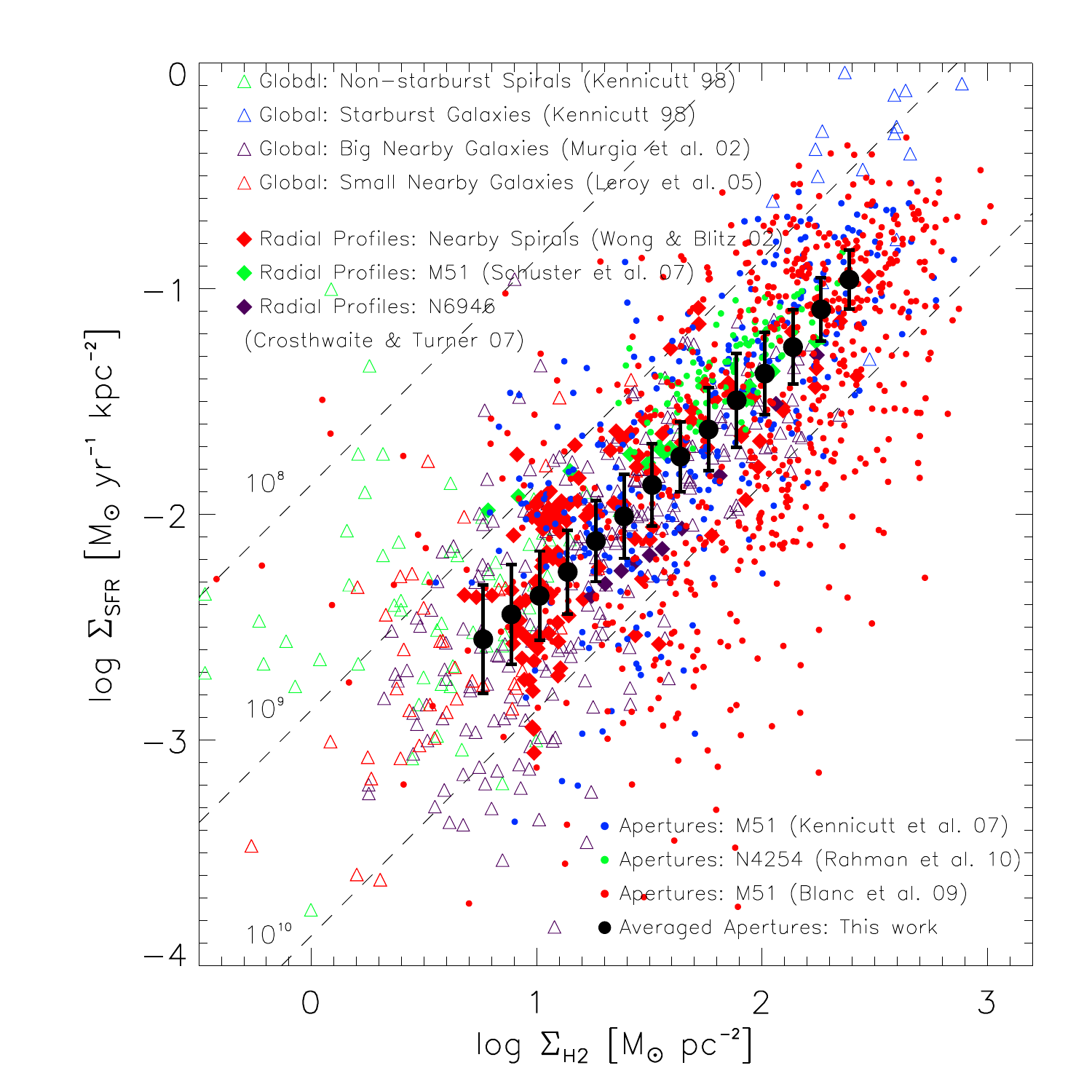}%
\includegraphics[width=0.44\hsize]{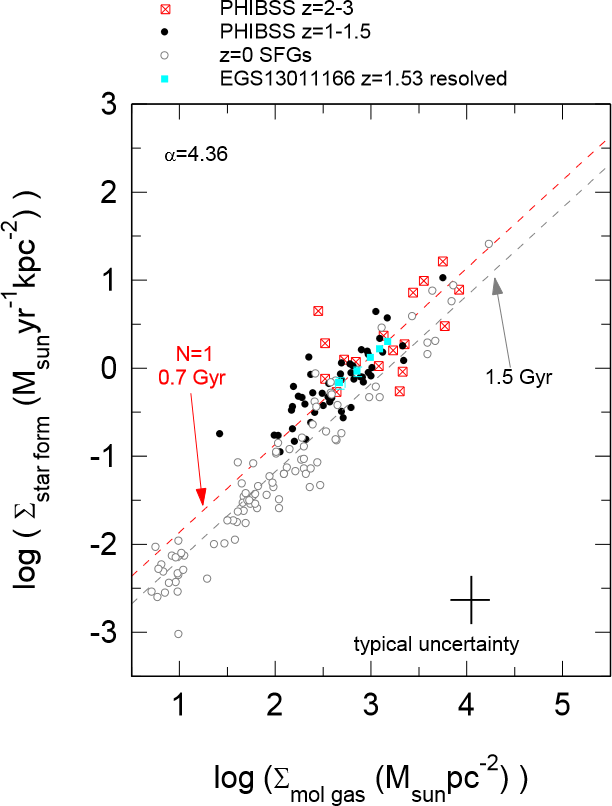}%
\caption{Star formation rate surface density versus the surface density of the molecular gas for local galaxies \protect\citep[left,][]{sfr:blwb11} and high redshift normal star forming galaxies \protect\citep[right,][]{sfr:tngc13}.\label{fig:tsf}}
\end{figure}

In figure \ref{fig:tsf} I show the molecular KS relation for normal star forming galaxies (i.e.\ not ULIRGs, with their complicated $\CO\rightarrow\H2$ conversion) in the local universe and at high redshift. These observations sample star formation on large scales (from many hundreds of parsecs to several kilo-parsecs). They all are consistent with roughly linear KS relation, 
\[
  \tsf\approx{\rm const}(L\gg100\dim{pc}),
\]
although with substantial \emph{intrinsic} (i.e.\ exceeding the formal observational error) scatter, and the actual value for $\tsf$ noticeably different at $z=0$ ($\tsf\approx2\dim{Gyr}$) and high redshift ($\tsf\approx0.7\dim{Gyr}$).

At the present moment (Sep 2013) it seems difficult to make any further inference from these measurements. For example, is $2=0.7$? In fact, they might, since local and high redshift measurements probe very different, non-overlapping ranges of gas surface density: few local observations reach $100\Msun/\dim{pc}^2$, while all high redshift measurement are way above that limit. The \citet{ng:fgk12a} model for the $X_\CO$ factor predicts that $X_\CO$/$\alpha_\CO$ factor increases gradually with the gas surface density; \citet{ng:fgk12b} present, as an example, a cosmological simulation with constant in time and space $\tsf=1.5\dim{Gyr}$, which is consistent with both the low and high redshift measurements. We do not yet know how accurate the \citet{ng:fgk12a} model is, but, at the very least, there exists a plausible  counterexample to any potential claim that low- and high-redshift KS relation are inconsistent with each other.

The same uncertainty applies to the scatter around the mean KS relation. We do know that the $X_\CO$ factor varies across single galaxies and between different galaxies, so some fraction of the scatter should be due to scatter in $X_\CO$. In addition, there is scatter from the time dimension that we have so far ignored: $\CO$ emission from the molecular gas is essentially instantaneous, but observational estimates of star formation are not. For example, \citet{slw10} show that the depletion time is systematically higher around peaks of $\CO$ emission (molecular clouds where star formation is just starting) than around peaks of H-$\alpha$ emission (star forming regions where star formation is well under way). 

This difference is purely due to the fact that we do not measure the instantaneous rate of star formation, but use observational indicators that return a time-averaged star formation rate over some characteristic time-scale ($\sim20\dim{Myr}$ for UV light, $\sim5\dim{Myr}$ for H-$\alpha$). Hence, if we point our telescope on a freshly formed molecular cloud, we will see a lower star formation rate than the actual instantaneous one - if the cloud has been forming stars for only $1\dim{Myr}$ and we use H-$\alpha$, we will measure a $5\dim{Myr}/1\dim{Myr}=5$ times lower star formation rate than the true one. Now, if we point it at a mature star forming region, we will measure a higher time-averaged star formation rate than the instantaneous one, because $5\dim{Myr}$ ago the region contained more molecular gas (and, hence, higher instantaneous star formation rate) than it has right now. 

The combined scatter due to variations in the $X_\CO$ factor and the finite time-averaging is easily quantifiable, though, and appears to be less than the actual observed scatter in the KS relation \citep{ng:fgk12b}. This should not be particularly surprising, though - it is hard to imagine that the nature is so kind to us as to make each region of space with the same surface density of molecular gas to have \emph{exactly} the same star formation rate, surely there must be random or systematic variation from place to place that affect star formation rate, and that will appear as the true intrinsic scatter in equation (\ref{eq:tsfL}).

There exist several other constraints we can place on $\tsf(L,\langle\rhomol\rangle_L,...)$. \citet{lla10} found that on the scale of individual star-forming cores ($\sim1\dim{pc}$) the depletion time is also constant (i.e.\ independent of density) and is about $20\dim{Myr}$, but only if the density is above $\rho_{\rm min}=700\Msun/\dim{pc}^{3}$. A threshold must exist in that case, since any small-scale relation must be consistent with the large-scale one. Namely, if on $1\dim{pc}$ scale we have
\[
  \langle\dot{\rho}_*\rangle_{1} =
	\left\{
    \begin{array}{ll}
	  \displaystyle\frac{\langle\rhomol\rangle_{1}}{20\dim{Myr}}, & \rhomol > \rho_{\rm min}, \\
	  0, & \rhomol < \rho_{\rm min}, \\
	\end{array}
	\right.
\]
and on $500\dim{pc}$ scale we have a usual molecular KS relation,
\[
  \langle\dot{\rho}_*\rangle_{500} = \frac{\langle\rhomol\rangle_{500}}{2\dim{Gyr}},
\]
then these two relations can be mutually consistent if and only if exactly 1\% of the molecular gas sits above the small-scale density threshold $\rho_{\rm min}$ - after all, 
\[
  \langle\dot{\rho}_*\rangle_{500} = \left\langle\langle\dot{\rho}_*^{\phantom{|}}\rangle_{1}\right\rangle_{500}.
\]

Another commonly used ansatz for the star formation rate is constant efficiency per free-fall time,
\[
  \tau_{\rm SF}(L,\langle\rhomol\rangle_L,...) = 
  \frac{\tau_{\rm ff}(\langle\rhomol\rangle_L)}{\epsilon_{\rm SF}} = 
  \epsilon_{\rm SF}^{-1}\sqrt{\frac{3\pi}{32G\langle\rhomol\rangle_L}},
\]
or, in a more familiar form,
\begin{equation}
  \langle\dot{\rho}_*\rangle_L = \epsilon_{\rm SF}\frac{\langle\rhomol\rangle_L} 
  {\tau_{\rm ff}} = \epsilon_{\rm SF} \frac{\langle\rhomol\rangle_L^{3/2}}{\sqrt{3\pi/(32G)}}.
  \label{eq:tffsf}
\end{equation}
The origin of that formula disappears in the depths of time; it is often used without any attention to the scale under consideration. In an influential paper, \citet{sfr:kt07} argued that many observational constraints are consistent with that ansatz\footnote{One should never forget that the "constant efficiency per free-fall time" model is no more than an ansatz; molecular clouds are turbulent and the free-fall time has no physical relevance on scales above the sonic length.} with $\epsilon_{\rm SF}\approx1-2\%$ for a wide array of molecular densities, from average molecular clouds to molecular cores. 

However, observational constraints used by \citet{sfr:kt07} sample not only various densities, but also \emph{various spatial scales}; namely, they all fall along a particular track $L^2\times\langle\rhomol\rangle_L \approx 10^4\dim{cm}^{-3}\dim{pc}^2$ in the two-dimensional plane $(L,\langle\rhomol\rangle_L)$. In other words, observational constraints that support the "constant efficiency per free-fall time" are equally well support the "constant efficiency per unit scale",
\[
  \langle\dot{\rho}_*\rangle_L = \epsilon_{\rm SF}\frac{\langle\rhomol\rangle_L} 
  {\tau_{\rm ff}} \approx \epsilon_{\rm SF} \frac{\langle\rhomol\rangle_L}{\tau_{\rm ff}(10^4\dim{cm}^{-3})(L/1\dim{pc})}.
\]
The two alternatives cannot be distinguished at present without additional observational constraints.

In fact, I am going to make a bold claim (and challenge anyone to refute it) that \emph{all of the existing observational constraints are consistent with the linear (in density) star formation ansatz} in which the depletion time is function of scale only,
\begin{equation}
  \langle\dot{\rho}_*\rangle_L =
	\left\{
    \begin{array}{ll}
	  \displaystyle\frac{\langle\rhomol\rangle_L}{\tsf(L)}, & \rhomol > \rho_{\rm min}(L), \\
	  0, & \rhomol < \rho_{\rm min}(L), \\
	\end{array}
	\right.
	\label{eq:linsf}
\end{equation}
with
\[
  \tsf(L) \sim 2\dim{Gyr}\times\mbox{min}\left(1,\frac{L}{L_0}\right),
\]
\[
  \rho_{\rm min}(L) \sim \rho_0\times\mbox{min}\left(1,\frac{L_0^2}{L^2}\right),
\]
and $L_0$ is in the range of a few hundred parsecs (for example, the scale height of the gaseous disk). In the Milky Way galaxy $\rho_0$ is such that the \citet{lla10} result is matched ($\rho_{\rm min}(L)\approx700\Msun/\dim{pc}^{3}$ at $L\sim1\dim{pc}$), but in other galaxies it may be different (for example, being proportional to the density of the atomic-to-molecular transition).

\begin{figure}[t]
\includegraphics[width=0.53794\hsize]{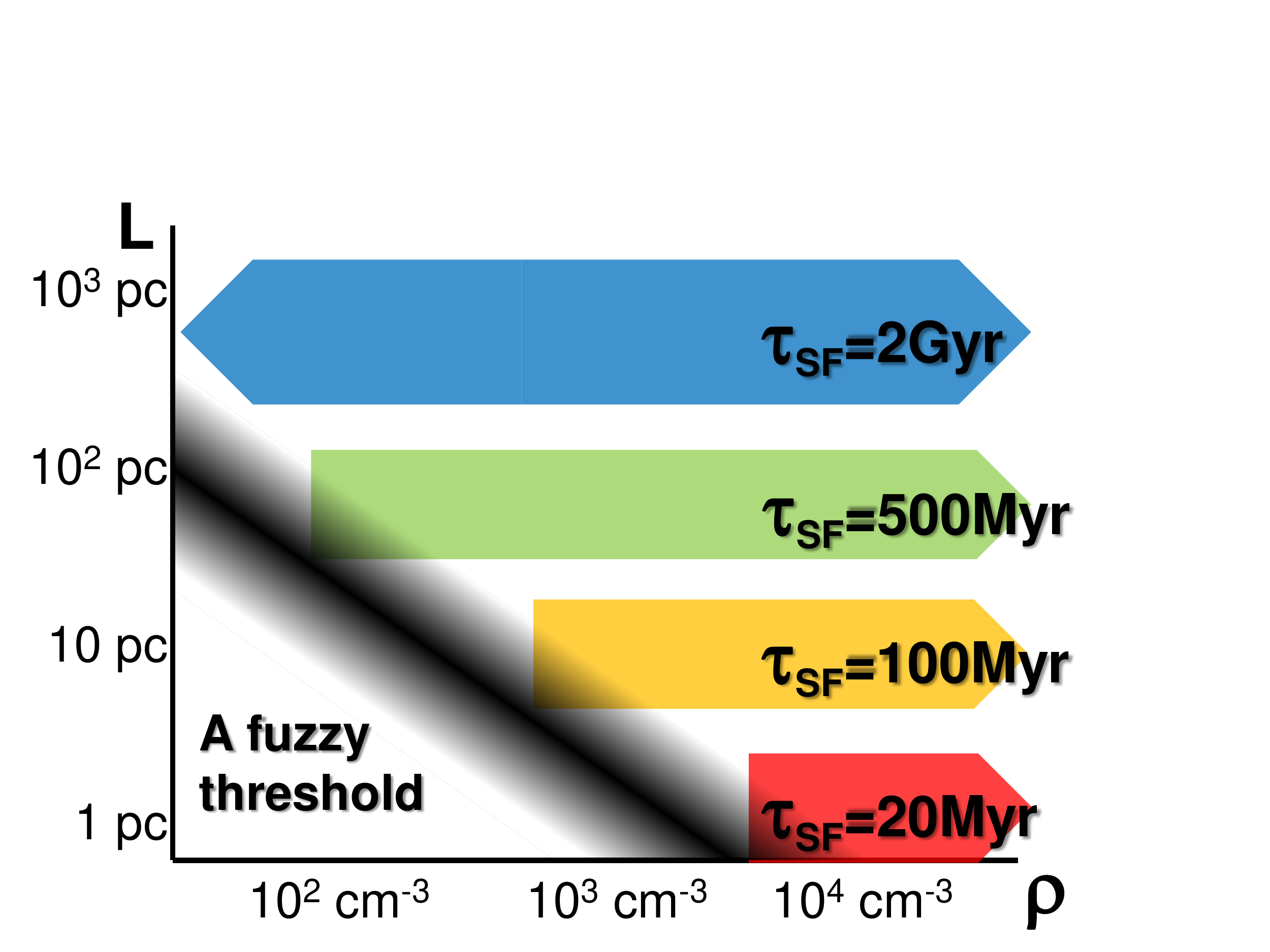}%
\includegraphics[width=0.46206\hsize]{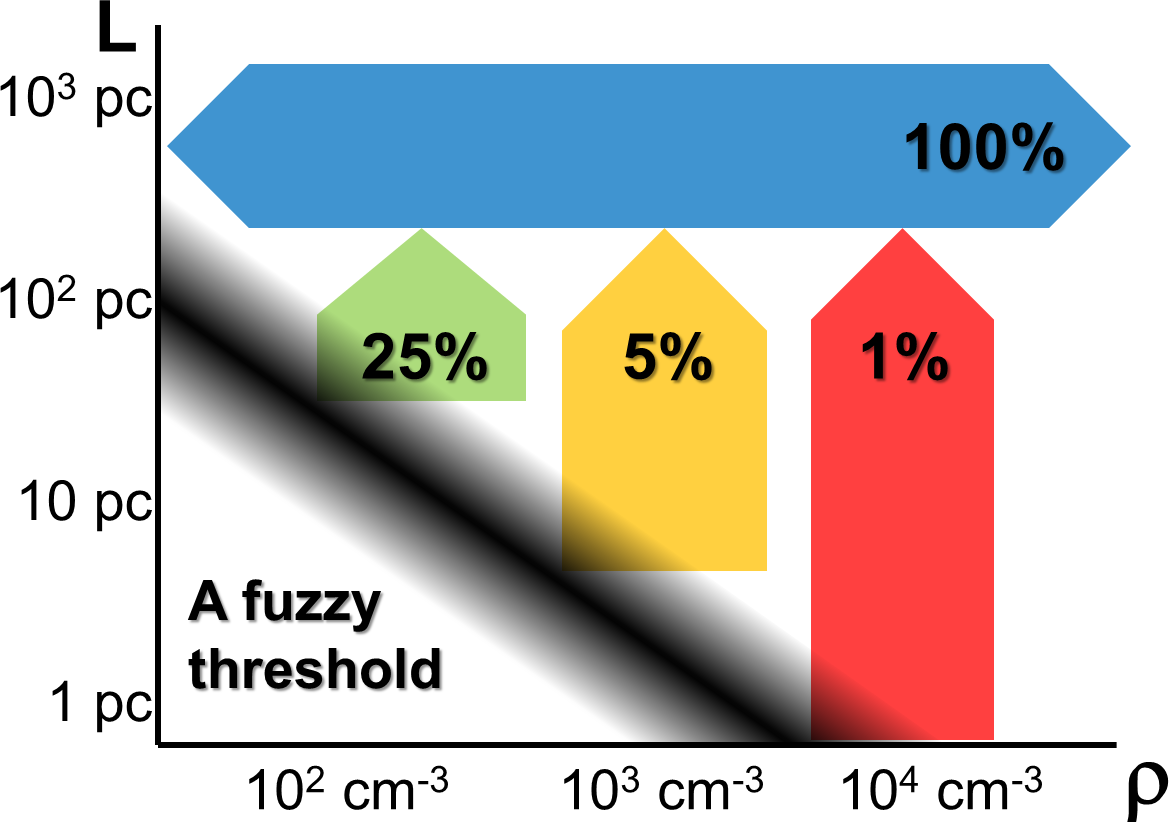}%
\caption{Cartoon version of contours of constant depletion time (shown as different colored bands) on the $(L,\langle\rhomol\rangle_L)$ plane. The left panel shows the linear star formation relation (\ref{eq:linsf}), while the right panel shows the constant-efficiency-per-free-fall-time star formation relation (\ref{eq:tffsf}). In the latter case the depletion function must transition to a constant on the largest scale somehow to be consistent with large-scale KS relation.\label{fig:linsf}}
\end{figure}

Figure \ref{fig:linsf} shows two alternatives (linear star formation relation \ref{eq:linsf} and constant-efficiency-per-free-fall-time star formation relation \ref{eq:tffsf}) in a cartoon fashion. At present, either one is a sensible model, as well as any other, intermediate or more complex model, that still matches the observational constraints.

\subsection{Excursion Set Formalism in Star Formation}

The idea of using Excursion Set formalism in star formation is based on a well established fact: in isothermal supersonic turbulence the density PDF is lognormal, in a direct analogy with the Gaussian distribution of the linear overdensity $\delta$. Such an approach was first attempted by \citet{pn02}, picked up later by \citet{hc08} and developed much further by Phil Hopkins in a recent series of papers \citep{h12a,h12b,h13}.

\subsubsection{Refresher: Excursion Set Formalism}

\begin{figure}[b]
\sidecaption[t]
\includegraphics[width=0.64\hsize]{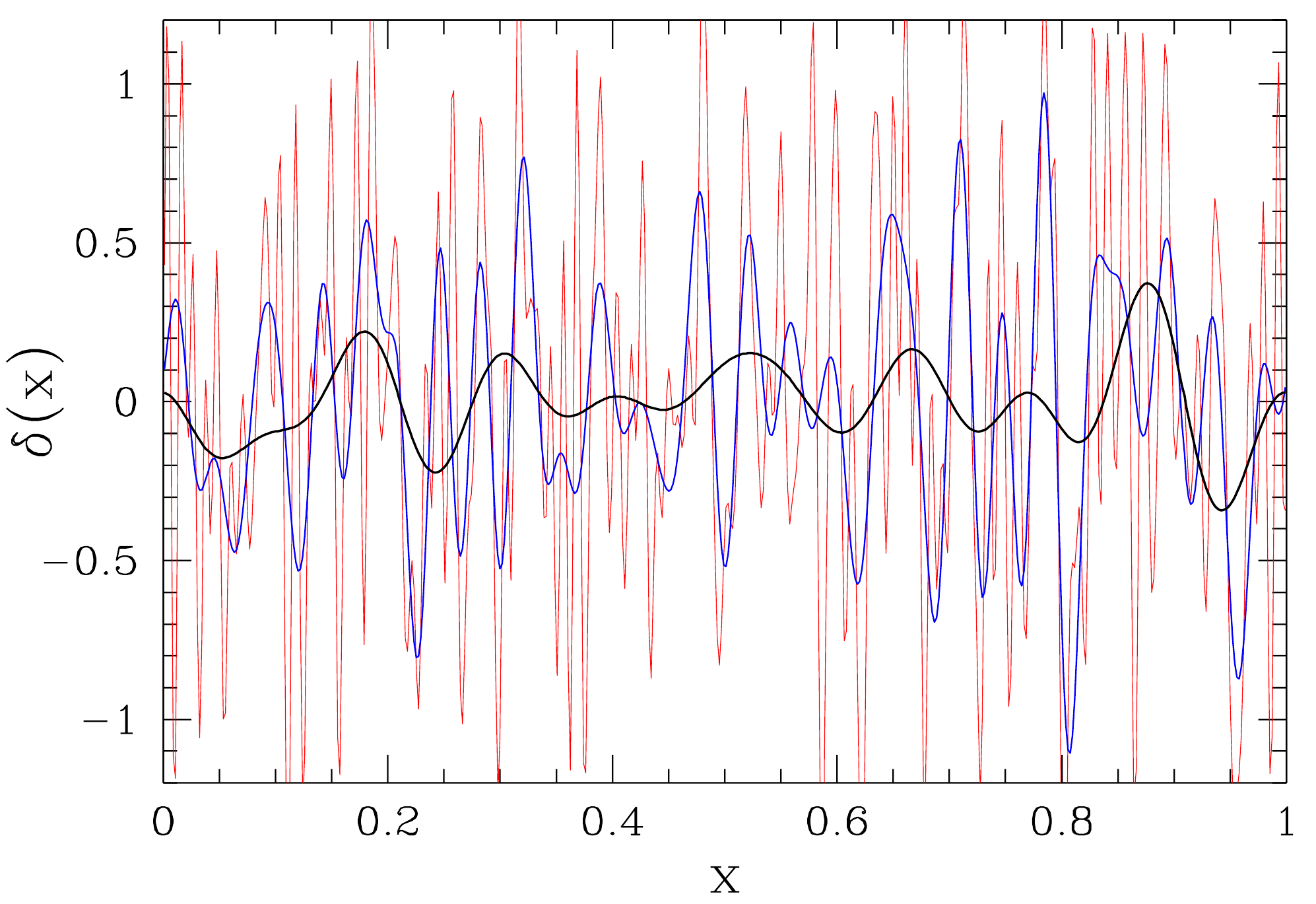}%
\caption{Example of the Gaussian random field at three different values for the smoothing scale $R$.\label{fig:grf}}
\end{figure}

Excursion Set formalism (sometimes also called "Press-Schechter formalism") deals with a \emph{Gaussian random field} $\delta(\vec{x})$ (and $\delta$ can be anything, for supersonic turbulence it will be $\ln(\rho/\rho_0)$). For a Gaussian random field different wavenumbers of the Fourier transform 
\[
  \delta_{\vec{k}}\equiv \int d^3x\delta(\vec{x})e^{i\vec{k}\vec{x}}
\]
of the field are uncorrelated,
\[
  \langle\delta^{\phantom{*}}_{\vec{k}_1}\delta^*_{\vec{k}_2}\rangle = P(k_1)\delta^3_D(\vec{k}_1-\vec{k}_2).
\]
One can reverse the Fourier transform,
\begin{equation}
   \delta(\vec{x}) = \int d^3k \sqrt{P(k)} \lambda_{\vec{k}} e^{-i\vec{k}\vec{x}},
  \label{eq:grf}
\end{equation}
with uncorrelated, normally distributed random numbers $\lambda_{\vec{k}}$ satisfying the relation $  \langle\lambda^{\phantom{*}}_{\vec{k}_1}\lambda^*_{\vec{k}_2}\rangle = \delta^3_D(\vec{k}_1-\vec{k}_2)$.
Equation (\ref{eq:grf}) should be considered in a generalized sense (similar to Dirac delta-function), because for some $P(k)$ the integral may actually diverge. In that case $\delta(\vec{x})$ should be considered as a limit of the smoothed density field,
\[
   \delta(\vec{x}) \equiv \lim_{R\rightarrow0}  \delta(\vec{x};R) = \int d^3k \sqrt{P(k)} \lambda_{\vec{k}} W(kR) e^{-i\vec{k}\vec{x}},
\]
where $W(kR)$ is a low-pass filter ($W(0)=1$, $W(\infty)=0$). An example of a Gaussian random field at 3 different resolutions is shown in figure \ref{fig:grf}.

Excursion Set formalism considers $\delta(\vec{x};R)$ as a function of the filter scale $R$ and compares it with some “barrier” function $b(R)$. Obviously, $\delta(\vec{x},R=\infty)=0$. As $R$ decreases, $\delta(\vec{x};R)$ starts deviating from zero. For some value of $R$ it may cross the barrier for the first time. The fraction of all $\delta(\vec{x};R)$ that cross the barrier at $R$ is called the “first crossing distribution”.  For example, in the canonical Press-Schechter formalism the barrier is constant, $b=\delta_L(t_f)=1.69$. Then the first crossing distribution becomes (half) the mass function of dark matter halos with $M_h=4\pi \bar\rho_m R^3/3$.

In modeling star formation Excursion Set formalism can be used for several goals:
\begin{itemize}
\item First crossing distribution gives the mass function of largest bound objects -– molecular clouds.
\item Last crossing distribution gives the mass function of smallest bound objects -– molecular cores/stars.
\item It is useful for other purposes too: distribution of holes in the ISM, clustering of stars, etc.
\end{itemize}

One only needs to define a barrier - but we have already considered it! After all, gas collapses when it becomes gravitationally unstable, hence the barrier is simply the stability criterion for the disk with finite thickness, our equation (\ref{eq:diskdrh}) - with a minor correction of adding turbulent velocity dispersion $\sigma_t^2$ to the gas sound speed $c_S^2$, since turbulence also provides support against gravitational collapse.

\begin{figure}[b]
\begin{minipage}[b]{0.50\hsize}
\caption{Several predictions of the Excursion Set formalism as a theory of star formation: GMC mass functions for the Milky Way, LMC, and M31 (right, adopted from \protect\citet{h12a}) and clump mass function (bottom, adopted from \protect\citet{h12b}). \label{fig:esf}}
\vspace{10ex}
\includegraphics[width=\hsize]{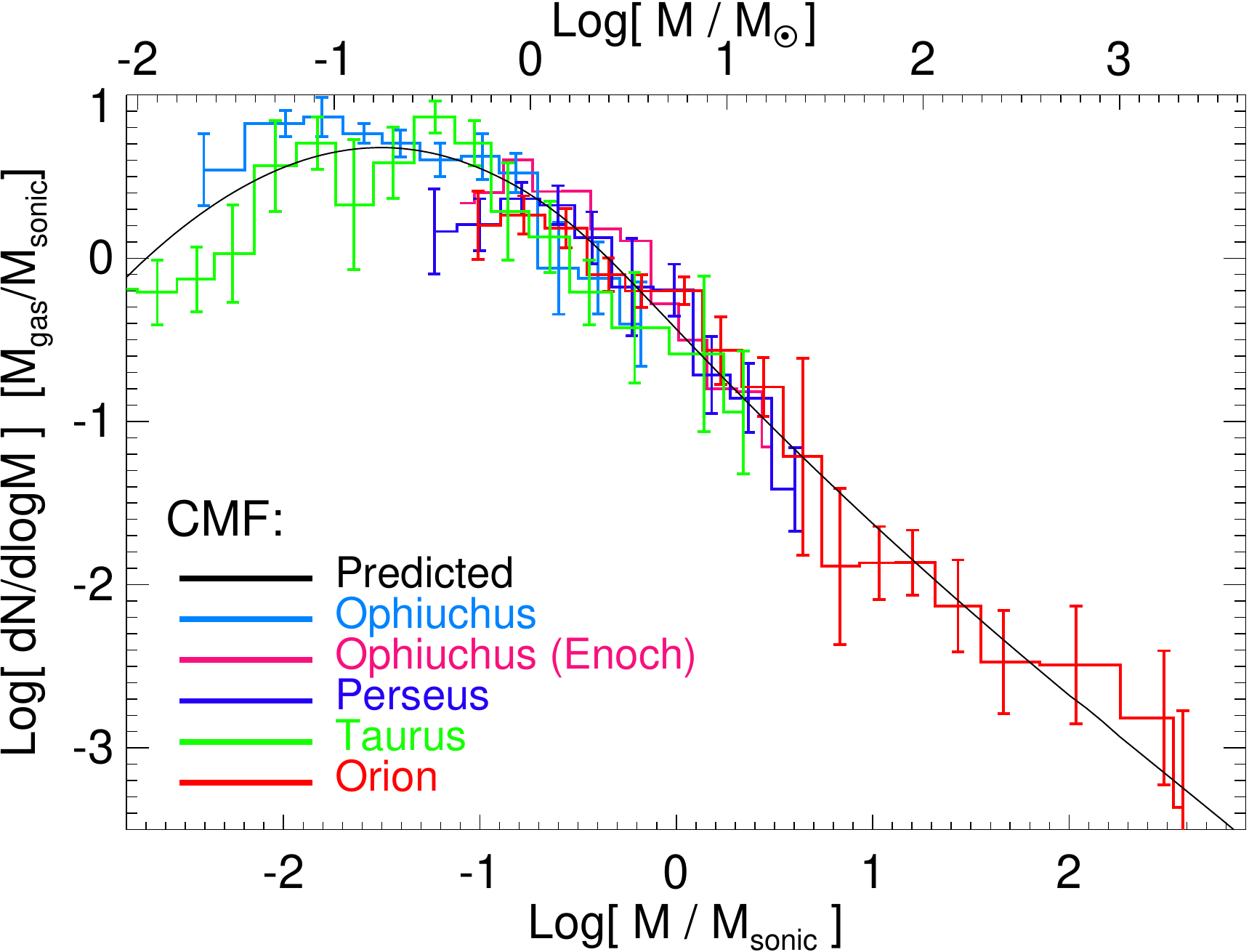}
\end{minipage}
\begin{minipage}[b]{0.04\hsize}
\hfill
\end{minipage}
\begin{minipage}[b]{0.45\hsize}
\includegraphics[width=\hsize]{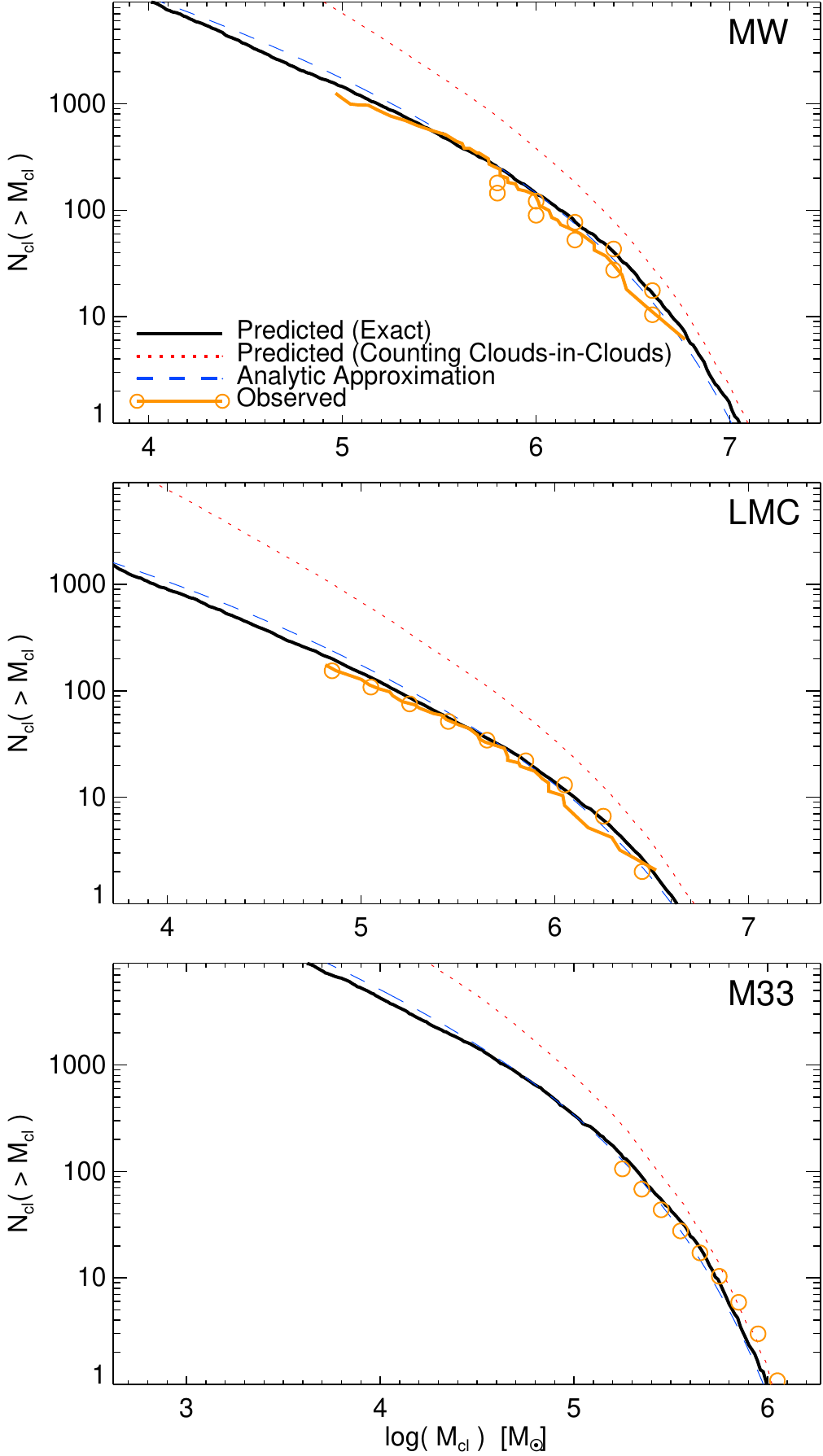}%
\end{minipage}
\end{figure}

Excursion Set formalism makes predictions that are computable “analytically” and match a large variety of observations unexpectedly well (see figure \ref{fig:esf}). The final verdict on this novel approach is still pending, with opinions ranging from "it should never work" to "it solves all the problems". So, if you are bold enough, make your bet...

\abstract*{TBWL}

\def\Ssfr{\Sigma_{\rm SFR}}
\def\Sgas{\Sigma_{\rm gas}}
\def\Sntr{\Sigma_{\HI+\H2}}
\def\Sh2{\Sigma_\H2}
\def\Shi{\Sigma_\HI}
\def\Shii{\Sigma_\HII}
\def\tsf{\tau_{\rm SF}}
\def\rhomol{\rho_{\rm mol}}

\section{Stellar Feedback}
\label{sfb}

Stars affect their environments by their \emph{feedback} - anyone reading these lines knows that well, without the stellar feedback we would not even exist (as there would not exist planets made out of heavy elements).

\subsection{What Escapes from Stars}

\begin{figure}[b]
\sidecaption[t]
\includegraphics[width=0.64\hsize]{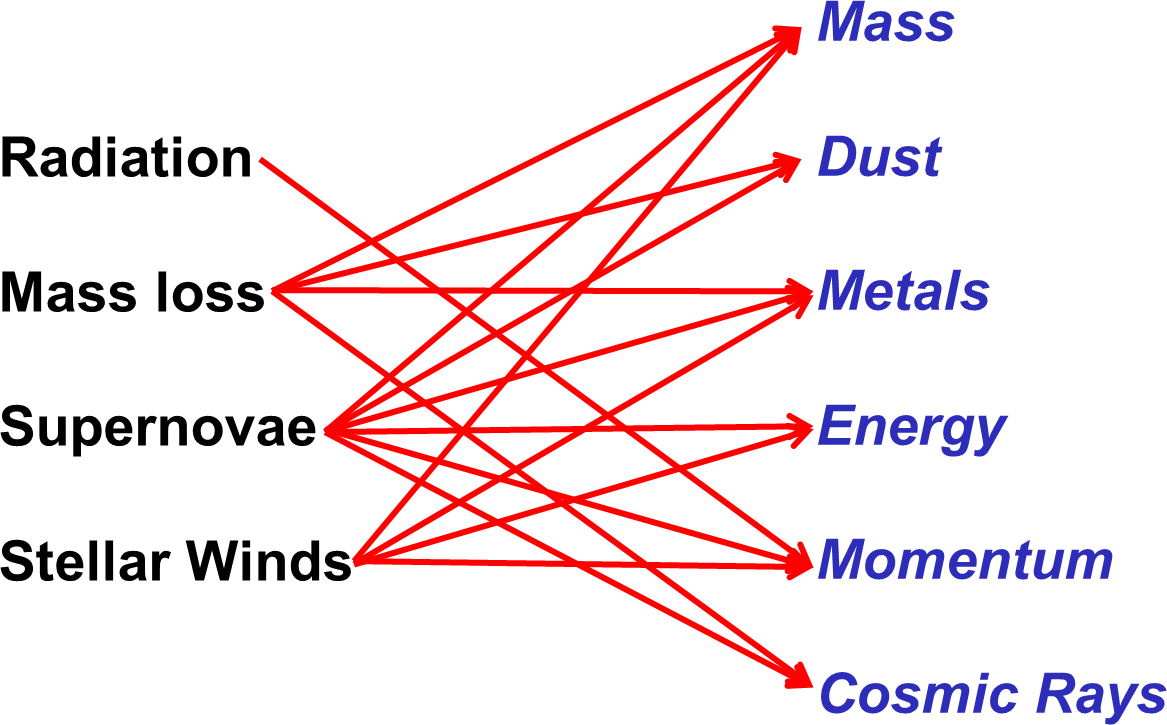}%
\caption{Various pathways for stellar feedback.\label{fig:map}}
\end{figure}

However, stellar feedback is not just supernovae (sometimes that's the impression one gets by reading simulation papers). Stars affect their environments in several ways: \emph{supernovae} (both type II and type Ia), \emph{stellar winds} from massive stars, \emph{mass loss} from AGB and planetary nebulae, and, of course, \emph{radiation}. Each of these modes inject into surrounding gas  \emph{energy}, \emph{momentum}, \emph{mass}, \emph{metals}, \emph{dust}, and \emph{cosmic rays}. The various pathways the inputs and outpust are connected are shown in figure \ref{fig:map}. If you think the feedback is complicated, then you are right!

\begin{figure}[t]
\sidecaption[t]
\includegraphics[width=0.64\hsize]{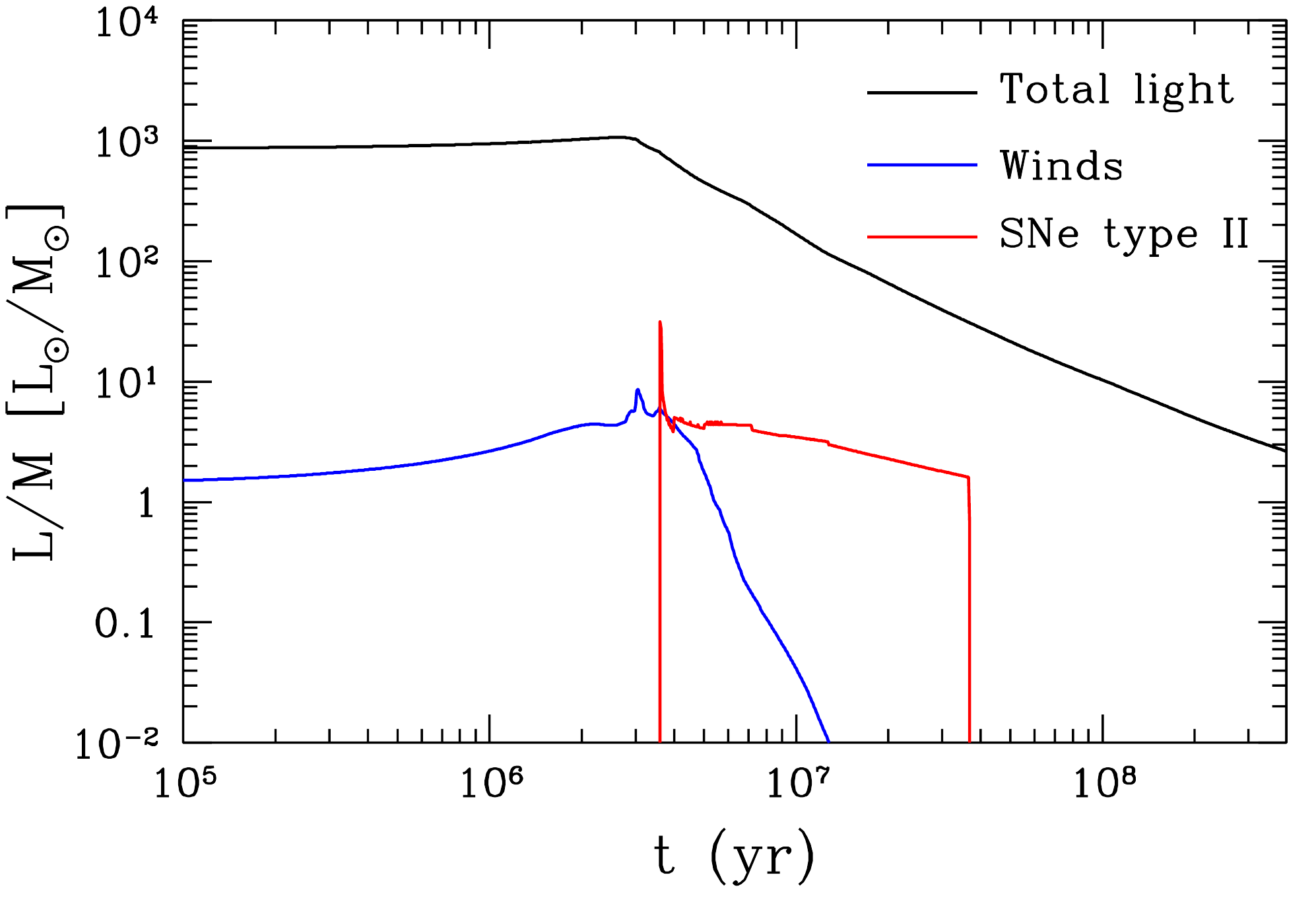}%
\caption{Energy injection rate (= luminosity) per unit mass for the total radiation, winds, and supernovae as a function of time for a normal stellar population at solar metallicity.\label{fig:loft}}
\end{figure}

It is easy to get lost in this maze of feedback pathways. But one important fact should light our way (literally) - by far the largest energy output of stars is light! Just as an illustration, I show in figure \ref{fig:loft} the energy production rate as a function of time for a normal stellar population at solar metallicity. The bolometric luminosity of stars dwarfs all other feedback channels at all times. And we know that at least half of that energy is re-radiated in the infrared by dust, hence a substantial fraction of stellar light is indeed absorbed by the surrounding gas. We should, therefore, consider that feedback channel very seriously.

\subsubsection{Radiation Pressure}

\begin{figure}[t]
\sidecaption[t]
\includegraphics[width=0.64\hsize]{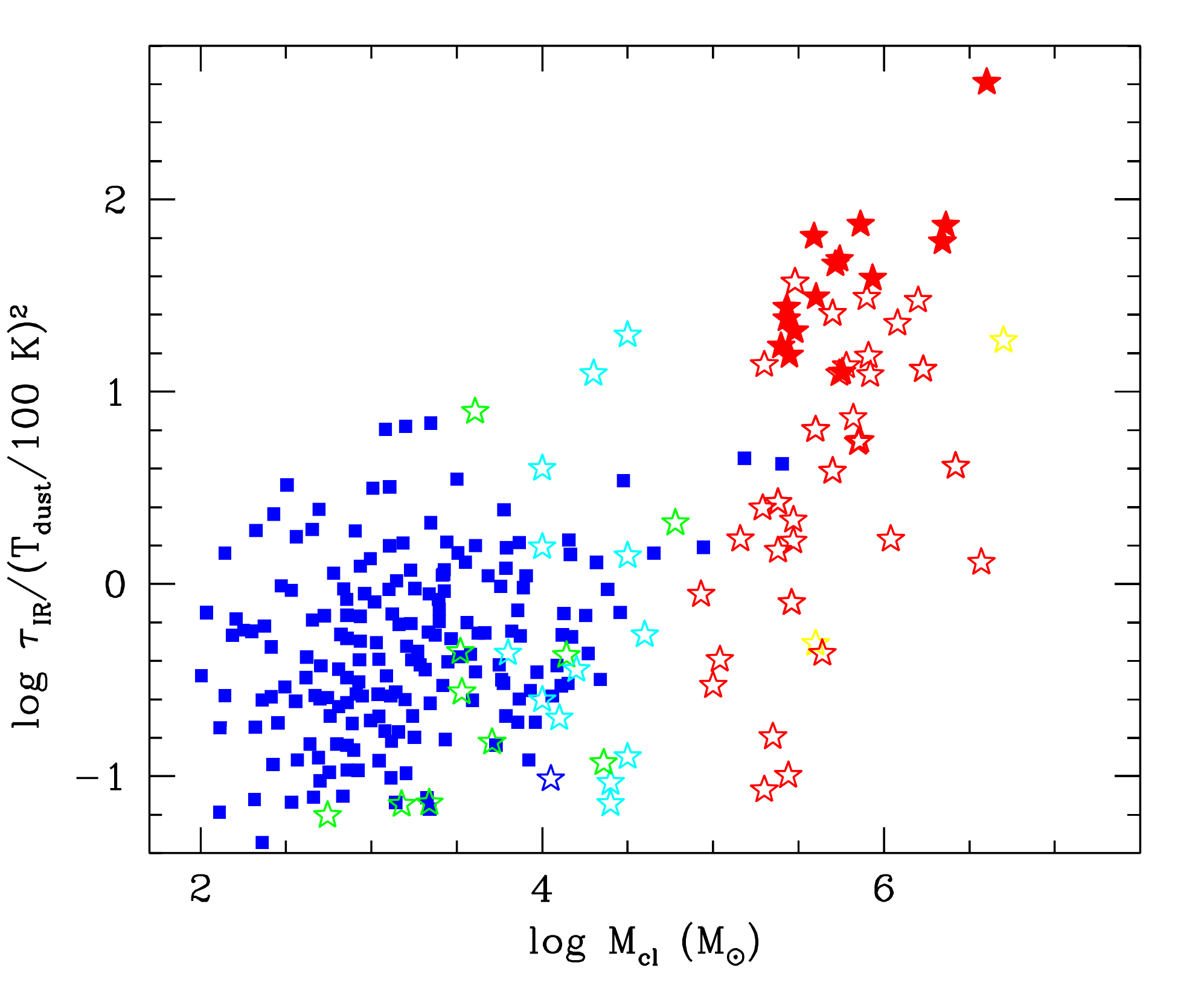}%
\caption{Observational estimates of $\tau_{\rm IR}$ for GMC clumps (blue squares) and young stellar clusters (stars). Adopted from \protect\citet{ng:akl13}.\label{fig:tauobs}}
\end{figure}

Massive (hence, young) stars spend a substantial fraction of their lives heavily embedded into the surrounding gas and dust; for heavily obscured stars most of their light is absorbed. Since photons have momentum, absorbing all light from a star/star cluster of luminosity $L$ injects momentum into the surrounding gas,
\[
  \dot{p}_1 = \frac{L}{c}.
\]
The energy, however, is conserved - the absorbed bolometric luminosity of the star must be re-emitted by dust in the infrared. If there is enough dust around a young massive star (and, at least in the Milky Way, there is), the dust will be optically thick to its own IR radiation. That radiation will do work on the surrounding gas, i.e.\ will inject extra momentum, so that the total momentum injection rate is
\begin{equation}
  \dot{p}_{\rm tot} = \left(1-f_{\rm esc}+\tau_{\rm IR}\right)\frac{L}{c},
  \label{eq:radp}
\end{equation}
where, in order to be completely general, I included the fraction $f_{\rm esc}$ of stellar radiation (of all frequencies) that escapes the star forming region. The new factor $\tau_{\rm IR}$ is easy to derive for a homogeneous medium \citep{goc95}. Since energy is conserved, the radiation flux at each radius $R$ from the star is still
\[
  F_R = \frac{L}{4\pi R^2}.
\]
Hence, the momentum (in the radial direction) imparted on the gas between $R$ and $R+dR$ is simply 
\[
  d\dot{p}_{\rm IR} = 4\pi R^2 \frac{F_R}{c} \kappa dR = \frac{L}{c}d\tau,
\]
and, hence,
\[
  \dot{p}_{\rm IR} = \tau_{\rm IR} \frac{L}{c}.
\]
In the infrared dust opacity is
\[
  \kappa_{\rm IR}\approx 3\frac{\mbox{cm}^2}{\mbox{g}}\left(\displaystyle\frac{T_d}{100\mbox{K}}\right)^2
\]
\citep{shh03}. Observational estimates of $\tau_{\rm IR}$ at $T_d=100\dim{K}$ are shown in figure \ref{fig:tauobs}; radiation pressure is particularly important for large stellar clusters.

\subsubsection{All Feedback, All the Time}

\begin{figure}[t]
\includegraphics[width=0.5\hsize]{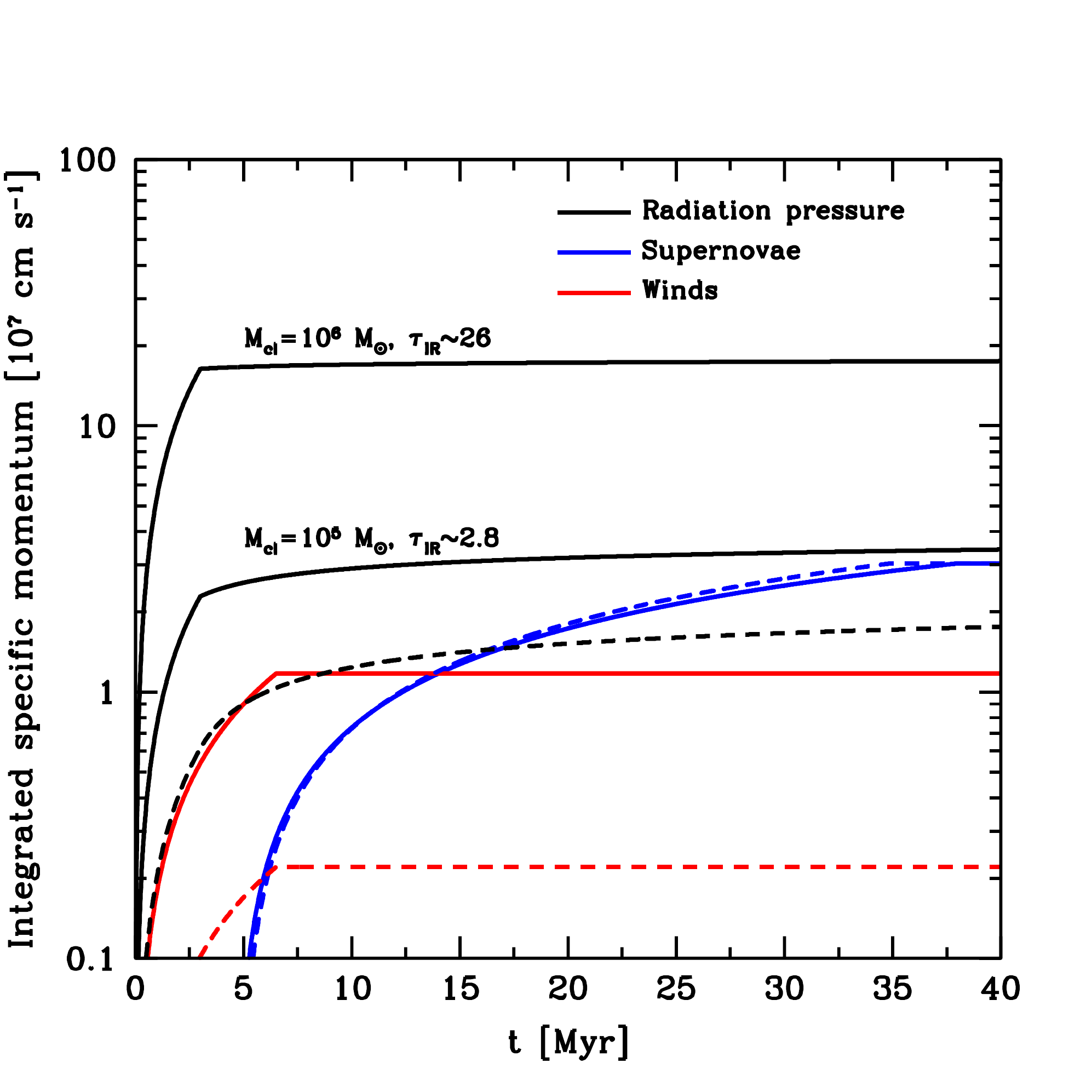}%
\includegraphics[width=0.5\hsize]{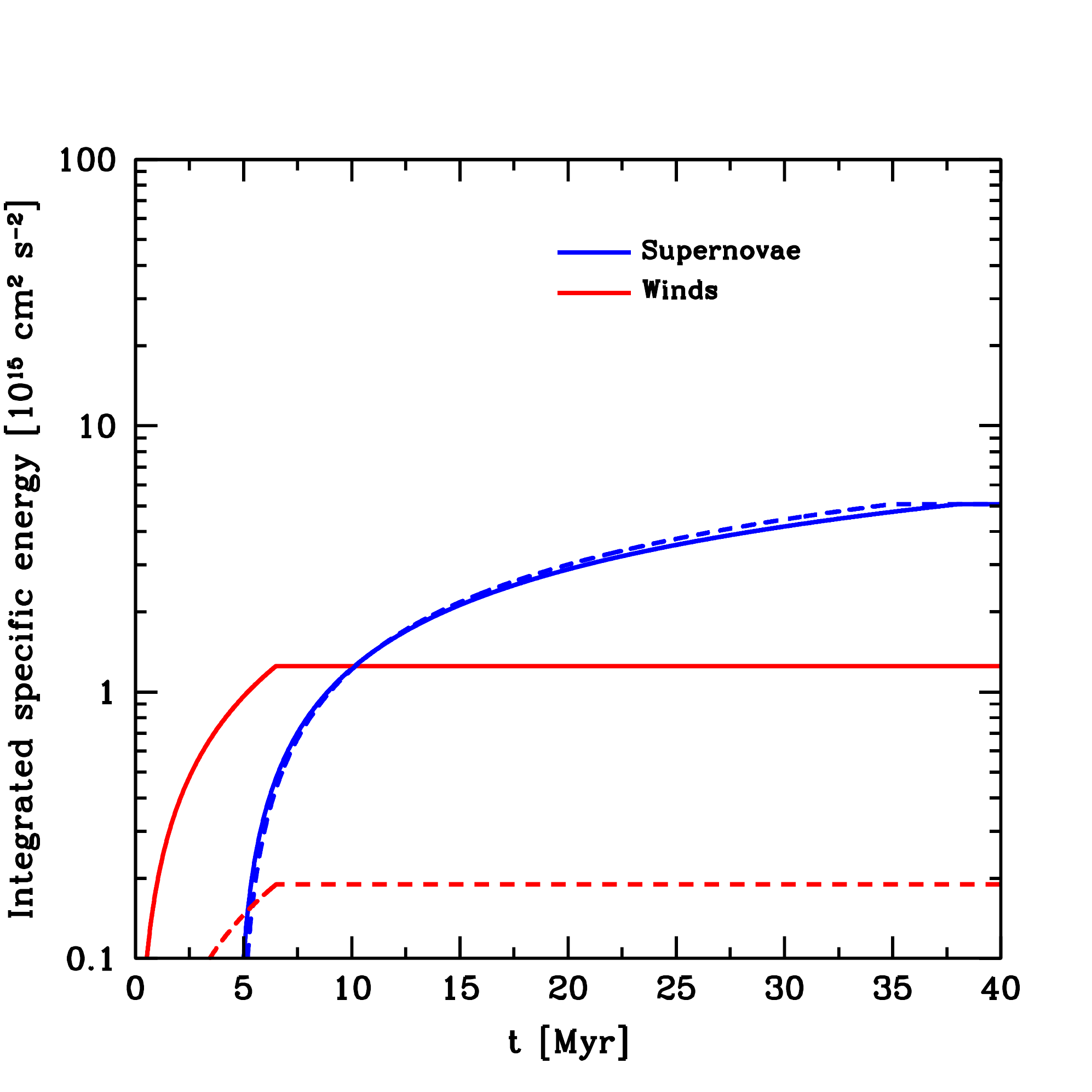}%
\caption{Time evolution of the cumulative injected specific momentum (left) and specific energy (right)  for various feedback channels at solar metallicity (solid lines) and at $Z=0.01\Zsun$ (dashed lines). Notice that supernovae kick in after all other feedback channels already fired. Adopted from \protect\citet{ng:akl13}.\label{fig:spec}}
\end{figure}

Stars, of course, do not have a freedom to selectively fire only some of their feedback channels, they all work all the time. In figure \ref{fig:spec} the time evolution of the momentum and energy injection is shown for several dominant feedback modes. Supernovae form the last episode in the feedback fireworks - by the time they start in earnest (after about $10\dim{Myr}$ after the onset of star formation) all other feedback channels have already finished.

In fact, with all likelihood, supernovae are not important to actually destroying molecular clouds \citep[and, hence, controlling the efficiency of star formation,][]{fkm10}. They may be important for heating the overall ISM, for stabilizing the disk, for driving galactic winds, but star formation they control not.

\begin{figure}[b]
\sidecaption[t]
\includegraphics[width=0.64\hsize]{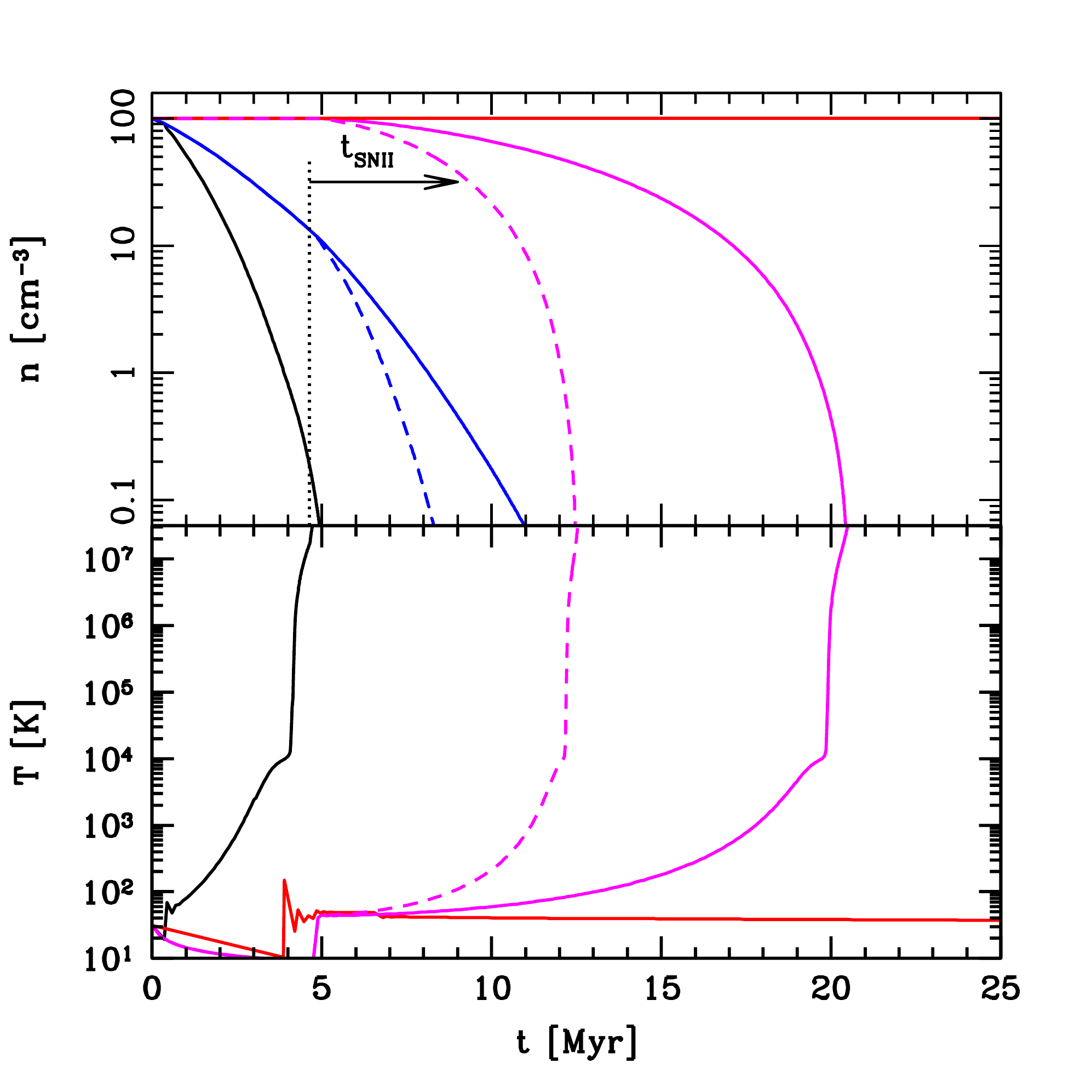}%
\caption{Density and temperature in a computational cell evolving in response to various feedback channels:  only energy from supernovae (red), energy and momentum from supernovae (magenta), only momentum feedback from all channels (blue), and all feedback channels acting together (black). Adopted from \protect\citet{ng:akl13}.\label{fig:evac}}
\end{figure}

The relative role of different feedback channels can be understood even better with a simple numerical exercise - a single computational cell "simulation". Figure \ref{fig:evac} shows the fate of a such a cell when various feedback channels are switched on and off.  

The first important lesson is that purely thermal feedback - injecting all of the supernova energy as thermal energy into the parent cell (or particle in case of SPH) of even a large stellar cluster does not do anything, the cooling times are always so short that the thermal energy is quickly radiated away. This is not a new result, it has been known since the dawn of numerical galaxy formation, and re-discovered independently by many research groups; but it does pose a dilemma for cosmological and even galactic-scale simulations - the only direct way of implementing stellar feedback does not work, and one has to use a \emph{sub-grid model}, i.e.\ a specific recipe about how to implement the feedback in a numerical code.

Figure \ref{fig:evac} may offer a clue how such a sub-grid model may be implemented: other feedback channels produce a large effect on the dynamics of a single cell, and, hence, may have a significant effect on the dynamics of larger scales as well. There is just one problem with that approach - actual stellar feedback like equation (\ref{eq:radp}) is operating on scales of molecular cores and their very vicinities, on sub-parsec scales. Whenever we use, say, the radiation pressure formula in cosmological (or even galactic-scale) simulations, we are injecting the momentum on scales of many tens, even hundreds of parsecs, well beyond the range of scales where it is actually operating. Hence, using  equation (\ref{eq:radp}) in a cosmological code is also a \emph{sub-grid model}, an ansatz that is a priori as good or bad as any other sub-grid model. Not surprising, then, that the radiation pressure is gradually falling out of fashion.

Before we trash all sub-grid models or pick one of them and place it on the throne, it is worth taking a step back and re-thinking what we are actually trying to achieve.

\subsection{Unconventional Marriage: Feedback and Star Formation}

If you did not skip the previous chapter, my dear reader, then you know that star formation is \emph{inefficient} - the depletion time $\tsf$ is of the order of a Gyr (give or take a factor of 2-3), while molecular clouds are short-lived ($10-20\dim{Myr}$). During their lifetimes molecular clouds convert only a small fraction, mere percents, of their gas into stars. A natural conclusion from that fact is, since star formation is inefficient, then so must be the feedback. And that conclusion is utterly wrong!

Do you recall a simulation of a Milky Way like galaxy that I used to illustrate the properties of the gaseous halo (figure \ref{fig:halo})? Have you wondered why I never showed you the circular velocity profile for that galaxy? There is a good reason I have not - I am ashamed to! While the gaseous halo for that galaxy may look ok, the disk is totally wrong, it has an extremely dense spike at the center, with the circular velocity peaking at $450\dim{km/s}$, more than twice the rotation velocity of the Milky Way. The reason for such a huge discrepancy is the absence of any feedback process in the simulation.

If we learned anything after 20+ years of modeling galaxy formation, then it is that the central spikes in circular velocity (caused by unrealistic central mass concentrations) can only be destroyed by strong feedback. No other physics can do the trick - in fact, as simulations grew more sophisticated, included more physics, and reached higher resolution, the central mass concentration problem became worse. It is a real, physical problem, not a numerical one - the high-redshift progenitors of normal galaxies are too dense, and these early dense gaseous concentrations survive all the subsequent adventures of galaxy evolution; if not blown out, they will become large stellar bulges. 

Indeed, that was commonly occurring in simulations until only a few years ago - for example, check out beautiful pictures of center-heavy galaxies in \citet{sbc10}. At the same time, as observers figured out the difference between the real bulges and pseudo-bulges (central features formed by secular evolution from barred disks), they realized that a significant fraction, perhaps as much as 50\%, of galaxies are actually \emph{bulgeless}, pure disks.

All these examples illustrate one crucially important conclusion about star formation and feedback - while start formation is \emph{inefficient}, the feedback is actually \emph{strong}. These two facts may be deeply connected, but we are not going to dive into the connection between star formation and feedback, for our purpose what is important is this apparent dichotomy in behavior.

\begin{figure}[t]
\sidecaption[t]
\includegraphics[width=0.64\hsize]{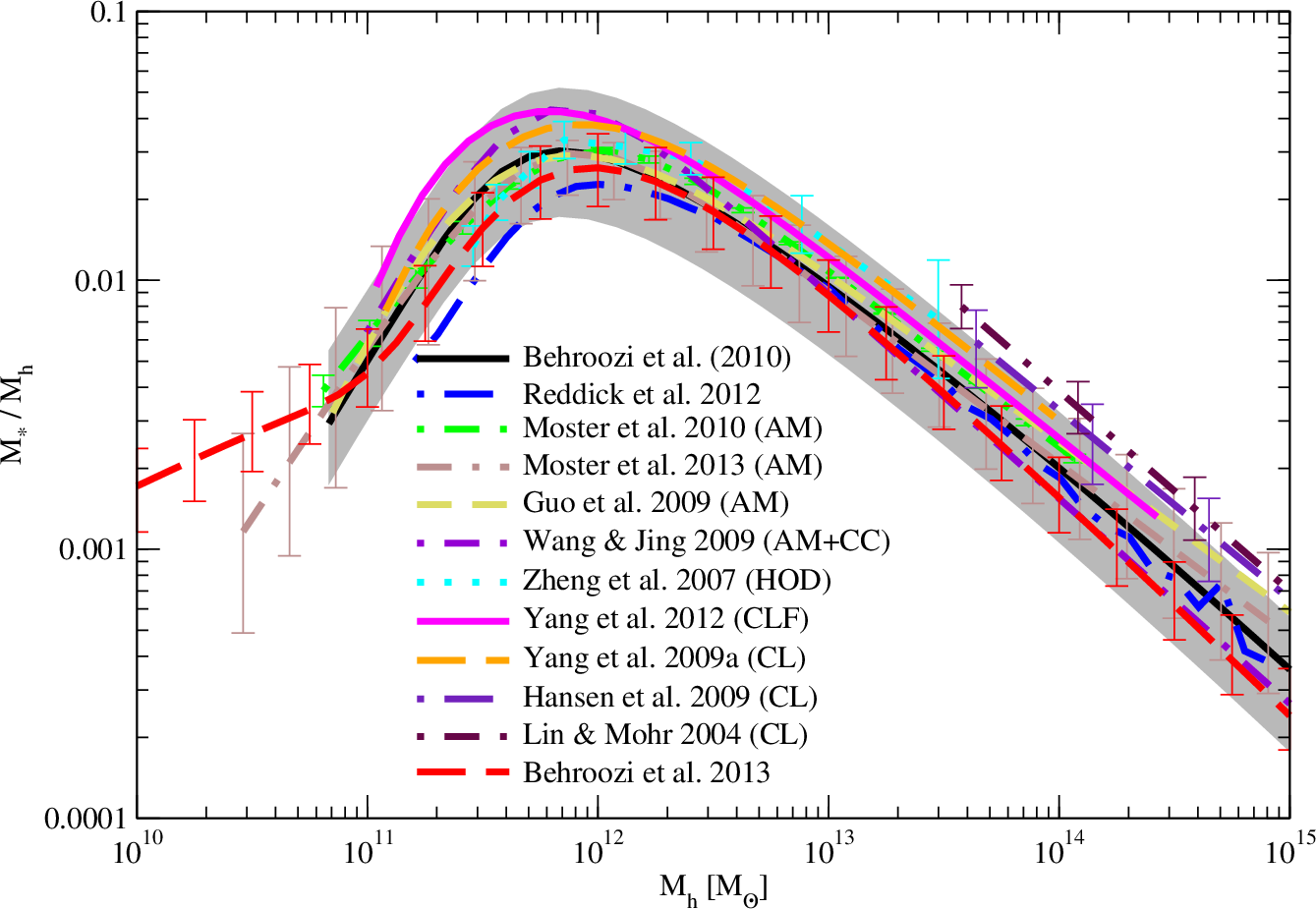}%
\caption{Abundance matching in action: stellar vs total mass for dark matter halos from several independent groups; they all agree that star formation is inefficient. Adopted from \protect\citet{bwc13}.\label{fig:msmt}}
\end{figure}

A good illustration of that dichotomy comes from the so-called "abundance matching" exercise - a match between the \emph{observationally derived}\footnote{Never forget that stellar masses are \emph{not} observed, they are always derived from observations of luminosity functions, with all the inherent in spectral synthesis uncertainties and biases.} stellar mass and the theoretically known mass function of dark matter halos. Such a match results in a one-to-one correspondence between the stellar mass and the halo mass for individual halos (or, in a more complex implementation of the abundance matching idea, a distribution of stellar masses for a given halo mass). Figure \ref{fig:msmt} shows a comparison (and uncertainty) of the stellar mass - halo mass relation for several independent applications of that approach from \citet{bwc13}.

The most important result of the abundance matching exercise is that stellar masses of low mass halos are very small, roughly $M_*\propto M_h^{2.5}$ for $M_h\ll10^{12}\Msun$. To get such a behavior, it is not enough to make star formation inefficient - that would still result in stellar mass being proportional to halo mass, the inefficiency would only make the coefficient of proportionality small - but it also requires the feedback to be progressively more efficient in lower mass galaxies to sculpt the inferred $M_*\propto M_h^{2.5}$ relation.

Attempts to model the feedback as a sub-grid model  are as old as the galaxy formation simulations themselves. It is not too instructive to review all of them, as until 2010 none of the sub-grid models were particularly successful. The important thing to remember about any sub-grid model is that it would only work over a finite range of spatial scales. If the model is good, that range would be sufficiently large (say, a decade in spatial scale); if the model is bad, the range may be zero. Even if the model is good, but its range of validity does not match the resolution of simulations, then it would not work well. 

Indeed, that is what have happened with one simple sub-grid feedback model. In 1997 in his PhD thesis, Jeroen Gerritsen proposed a simple way to make feedback strong - simply to disable cooling in star forming regions for several tens of Myr (we now call this method "delayed cooling"). The model did not work too well with the spatial resolution simulations were able to reach in 1997. However, miracles do happen - as the resolution improved, the delayed cooling model appeared to work better and better, until, finally, in 2010 it was declared to be a panacea for galaxy formation \citep{gbm10}!

\begin{figure}[t]
\includegraphics[width=0.67\hsize]{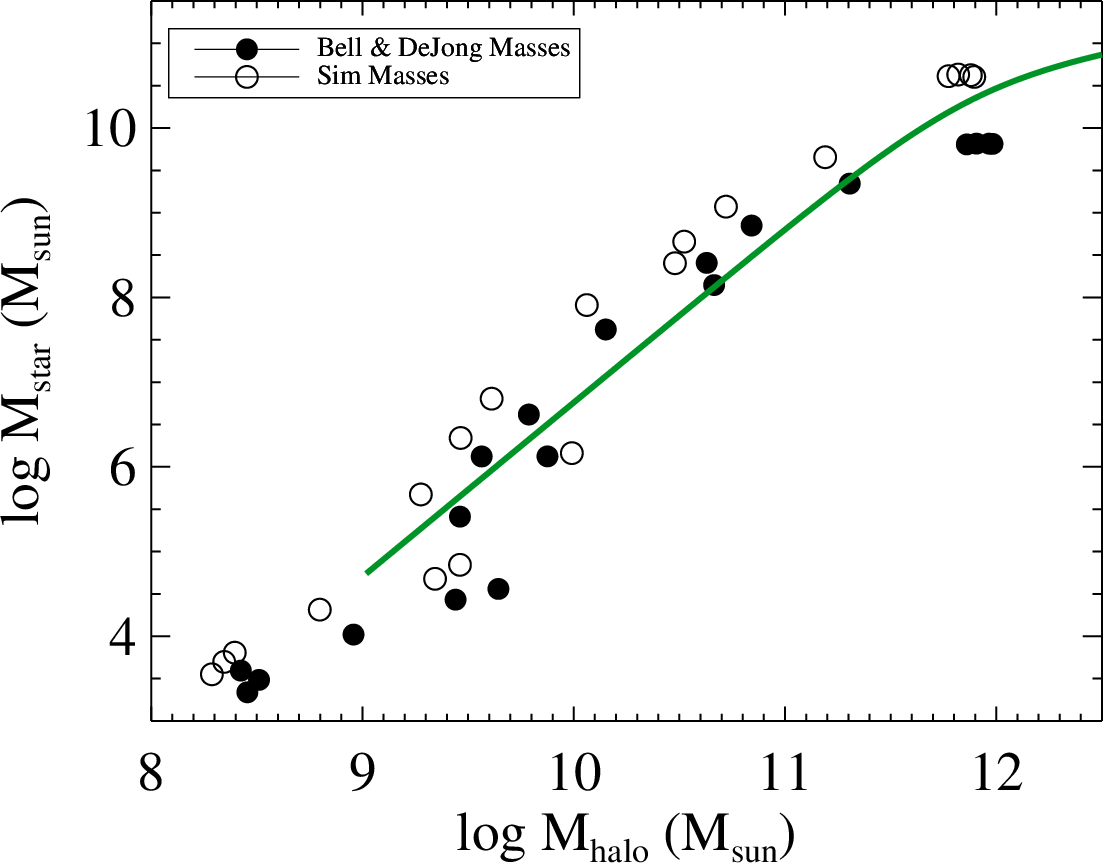}~~%
\includegraphics[width=0.315\hsize]{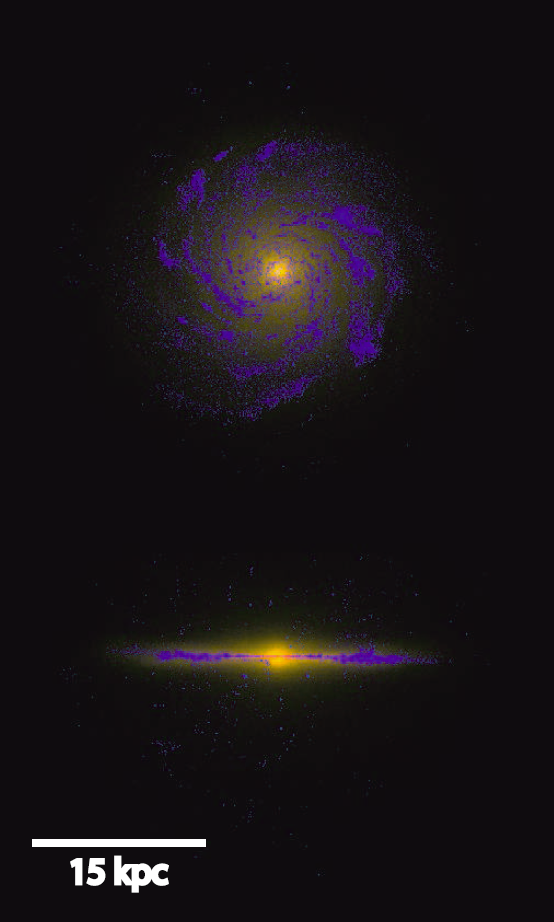}%
\caption{\emph{Left:} stellar versus total mass from abundance matching (green line) and modern galaxy formation simulations (open and filled circles - adopted from \protect\citet{mgb13}). \emph{Right:} face- and edge-on projections of Eris simulation of the Milky Way galaxy (adopted from \protect\citet{gcm11}).\label{fig:simmsmt}}
\end{figure}

Figure \ref{fig:simmsmt} gives two examples of how well modern simulations with delayed cooling feedback reproduce observations, but similarly impressive examples for various observational constraints are abound.

\subsubsection{Why Delayed Cooling Works}

While it is easy to declare the delayed cooling a success, it is much harder to understand what it actually means. As such, it is just a numerical trick, without any serious physical justification. The fact that it works may be a pure coincidence; alternatively, it can be a manifestation of a real physical process that operates on sub-parsec scales, but its consequences on $\sim100\dim{pc}$ scales appear as if cooling was switched off.  In fact, it is easy to come up with several real physical processes that will all manifest themselves as delayed cooling on large scales:
\begin{itemize}
\item radiation pressure from massive stars (we now know it is important) provides support for gas that "does not cool", i.e. if treated as an effective additional pressure, that pressure would not be affected by the cooling processes in the gas, but will diminish after about $10\dim{Myr}$;
\item coronal gas - the hot, million-degree gas produced in supernova explosions may accumulate in regions of low density in a supersonically turbulent IGM; cooling times in such gas will depend on its density, but generally will be of the order of several to several tens of Myr;
\item as stellar feedback continue to stir supersonic turbulence in molecular clouds on small scales, the energy of the kinetic motions will accumulate to the point at which the dissipation rate will approximately equal the production rate; while the dissipation time-scale is likely to be short, the supersonic turbulence (i.e.\ highly super-thermal additional pressure in the gas) will be maintained for the duration of stellar feedback, several tens of Myr;
\item cosmic rays produced in supernova explosions are observationally known to provide significant additional support in the magnetized molecular clouds; cosmic rays diffuse out of GMC on time-scales of tens of Myr.
\end{itemize}
I am sure that list can be easily extended, but it already serves our purposed well - numerous real small-scale physical processes may hide themselves under the large-scale mask of "delayed cooling", and one, several, all of them, or different combinations of them in different environments may be the actual feedback process(es) that is/are responsible for making the real galaxies as they are...

\subsection{Toward The Future}

\begin{figure}[t]
\includegraphics[width=1\hsize]{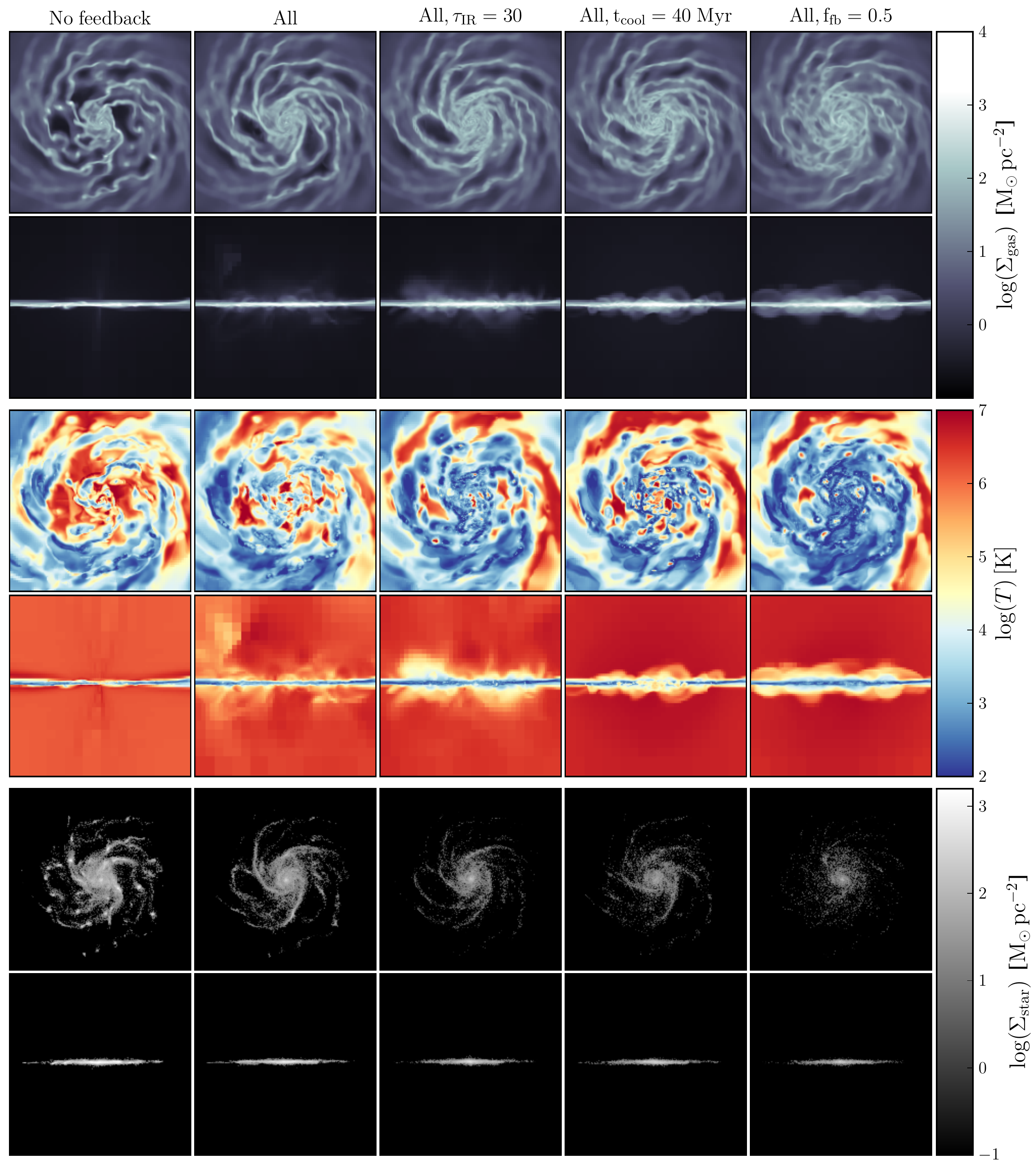}%
\caption{Face-on and edge-on maps of simulated galactic disks with different small-scale feedback models. Separate rows show gas surface density (top), mass weighted average gas temperature (middle), and stellar surface density (bottom).  Columns from left to right are no feedback,  all feedback channels from figure \ref{fig:map}, all feedback with extra radiation pressure, delayed cooling feedback, feedback model with extra energy variable
(adopted from \protect\citet{sims:aklg13}).\label{fig:maps}}
\end{figure}

So, where do we go from there? If only we could figure out which of the actual feedback channels hides behind the mask of delayed cooling, the galaxy formation (of normal disk galaxies - the AGN feedback is entirely different story) will be essentially solved (well, hopefully you do not take me as being too optimistic). 

These feedback processes are more-or-less understood, as, hopefully, I persuaded you in the beginning of this chapter. Actually modeling them in the cosmological and galactic-scale simulations is not trivial, but big strides in that directions have been already made. We may still argue occasionally how to do it better, or what the most appropriate value for, say, $\tau_{\rm IR}$ should be, but the importance of the modeling feedback correctly is not a subject of debate any more. 

The good piece of news is that even if various feedback channels are tuned to match the basic observational constraints like the stellar mass vs total halo mass, Kennicutt-Schmidt relation, rotational velocity curves, etc, simulated galaxies in runs with different feedback channels still look amazingly different (figure \ref{fig:maps}), and there lies the key to the eventual success.

Hence, the plan for the future is to identify the best observational probes that will help us understand which of all of the potential feedback channels are important in which environments and on what spatial scales. This is left as an exercise for the reader...

\appendix

\section{Answers To Brain Teasers}
\label{bt}

\begin{enumerate}
\item Sound waves indeed do not dissipate. However, they also do not grow with time, since they are stable perturbations, while large-scale, unstable perturbations in both dark matter and gas grow. Hence, relative to the large-scale perturbations, the small-scale sound waves become smaller and smaller, i.e.\ they appear to be "suppressed".
\item The proper term for "Lyman-$\alpha$ absorption" is \emph{resonant scattering}. A Lyman-$\alpha$ photon is re-emitted by the atom, but in the meantime that atom experienced a large number of collisions with other atoms and ions, so its momentum is now unrelated to the momentum it had at the moment of absorption. Hence, the re-emitted Lyman-$\alpha$ photon will be send out into a random direction in the frame of the atom, and will not reach our telescope. For us, that photon is lost, hence we, sometimes, call it absorption.
\item The term "equation of state" relates the perturbations in the gas pressure (or temperature) to those of the density. If we impose (adiabatically) a perturbation $\delta\rho$ to the gas density, the instantaneous response to the pressure will be identical to the ideal gas,  $\delta P = c_S^2 \delta\rho$. Only with time adiabatic expansion and photoheating will bring that perturbation back to the temperature-density relation.
\item A typical ionizing photon is not sitting at the Lyman edge, it has the energy of $E_0+\langle\Delta E\rangle$ (see equation \ref{eq:deltae}), which is about $40-50\dim{eV}$ for the cosmic background. The ionizing cross-section falls off with energy as $E^{-3}$, hence the typical cross-section is $\sim(1-2)\times10^{-19}\dim{cm}^2$ instead of $6.3\times10^{-18}\dim{cm}^2$. In addition, the typical ionization level in the forest is $10^{-5}$, which requires $\tau=\ln(10^5)\approx10$ to neutralize. Hence, hydrogen absorbers only become fully neutral at column densities of $N_H\sim(0.5-1)\times10^{20}\dim{cm}^{-2}$. 
\item This one is really tricky. In fact, I do not know the full answer to it. One possible reason why Lyman-$\alpha$ forest is not turbulent was suggested to me by Andrea Ferrara: for turbulence to develop, the gas needs to have vorticity, but in the linear regime vorticity in cosmic gas decays, so there should be no vorticity at $\delta\approx0$ in the forest. Non-linear evolution will generate some vorticity, but since most of the forest is not extremely non-linear, it is plausible that the vorticity generated in the forest may not be enough to create a full turbulent cascade.
\end{enumerate}

\bibliographystyle{apj}
\bibliography{igm,cgm,ism,sf,sfb,ng-bibs/self,ng-bibs/ism,ng-bibs/igm,ng-bibs/gals,ng-bibs/sims,ng-bibs/sfr,ng-bibs/misc}

\end{document}